\documentclass[12pt]{article}
\usepackage{epsfig}
\usepackage{amssymb}

\newcommand{\sect}[1]{\setcounter{equation}{0}\section{#1}}
\renewcommand{\theequation}{\arabic{section}.\arabic{equation}}

\def\be{\begin{equation}}
\def\ee{\end{equation}}
\def\ba{\begin{eqnarray}}
\def\ea{\end{eqnarray}}

\title{\vspace{-1in}
\parbox{\linewidth}
{\small\hfill HUTP-01/A041}\\
{\small\hfill ITEP-TH-47/01}\\
{\small\hfill DAMTP-2001-77}\\
{\small\hfill hep-th/0109025}\\
\vspace{0.6in}
{\bf M-Theory on Spin(7) Manifolds}}
\author{\textsc{Sergei Gukov}\thanks{email:
gukov@democritus.harvard.edu}\\
\emph{Jefferson Physical Laboratory, Harvard University}\\
\emph{Cambridge, MA 02138, USA}\and
\textsc{James Sparks}\thanks{email: J.F.Sparks@damtp.cam.ac.uk}\\
\emph{Centre for Mathematical Sciences} \\
\emph{University of Cambridge} \\
\emph{Wilberforce Road, Cambridge CB3 0WA, UK.}}

\begin{document}
\pagestyle{plain}
\setcounter{page}{1}
\newcounter{bean}
\baselineskip16pt

\maketitle

\begin{abstract}


We study M-theory on two classes of manifolds of Spin(7) holonomy that are developing an isolated conical singularity. We construct explicitly a new class of $\mathrm{Spin}(7)$ manifolds and analyse in detail the topology of the corresponding classical spacetimes. We discover also an intricate interplay between various anomalies in M-theory, string theory, and gauge theory within these models, and in particular find a connection between half-integral $G$-fluxes in M-theory and Chern-Simons terms of the $\mathcal{N}=1$, $D=3$ effective theory.

\end{abstract}

\vspace{0.5in}

\newpage

\tableofcontents

\sect{Introduction and Summary}

Recently, M-theory compactifications on manifolds
of exceptional holonomy have attracted considerable attention. These models
allow one to geometrically engineer various minimally supersymmetric gauge
theories, which typically have a rich dynamical structure. A particularly
interesting aspect of such models is
the behaviour near a classical singularity, where
one might expect extra massless degrees of freedom,
enhancement of gauge symmetry, or a phase transition
to a different theory.

In the case of $G2$ compactifications one obtains an
$\mathcal{N}=1$ supersymmetric field theory
in four dimensions, where certain properties of
the IR quantum theory can be obtained from
non-renormalization theorems, holomorphy, and R-symmetry.
Using holomorphy constraints,
Atiyah, Maldacena, and Vafa \cite{AMV} argued that
in the quantum theory one can smoothly
interpolate between certain spacetime manifolds
of $G2$ holonomy which have three classical limits.
Each of these classical limits can be understood as
an M-theory lift of Type IIA string theory on a deformed/resolved
conifold with D6-branes/RR-flux in the background.
Therefore, the smooth geometric transition found in M-theory implies
that in Type IIA one has a continuous transition from a vacuum
with D6-branes to another vacuum where the branes have disappeared and
have been replaced with RR-flux.

More evidence in favour of a smooth transition in
this model was presented in the recent work of
Atiyah and Witten \cite{AW}, where M-theory dynamics
on other known $G2$ holonomy manifolds was also discussed.
Specifically, these are resolutions of the cones on ${\bf \mathbb{C}P}^3$
and on $SU(3)/U(1)^2$. In both cases one has a collapsing four-cycle
(${\bf S}^4$ and ${\bf \mathbb{C}P}^2$, respectively) in the limit that the $G2$
manifold develops a conical singularity. Unlike the model considered in \cite{AMV}, these
manifolds do not have an interpretation as M-theory
lifts of D6-branes wrapped on non-compact,
topologically non-trivial Calabi-Yau manifolds.
However, the dynamics of these models can be obtained from
a different reduction to Type IIA theory with D6-branes
in a (topologically) flat spacetime \cite{AW}.
Via this reduction, the problem of studying
M-theory on a $G2$ manifold developing a conical singularity
can be translated into the simpler problem of studying
configurations of intersecting D6-branes in flat spacetime.
In particular, for the cone on ${\bf \mathbb{C}P}^3$
one finds restoration of a global $U(1)$ symmetry
at the conifold point, whereas for the $SU(3)/U(1)^2$ model
there are three different branches related by a ``triality'' symmetry.
In fact, there is a close relation between the $\mathrm{Spin}(7)$
examples in the current paper and the $G2$ models of \cite{AW}.

In the case of $\mathrm{Spin}(7)$ compactifications there are fewer
constraints from supersymmetry. Namely, compactification of
M-theory on a $\mathrm{Spin}(7)$ manifold gives $\mathcal{N}=1$
supersymmetric field theory in three dimensions. This theory cannot
be obtained via dimensional reduction from four dimensions.
One might hope to use this fact to explain the vanishing of
the cosmological constant in four dimensions in the absence
of supersymmetry, along the lines of \cite{wittensusy,wittencc}.
Scalar fields in $\mathcal{N}=1$ multiplets are real in
three dimensions. So, there is no holomorphy, and in general
one would not expect smooth transitions between different
branches similar to the phase transitions in the $G2$
case \cite{AMV,AganagicV,AW}.
Moreover, there are no non-renormalization theorems
and no R-symmetries in $\mathcal{N}=1$ three-dimensional theories.
However, certain constraints may be obtained from
the discrete parity symmetry \cite{AHW,Gremm,BKS}:
\be
P \colon (x^0, x^1, x^2) \to (x^0, - x^1, x^2)
\label{parityP}
\ee
For example, in a parity-invariant theory the superpotential
is odd under this transformation \cite{AHW}.
Important questions, such as dynamical supersymmetry breaking
in $\mathcal{N}=1$ three-dimensional theories, may also be
addressed by studying the supersymmetric index \cite{wittengauge}
and supergravity duals of these theories \cite{MN}.

Motivated by \cite{AMV,AW}, in this paper we study M-theory
dynamics on manifolds of $\mathrm{Spin}(7)$ holonomy which are developing
an isolated conical singularity. Until recently, only one example
of this type was known, corresponding to a cone on ${\bf S}^7=SO(5)/SO(3)$ \cite{gary}.
Existence of many other complete metrics of $\mathrm{Spin}(7)$ holonomy
can be conjectured, as in the $G2$ case \cite{AMV}, by lifting
D6-brane configurations to M-theory \cite{gomis}.
Specifically, one starts with Type IIA string theory
on $M^3 \times M^7$, where $M^3$ is a $(2+1)$-dimensional
spacetime\footnote{$M^3$ is usually assumed to be either
$\mathbb{R}^3$ or $\mathbb{R}^1 \times T^2$, unless otherwise stated.},
and $M^7$ is a (non-compact) 7-manifold with $G2$ holonomy.
This gives $\mathcal{N}=2$ supersymmetric field theory
(without gravity) on $M^3$.
Let us further assume that $M^7$ has a topologically
non-trivial supersymmetric 4-cycle $B$, known as a coassociative cycle
\cite{HL,BBMOOY}, and let us introduce a space-filling D6-brane
with world-volume $M^3 \times B$.
Since $B$ is supersymmetric, we obtain
an $\mathcal{N}=1$ effective field theory
in three dimensions. Now consider the M-theory lift of this system.
The eleven-dimensional metric should look like
$M^3 \times X$, where $X$ is a (degenerate) circle bundle over $M^7$.
Moreover, since a D6-brane lifts to a purely geometric
background (Taub-NUT space), one can roughly speaking think of $X$
as a Taub-NUT bundle over $B$, such that $X$ preserves
only two real supercharges,
{\it i.e.} $\mathrm{Hol} (X) = \mathrm{Spin}(7)$.
A metric with $\mathrm{Spin}(7)$ holonomy obtained in this way
should be asymptotically locally conical (ALC), since the size of the ${\bf S}^1$ fiber at large distance is related
to the Type IIA coupling constant and, therefore, should be finite.
Furthermore, the circle should degenerate on a codimension
four submanifold $B$, describing the D6-brane locus.

A complete asymptotically locally conical metric with these
properties was constructed in the case $B={\bf S}^4$ in \cite{gary7}.
This solution describes D6-branes wrapped on a coassociative
4-sphere in the total space of $\Lambda^- {\bf S}^4$,
the bundle of anti-self-dual two-forms over ${\bf S}^4$.
In this paper we explicitly construct another family
of new ALC metrics with $B={\bf \mathbb{C}P}^2$.
After reduction to Type IIA these metrics represent D6-branes
wrapped on the supersymmetric ${\bf \mathbb{C}P}^2$ inside $\Lambda^- {\bf \mathbb{C}P}^2$.
These are the two main examples of non-compact $\mathrm{Spin}(7)$
manifolds that we analyze in this paper.
Note that in both examples we have D6-branes on
one of the asymptotically conical $G2$ holonomy manifolds
studied in \cite{AW}.

In both cases ($B={\bf S}^4$ and $B={\bf \mathbb{C}P}^2$)
the non-compact $\mathrm{Spin}(7)$ manifold is homotopy equivalent
to $\mathbb{R}^4 \times B$ with level surfaces (constant $r$
surfaces) $Y=SO(5)/SO(3)$ and $Y=SU(3)/U(1)$, respectively.
In the limit when the 4-cycle $B$ shrinks to zero size,
the 8-manifold $X$ develops an isolated conical singularity.
Since in this limit the physics is described by the local
behaviour near the singularity, we usually take $X$
to be a cone over the appropriate weak $G2$ manifold $Y$.
However, it is useful to bear in mind that M-theory on $X$
can be thought of as a configuration of D6-branes wrapped
on the coassociative 4-cycle in the corresponding
topologically non-trivial $G2$ manifold $M^7$.

There is another reduction to Type IIA theory that will
be very useful in our discussion. As for the $G2$ case \cite{AW},
one can find a semi-free $U(1)$ action on $X$
such that $X/U(1)$ is topologically trivial,
{\it i.e.} $X/U(1) \cong \mathbb{R}^7$.
Following \cite{AW}, we denote the fixed point set of
such a $U(1)$ action as $L$. The space $L$ has real
dimension four and represents the location of space-filling
D6-branes.

To summarise, one may think of M-theory on the non-compact
$\mathrm{Spin}(7)$ manifolds discussed here in several equivalent ways:

\begin{itemize}

\item
M-theory on a manifold $X$ of $\mathrm{Spin}(7)$ holonomy;

\item
Type IIA theory on a $G2$-holonomy manifold
$M^7 = \Lambda^- B$ with D6-branes wrapped over the supersymmetric 4-cycle $B$;

\item
A supersymmetric configuration of D6-branes in Type IIA theory
with world-volume:
\be
M^3 \times L \subset M^3 \times \mathbb{R}^7
\ee

\end{itemize}

The paper is organised as follows.
In order to make the paper self-contained,
we begin in section 2 with a brief review of special
holonomy manifolds and describe in outline the existing examples
of explicitly known metrics relevant to our discussion.

In section 3 we describe the reduction of our models
to configurations of D6-branes in flat spacetime.
We study the spectrum of topologically stable objects,
such as solitons, domain walls, {\it etc.}, in Type IIA
and in M-theory. Identifying the corresponding states
in the spectrum, we find a simple relation between
the homology groups of the D6-brane locus $L$ and those of the $\mathrm{Spin}(7)$ manifold $X$.
Global world-sheet anomalies and knotted 3-spheres
inside ${\bf S}^7$ enter our discussion in a natural way.

\begin{table}\begin{center}
\begin{tabular}{|c|c|c|}
\hline
$\mathrm{Spin}(7)$ Manifold $X$ &
$\Sigma_-{\bf S}^4 \cong \mathbb{R}^4 \times {\bf S}^4$ &
$\mathcal{Q} \cong \mathbb{R}^4 \times {\bf \mathbb{C}P}^2$ \\
\hline
\hline
Principal Orbit $Y$    & $SO(5)/SO(3)$    & $SU(3)/U(1)$  \\
\cline{1-3}
Collapsing Cycle $B$   & ${\bf S}^4$            & ${\bf \mathbb{C}P}^2$  \\
\cline{1-3}
$U(1)$ Bundle over $G2$ Manifold $M^7$ &
$\Lambda^- {\bf S}^4$ & $\Lambda^- {\bf \mathbb{C}P}^2$ \\
\cline{1-3}
D6-Brane Locus $L$ &  $\mathbb{R}^2 \times {\bf S}^2$ &
$\mathbb{R}^4 \cup \mathbb{R}^2 \times {\bf S}^2$ \\
\cline{1-3}
Global Symmetry        & $Sp(2)\times_{\mathbb{Z}_2} Sp(1) \times \mathbb{Z}_2$ &
$SU(3) \times U(1) \times U(1)$ \\
\cline{1-3}
Is Modulus Dynamical?     &  No              &     No   \\
\cline{1-3}
Background Flux, $\int_B G^{(4)}$     &  $\mathbb{Z} + 1/2$  &    $\mathbb{Z} + 1/2$  \\
\cline{1-3}
Number of Massive Vacua     &  1              &   1 or 2  \\
\hline
\end{tabular}\end{center}
\caption{The two examples of non-compact manifolds of $\mathrm{Spin}(7)$
holonomy studied in this paper.}
\end{table}

In section 4 we explicitly construct a new family of
complete $\mathrm{Spin}(7)$ metrics on a certain $\mathbb{R}^4$ bundle over
${\bf \mathbb{C}P}^2$, known as the universal quotient
bundle $\mathcal{Q}$. This solution has recently been extended
\cite{garychris}. We study in detail the global topology of these
solutions, whose level surfaces are various so-called Aloff-Wallach spaces $N_{k,l} = SU(3)/U(1)$,
where the integers $k$ and $l$ (such that $kl \ne 0$)
parametrise the embedding of $U(1)$ in $SU(3)$.
Every pair $(k,l)$ corresponds to a distinct $\mathrm{Spin(7)}$
manifold, which, modulo discrete identifications, is the total space
of a $\mathrm{spin}^c$ structure on ${\bf \mathbb{C}P}^2$. We also discuss the
action of the ``triality'' group $\Sigma_3$ of permutations of three
elements on these spaces. A systematic approach to the construction of new exceptional holonomy metrics can be found in \cite{Hitchin}.

In sections 5 and 6 we discuss various M-theoretic aspects of our work.
In section 5
we describe how the M-theory lift of a configuration of D6-branes
wrapping a coassociative cycle is related to $\mathrm{spin}^c$ bundles,
and also discuss $G$-flux quantisation \cite{witten1}.
In particular, we find that the $G$-flux obeys a shifted
quantisation condition, and has to be half-integer in our models.
This shift is related to the K-theory classification of RR-fields in Type IIA string theory. In section 6 we explain the relation between
the anomalous shift in the $G$-flux quantisation
condition and the shift of the Chern-Simons
coefficient in the effective $\mathcal{N}=1$
gauge theory in $(2+1)$ dimensions.
We also study certain dynamical aspects of M-theory
on (singular) $\mathrm{Spin}(7)$ manifolds, which
we summarise in table 1.


{\bf Note Added:}
Recently we received a preprint
\cite{kanno}
that significantly overlaps with sections 4.3 and 4.4 of the present
paper, which were completed some time ago
. This has prompted us to publish the paper in two parts, of which this is the first. A second paper will contain a more detailed analysis of various aspects of M-theory on $\mathrm{Spin}(7)$ manifolds \cite{sergeijames2}. An extension of some of the results in section 4 of this paper may also be found in the recent publication \cite{garychris}.

\sect{Riemannian Manifolds of Special Holonomy}

In this section we review the metrics of $G2$ and
$\mathrm{Spin}(7)$ holonomy constructed in \cite{gary}, together with the
recent examples of $\mathrm{Spin}(7)$ manifolds constructed in \cite{gary7}.

\subsection{The Holonomy Groups G2 and Spin(7)}

The holonomy group $H$ of a generic oriented Riemannian $n$-manifold
$Y$ is the special orthogonal group, $SO(n)$. However, if $H$ is a proper
subgroup of $SO(n)$ then the manifold $Y$ will inherit special geometric
properties. These properties are typically characterised by the
existence of non-degenerate (in some suitable sense) $p$-forms which are covariantly constant. Such $p$-forms also
serve as calibrations, and are related to the subject of minimal varieties.

The possible choices for $H\subset SO(n)$ are limited. Specifically, Berger's Theorem tells us that, for $Y$
simply-connected and neither locally a product nor symmetric, the only
possibilities for $H$, other than the generic case of $SO(n)$, are $U\left(\frac{n}{2}\right)$,
$SU\left(\frac{n}{2}\right)$, $Sp\left(\frac{n}{4}\right)Sp(1)$,
$Sp\left(\frac{n}{4}\right)$, $G2$ \footnote{The fourteen-dimensional simple
Lie group $G2\subset\mathrm{Spin(7)}$ is precisely the automorphism group of the
octonions, $\mathbb{O}$.}, $\mathrm{Spin}(7)$ or
$\mathrm{Spin}(9)$. The first four of these correspond, respectively, to a K\"ahler, Calabi-Yau, Quaternionic K\"ahler or
hyper-K\"ahler manifold. The last three possibilities are
the exceptional cases, occuring only in dimensions $7$, $8$ and $16$,
respectively. The latter case is in some sense trivial in that any
$16$-manifold of $\mathrm{Spin}(9)$ holonomy is locally isometric to
the Cayley projective plane, ${\bf \mathbb{O}P}^2$ (or its dual).

In the present paper, we shall be interested in both $G2$ and $\mathrm{Spin}(7)$ manifolds; that is, Riemannian manifolds with holonomy group $G2$ and $\mathrm{Spin}(7)$, respectively. The local
existence of such manifolds was
first demonstrated by Bryant, although a more thorough treatment, which we
review briefly in the next two subsections, was given in
\cite{gary}. The first examples of metrics with $G2$ and
$\mathrm{Spin}(7)$ holonomy on compact manifolds were constructed by
Joyce \cite{joyce}. We note in passing that $G2$ and $\mathrm{Spin}(7)$ manifolds are always Ricci flat.

On a $G2$ manifold $M^7$, there exists a distinguished harmonic three-form
$\Psi$, the associative three-form, which locally determines the
reduction of the structure group $\mathrm{Spin}(7)$ to $G2$. The Hodge
dual form $*\Psi$, is therefore also harmonic, and is
referred to as the coassociative four-form. Similarly, on a $\mathrm{Spin}(7)$ manifold $X$, there exists a distinguished
self-dual harmonic four-form, the Cayley form, $\Phi=*\Phi$, that locally determines
the reduction of the structure group $\mathrm{Spin}(8)$ to
$\mathrm{Spin}(7)$.

The $G2$ and $\mathrm{Spin}(7)$ conditions may also be characterised by
examining the behaviour of spinors under the decomposition of the
structure group. Specifically, for a $G2$ manifold, the decomposition
of the Majorana $\mathbf{8}$ of $\mathrm{Spin}(7)$ under $G2$ is

\be
\mathbf{8}\rightarrow\mathbf{7}+\mathbf{1}\ee

The singlet $\mathbf{1}$ corresponds to a parallel spinor; that is, a
covariantly constant section of the appropriate spin bundle. Similarly, for a
$\mathrm{Spin}(7)$ manifold, the decomposition of the Majorana-Weyl
$\mathbf{8}_-$ of $\mathrm{Spin}(8)$ under $\mathrm{Spin}(7)$ is

\be
\mathbf{8}_-\rightarrow\mathbf{7}+\mathbf{1}\ee

The singlet is again a parallel spinor. Note that the $\mathbf{8}_+$
then decomposes irreducibly.

We conclude this subsection by reminding the reader of the
definition of a calibration. A closed $p$-form $\phi$ is said to be a
calibration if the restriction of $\phi$ to each tangent $p$-plane is
less than or equal to the volume form of that $p$-plane. A
$p$-dimensional submanifold on which equality is obtained, at each point, is then referred to as a calibrated submanifold (with respect to
the calibration $\phi$). It is then a trivial exercise to show that a
calibrated submanifold is volume-minimising within its homology
class, and is therefore stable. The forms $\Psi$ and $*\Psi$
both serve as calibrations on a $G2$ manifold, the calibrated submanifolds
being respectively referred to as associative or coassociative
submanifolds. Likewise, the Cayley form $\Phi$ is a calibration for a $\mathrm{Spin}(7)$ manifold, the calibrated submanifolds then
being referred to as Cayley submanifolds. The deformability of
calibrated submanifolds was studied by McLean \cite{mclean} and will be relevant in the present paper. The
calibrations themselves may be constructed using the parallel spinors, essentially by "squaring" them.

\subsection{G2 Manifolds}

In this section, we briefly summarise the properties of the two known
complete non-compact $G2$
manifolds that contain a coassociative submanifold, \cite{gary}.

Let us start by considering the consequences of the existence of a
coassociative submanifold, $B$. We begin with some preliminaries. If $B$ is a
closed oriented Riemannian four-manifold, we denote the
bundle of $p$-forms over $B$ as $\Lambda^p\equiv\Lambda^pT^*B$. The
Hodge map on $B$ induces a direct sum decomposition
$\Lambda^2=\Lambda^+\oplus\Lambda^-$, where the rank three vector
bundles $\Lambda^{\pm}$ are the bundles of self-dual and
anti-self-dual two-forms on $B$. Note that since the Hodge map acting
on middle-dimensional forms is invariant under conformal rescalings of
the metric, the decomposition only depends on the conformal
class of the metric on $B$. In \cite{mclean}, it was shown that the
normal bundle $NB$ of a coassociative submanifold $B$ in $M^7$ is
isomorphic to the bundle of anti-self-dual two-forms over $B$,
$NB\cong\Lambda^-B$.

In \cite{gary}, complete Ricci-flat metrics of $G2$ holonomy were constructed on the total
spaces of the bundles of anti-self-dual two-forms over ${\bf S}^4$ and
${\bf \mathbb{C}P}^2$, the zero-section, or bolt, being a coassociative
submanifold in each case. The authors chose a cohomogeneity one ansatz for the
metric, so that the Ricci-flatness condition reduces to a coupled
system of second order differential equations for the metric functions, in terms of
the the radial variable $r$. In light of the above comments, the level
surfaces $\{r=\mathrm{constant}\}$ must be topologically the bundle
of unit vectors $\mathcal{S}\Lambda^-B$, in $\Lambda^-B$. This is
known as the twistor space of $B$ \cite{besse} and is an ${\bf S}^2$ bundle
over $B$.

The one-paramater family of metrics are given explicitly by

\be
ds^2 =
\left(1-\left(\frac{a}{r}\right)^4\right)^{-1}dr^2+\frac{1}{4}r^2\left(1-\left(\frac{a}{r}\right)^4\right)(D\mu^i)^2+\frac{1}{2}r^2d\Omega_4^2\label{G2}\ee

where $\mu^i$ are coordinates on $\mathbb{R}^3$, subject to the
constraint $\mu^i\mu^i=1$ (thus yielding the ${\bf S}^2$ fibre), and the covariant derivative is
$D\mu^i=d\mu^i+\epsilon_{ijk}A^j\mu^k$ where $A^i$ is the $SU(2)$
connection on the four-dimensional (Quaternionic K\"ahler) Einstein
manifold with metric $d\Omega_4^2$; that is, the field strengths
$J^i=dA^i+\frac{1}{2}\epsilon_{ijk}A^j\wedge{A}^k$ satisfy the unit
  quaternion algebra. We may take $d\Omega_4^2$ to be either ${\bf S}^4$ or
${\bf \mathbb{C}P}^2$, so that the conformal class of the metric on $B$ is the
canonical one in each case. The $G2$ metric is complete on the region
$r\ge{a}$, with $r=a$ the coassociative submanifold, and the
principal orbits $\{r=\mathrm{constant}\}$ are respectively
${\bf \mathbb{C}P}^3$ and $SU(3)/T^2$ where $T^2$ is a maximal torus in
$SU(3)$; these are precisely the twistor
spaces of ${\bf S}^4$ and ${\bf \mathbb{C}P}^2$. Indeed, the metric is asymptotic
to the cone over the squashed (nearly K\"ahler, rather than the K\"ahler Fubini-Study) Einstein
metric on ${\bf \mathbb{C}P}^3$, or the squashed (nearly K\"ahler) metric on $SU(3)/T^2$, respectively.

\subsection{Spin(7) Manifolds}

Until recently only one complete non-compact $\mathrm{Spin}(7)$
manifold was explicitly known, and was originally constructed along with the above G2 manifolds in \cite{gary}. This construction has recently been extended \cite{gary7} to yield a new family of $\mathrm{Spin}(7)$ manifolds. We also discuss the $\mathrm{Spin}(7)$ orbifold discovered in \cite{gary2}. The construction of this particular solution is in fact a special case of the construction used in the present paper to find a new family of $\mathrm{Spin}(7)$ metrics on a certain $\mathbb{R}^4$ bundle over ${\bf \mathbb{C}P}^2$.

The $\mathrm{Spin}(7)$ manifold presented in \cite{gary} contains a
Cayley submanifold, which is an ${\bf S}^4$. In \cite{mclean}, it was shown
that the normal bundle of an ${\bf S}^4$ Cayley submanifold in a
$\mathrm{Spin}(7)$ manifold $X$ is topologically $N{\bf S}^4=\Sigma_-{\bf S}^4$, the bundle of
negative chirality spinors over ${\bf S}^4$. The $\mathrm{Spin}(7)$
manifold presented in \cite{gary} is in fact the total space of this normal bundle, as was the case for the $G2$ metrics in the last subsection. The metric is
again cohomogeneity one, with level surfaces $\{r=\mathrm{constant}\}$
being topologically ${\bf S}^7$, described as an ${\bf S}^3=SU(2)$ bundle over
${\bf S}^4$. This is the quaternionic Hopf map. $G$-bundles over a four-sphere
are classified by an element of $\pi_3(G)$. In this case,
$\pi_3(SU(2))\cong\mathbb{Z}$ and the transition funtions of the
quaternionic Hopf map correspond to the generator of $\pi_3(SU(2))$. Moreover, the Euler
class of the negative chirality spin bundle $\Sigma_-{\bf S}^4$ is the
generator of $H^4({\bf S}^4;\mathbb{Z})\cong\mathbb{Z}$. The
one-parameter family of $\mathrm{Spin}(7)$ metrics on the total space
of $\Sigma_-$ are given explicitly by

\be
ds^2 = \left(1-\left(\frac{a}{r}\right)^{10/3}\right)^{-1}dr^2+\frac{9}{100}r^2\left(1-\left(\frac{a}{r}\right)^{10/3}\right)(\sigma_i-A^i)^2+\frac{9}{20}r^2d\Omega_4^2\label{s4}\ee

Here, the $\sigma_i$ are a set of left-invariant one-forms on $SU(2)$,
and the connection $A^i$ is the BPST Yang-Mills instanton on the unit
four-sphere, whose metric we denote $d\Omega_4^2$. The $\mathrm{Spin}(7)$ metric is complete on
$r\ge{a}$, with $r=a$ being the Cayley ${\bf S}^4$. At large distance, the
metric is asymptotic to the cone over the squashed (weak $G2$)
Einstein seven-sphere.

The construction of this metric has recently been extended in \cite{gary7}. The idea is simple. In (\ref{s4}) the level surfaces are an ${\bf S}^7$, described as an ${\bf S}^3$ bundle over ${\bf S}^4$ with the ${\bf S}^3$ fibres being "round". One may take a similar ansatz, but this time allow the ${\bf S}^3$ fibres themselves to become squashed. This allows for the possibility that the $U(1)$ fibres of $U(1)\hookrightarrow {\bf S}^3 \rightarrow {\bf S}^2$ approach a constant length asymptotically, rather like the Taub-NUT metric. Indeed, this is precisely what happens. The $\mathrm{Spin}(7)$ manifold of this form is given by

\be
ds^2 = \frac{(r-a)^2}{(r-3a)(r+a)} dr^2 + a^2 \frac{(r-3a)(r+a)}{(r-a)^2}\sigma^2 + \frac{1}{4}(r-3a)(r+a)(D\mu^i)^2 + \frac{1}{2}(r^2-a^2)d\Omega_4^2 \label{alcs4}
\ee

The metric $d\Omega_4^2$ is again the round ${\bf S}^4$ and, roughly speaking, the one-form $\sigma$ corresponds to the Hopf $U(1)$ fibre over ${\bf S}^2$, where the ${\bf S}^2$ has metric $(D\mu^i)^2$. The reader is referred to \cite{gary7} for the precise definitions. As $r\searrow 3a$, the level surfaces ${\bf S}^7$ collapse smoothly down to a Cayley ${\bf S}^4$. Thus the global topology of this space is the same as (\ref{s4}). However, there is an important difference between the two. At large radius, the $U(1)$ fibres of (\ref{alcs4}) (parametrised by $\sigma$) tend to a constant length as $r$ tends to infinity. The remainder of the metric asymptotes to a cone over ${\bf \mathbb{C}P}^3$, with its nearly K\"ahler Einstein metric (rather than the usual K\"ahler Fubini-Study metric). Thus this solution is asymptotically locally conical (ALC) rather than asymptotically conical (AC). The new $\mathrm{Spin}(7)$ metric in the present paper (\ref{spin7}) closely resembles this solution.

We should also point out that the same local solution (\ref{alcs4}) also describes a $\mathrm{Spin}(7)$ metric on $\mathbb{R}^8$, simply by taking the range of $r$ to be negative.

Finally, in \cite{gary2}, a $\mathrm{Spin}(7)$ metric was found on a
$\mathbb{Z}_2$ quotient of the cotangent bundle of
${\bf \mathbb{C}P}^2$. This construction will be explained in detail in section 4, and the reader should refer back to this section at the appropriate
points. The solution in \cite{gary2} solves the system (\ref{generalsys})
with $k=l=1$

\be
ds^2 =
\left(1-\left(\frac{a}{r}\right)^{10/3}\right)^{-1}dr^2+\frac{9r^2}{100}\left(1-\left(\frac{a}{r}\right)^{10/3}\right)\left(\lambda^2+4\nu_1^2+4\nu_2^2\right)+\frac{9r^2}{10}(\sigma_1^2+\sigma_2^2+\Sigma_1^2+\Sigma_2^2)\label{spinorb}\ee

Careful analysis \cite{gary2} shows that the topology of the three-dimensional
fibres with metric $(\lambda^2+4\nu_1^2+4\nu_2^2)$ is
${\bf \mathbb{R}P}^3$, which collapse down to ${\bf \mathbb{C}P}^2$ at $r=a$, with metric $\frac{9a^2}{10}(\sigma_1^2+\sigma_2^2+\Sigma_1^2+\Sigma_2^2)$. This
solution corresponds to viewing the level surfaces $N_{1,1}$ as an
$SO(3)={\bf \mathbb{R}P}^3$ bundle over ${\bf \mathbb{C}P}^2$. This will be
explained in section 4. The solution is
therefore defined on the orbifold $T^*{\bf \mathbb{C}P}^2/\mathbb{Z}_2$, and is asymptotically conical.

For completeness, we also mention that the ansatz (\ref{ansatz}) contains the hyper-K\"ahler Calabi metric \cite{calabi}

\be
ds^2 =
\left(1-\frac{1}{r^4}\right)^{-1}dr^2+\frac{1}{4}r^2\left(1-\frac{1}{r^4}\right)\lambda^2+r^2(\nu_1^2+\nu_2^2)+\frac{1}{2}(r^2-1)(\sigma_1^2+\sigma_2^2)+\frac{1}{2}(r^2+1)(\Sigma_1^2+\Sigma_2^2)\label{hyper}\ee

which was also explicitly constucted in \cite{gary2}. The manifold is
$T^*{\bf \mathbb{C}P}^2$, which corresponds to viewing $N_{1,1}$ as an ${\bf S}^3$
bundle over ${\bf \mathbb{C}P}^2$. Again, these comments should become transparent later.

\sect{Topological Charges
And Relation To Singularities Of Calibrated Cycles}

\subsection{Circle Quotients}

As pointed out by Atiyah and Witten \cite{AW},
one can often view M-theory on a non-compact manifold $X$
of special holonomy as a certain configuration of
D-branes in a (topologically) Minkowski space.
More precisely, given a suitable $U(1)$ action on $X$
such that $X/U(1) \cong \mathbb{R}^n$, one can identify
this $U(1)$ with the so-called "M-theory circle".
Then, the geometry of $X$ is mapped to the geometry
of the fixed point set $L$ of the $U(1)$ action.
In particular, when $X$ develops a conical singularity,
so does $L$.
Indeed, if $X$ is a cone on $Y$
such that
\be
Y/U(1) \cong {\bf S}^{n-1},
\ee
then $L$ is a cone on $F$,
where $F$ is the fixed point set in $Y$,
in the notations of \cite{AW}.

Since $Y$ is a smooth closed manifold,
the fixed point set $F \subset Y$ of a semi-free $U(1)$
action\footnote{This means that $U(1)$ acts freely on
the complement of the fixed point set $F$.}
on $Y$ is a smooth closed submanifold of even codimension
\cite{Bredon,MZ}. Consider a fixed point $p\in Y$, where
$Y$ is an oriented $n$-manifold equipped with a semi-free circle
action, which preserves the orientation. Then the circle group action maps
the tangent space $T_pY$ at $p$ into iteself. Hence $T_pY$ is a real
$U(1)$-module, which we may decompose into its real irreducible
representations. These are either one or two-dimensional:

\begin{eqnarray}
\pm1 & & \nonumber \\
R(\theta) & = & \left(\begin{array}{cc}\cos{\theta}
& -\sin{\theta}\\ \sin{\theta} &
\cos{\theta}\end{array}\right)\end{eqnarray}

Hence one may decompose the circle action on $T_pY$ into $r$
$2\times2$ rotations with parameters say $\theta_j = \kappa_j \tau$, ($j=1,\ldots,r$), together with $(n-2r)$ trivial $\pm1$
representations. Here, $\tau$ is the $U(1)$ group parameter, and
$\kappa_j$ are the skew eigenvalues of the matrix $k_{a;b}$, which are
the orthonormal-frame components of the covariant derivative of the
Killing covector $k$ associated with the $U(1)$ isometry. The fact
that the action preserves the orientation means that we must have an
even number of ``-1''s.

What we have just done is to decompose the tangent space $T_pY$ into
directions tangent to the codimension $2r$ fixed point set $F\subset Y$ containing the
point $p\in F$
(these are the trivial ''1'' representations), and directions normal to
$F$. The circle action acts orthogonally on this normal space, and
decomposes into $r$ $2\times2$ rotations in $r$ orthogonal
2-planes. Of course, for the orbits to close, the eigenvalues
$\{\kappa_j\mid j=1,\ldots,r\}$ must be rationally related. This means
that, after rescaling $\tau$ appropriately, the action on the
$j^{\mathrm{th}}$ normal 2-plane is by multiplication by $e^{i n_j
\tau}$, where the integers $n_j$ are relatively prime (in order that
the action be effective), and $\tau$ has period $2\pi$.

This action is not semi-free in general. Consider the unit
$(2r-1)$-sphere in the normal space. Its quotient under the circle
action gives the so-called weighted projective space
${\bf \mathbb{C}P}^{[n_1,\ldots,n_r]}$. This is a complex orbifold for
general integers $\{n_j\}$. Only if all the $n_j=\pm1$ do we get a
free action on the sphere, the projection then being the Hopf (or
anti-Hopf) map. Thus, a necessary condition for the circle action to be
semi-free is that all the integers $n_j=\pm1$, for every connected
component of the fixed point set.

If we require the orbit space to be a smooth
manifold, $F$ must be either codimension two or four in $Y$. Codimension two
corresponds to $r=1$. In this case, the ``unit sphere'' in the normal space
is just a circle. Indeed, we can define polar coordinates $(r,\phi)$
on the normal space. Taking the quotient obviously yields a half-line
$\mathbb{R}^+$, parametrised by the radial coordinate $r\geq0$. So, the orbit space looks locally (in a neighbourhood
of the fixed point set) like $F\times\mathbb{R}^+$. In this way, $F$
becomes a boundary of the orbit space. An illustrative example is a two-dimensional disk, $D^2 \cong {\bf S}^3 / U(1)$,
which may be viewed as a quotient
of ${\bf S}^3$ by a semi-free $U(1)$ action.
In this example the $U(1)$ acts freely at a general point
on the 3-sphere, except for the ``equator'' $F = {\bf S}^1$,
which is clearly a subspace of codimension two in ${\bf S}^3$.

On the other hand, in the case of codimension four,
the orbit space is a smooth \emph{closed} manifold containing
$F$ as a submanifold of codimension three. This corresponds to $r=2$. The
unit sphere in the normal space is a 3-sphere, and the quotient by the
$U(1)$ action is the Hopf map. Hence this 3-sphere projects down to a
2-sphere in the orbit space. We may now ``fill in'' this 2-sphere bundle
over $F$ with the associated 3-disc bundle over $F$, obtaining a
smooth closed manifold which contains $F$.

We do not obtain a smooth orbit space for higher codimension. In
particular, when $r=3$, the 5-sphere in the normal space projects down
to a ${\bf \mathbb{C}P}^2$ - which is not the boundary of anything! Since we are interested in the case when $Y/U(1)$
is a homotopy sphere, in particular, when it is a space without
a boundary, we should therefore restrict ourselves to the case of codimension four,
which is also the most interesting case in physics.
Given this motivation, we shall focus on the case
when $F$ is codimension four in $Y$, which implies
that $L$ is codimension four in $X$.

Semi-free $U(1)$ actions with fixed points of codimension four are
very familiar in string theory --- they correspond to D6-branes.
For instance, $Y={\bf S}^{10}$ admits 8 topologically different $U(1)$
actions with fixed point set being the standard ${\bf S}^6$,
among which there is a semi-free action corresponding
to $Y/U(1) \cong {\bf S}^9$ \cite{Levine}.
Building cones on all of these spheres, we find that
$X=\mathbb{R}^{11}$ admits a $U(1)$ action such that
$X/U(1) \cong C({\bf S}^9) = \mathbb{R}^{10}$, with fixed point set
\be
L \cong C({\bf S}^6) = \mathbb{R}^7
\ee
This gives a mathematical construction of a flat D6-brane
with world volume $L \subset \mathbb{R}^{10}$ as a fixed
point set of a $U(1)$ action on the eleven-dimensional
space-time $X \cong \mathbb{R}^{11}$.

Now let us implement the fact that $X$ has a reduced
holonomy group. This means that there is at least one
covariantly constant spinor on $X$ and, therefore, M-theory
compactification on $X$ is supersymmetric.
Hence, the same should be true about the equivalent
configuration of D6-branes on $L \subset \mathbb{R}^{10}$
in Type IIA string theory.
In general, in such a reduction from M-theory down
to Type IIA one does not obtain the standard flat
metric on $X/U(1) \cong \mathbb{R}^{n-1}$ due
to non-constant dilaton and other fields in the background.
However, one would expect that near the singularities of
the D-brane locus $L$ these fields exhibit a regular behavior,
and the metric on $X/U(1)$ is approximately flat, {\it cf.} \cite{AW}.
In this case the condition for the Type IIA background
to be supersymmetric can be expressed as a simple geometric
criterion: it says that the D-brane locus\footnote{The part of
the D-brane world-volume that is transverse
to $X$ is flat and does not play an important r$\mathrm{\hat{o}}$le in our
discussion here.}
$L$ should be a calibrated submanifold in $X/U(1)$
\cite{BBMOOY,calibr1,calibr2,AFFS,FFcal,PTcal,GPT}.

Even though our discussion in this subsection is quite general,
in order to be specific, let us focus on manifolds $X$ of $\mathrm{Spin}(7)$
holonomy -- the main theme of this paper.
One can easily extend all of the results obtained for
$\mathrm{Spin}(7)$ manifolds to other cases,
see {\it e.g.} (\ref{hlgeneral}) - (\ref{hl0general}) below.

In the case when an 8-manifold $X$ admits a $\mathrm{Spin}(7)$
structure $\Phi$, then away from the fixed point set $L$
the quotient space $X/U(1)$ has a 3-form $\Psi = \pi_* \Phi$,
where $\pi$ is the projection $\pi \colon X \to X/U(1)$.
The form $\Psi$ defines an ``approximate'' $G2$ structure
on $X/U(1)$, which becomes a $G2$ structure in the limit when
all $U(1)$ orbits on $X$ have the same length, {\it cf.} \cite{AW}.
In this approximation, the fixed point set $L$ representing
the location of a D6-brane must be a supersymmetric cycle in $X/U(1)$,
namely a coassociative submanifold, calibrated
with respect to $\star \Psi$.
Therefore, the problem of studying dynamics of M-theory on
$\mathrm{Spin}(7)$ singularities can be restated as a problem
of studying D6-brane configurations on singular coassociative
submanifolds in flat space.
When $X$ develops a conical singularity, the D-brane locus
$L$ also becomes a (singular) cone on $F \subset Y = \partial X$.

\subsection{Identification of Topological Charges}

The topology of $L$, which determines the dynamics of the D6-branes,
can be deduced from the topology of the 8-manifold $X$.
In the remainder of this subsection we obtain relations between
various homology groups of $X$ and $L$, identifying domain walls and other
topologically stable objects in M-theory on $\mathbb{R}^{3} \times X$
and in Type IIA theory with D6-branes on
\be
\mathbb{R}^3 \times L \subset \mathbb{R}^{10}
\ee
For simplicity, let us assume that there is only one D6-brane
on every connected component of $L$. Then, from the M-theory perspective, topological charges
in the effective ${\cal N}=1$ three-dimensional theory
correspond to membranes and five-branes wrapped
on topologically non-trivial cycles in $X$.
On the other hand, in the Type IIA picture with a D6-brane
these topological charges are represented by strings
and D4-branes which end on the D6-brane.

The fact that only fundamental strings and D4-branes
are allowed to have their boundaries on a D6-brane
follows from the structure of the Chern-Simons terms
on D6-brane world-volumes
in Type IIA theory \cite{BraneSurgery}. For example, for a D4-brane this follows from
the modified Bianchi identity:
\be
d (G^{(4)} - B \wedge G^{(2)} ) =0
\ee
which leads to the following modification of
the 4-brane charge:
\be
Q_4 = \int_{{\bf S}^4} \big( G^{(4)} - G^{(2)} \wedge B \big)
\ee
Now, sliding the 4-sphere to the end of the D4-brane and deforming
it into a product ${\bf S}^2 \times {\bf S}^2$, with the last
${\bf S}^2$ factor embedded into the D6-brane world-volume,
we come to the region where $Q_4$ has a contribution mainly from
the second term:
\be
Q_4 \approx - \int_{{\bf S}^2} G^{(2)} \times \int_{{\bf S}^2} B
= - Q_6 \times \int_{{\bf S}^2} B
\ee
In a similar way one can show that a D2-brane cannot end
on a single D6-brane. Indeed, there is no Chern-Simons
term like $\int G^{(4)} \wedge G^{(2)} \wedge C^{(4)}$
in the Type IIA effective action.

This also suggests that the boundary of a D4-brane inside
a D6-brane is magnetically charged with respect to
the $U(1)$ gauge field on the D-brane world volume.
The best way to see this is from the Wess-Zumino
term \cite{green}:
\be
I_{\mathrm{WZ}}
=\int_{\mathbb{R}^{3} \times L} e^{(\mathcal{F} - B)/2\pi} \wedge {C_*}
\ee

which includes the term

\be
\int_{\mathbb{R}^{3} \times L} \frac{(\mathcal{F} - B)}{2\pi} \wedge {C_5}
\label{wzfbterm}
\ee
We will use the fact that the first Chern class of $\mathcal{F}$
jumps once a D4-brane ends on a D6-brane further below.

Now we want to construct various (extended) objects in
the effective ${\cal N}=1$ three-dimensional theory and
compare their charges with the corresponding objects in M-theory.
In fact, the correspondence has to be one-to-one, so that
the charges computed both ways must be the same.
Note that we are not only saying that the charge
spectrum should be the same, but that every object
actually has its counterpart.
Namely, a D4-brane ending on D6-brane lifts in M-theory
to an M5-brane wrapped on a certain cycle $\Sigma \subset X$.
Since the D6-brane configuration lifts to
a purely geometrical background in M-theory,
{\it viz.} to the geometry of the space $X$,
the cycle $\Sigma$ should be closed in $X$,
for otherwise the five-brane would end on ``nothing''.
Similarly, a string ending on a D6-brane lifts
to a membrane wrapped on a closed submanifold in $X$.
This general rule is summarised in Table 2.
We also remark that this argument does not depend
on the amount of supersymmetry and, therefore,
can be used in compactifications on manifolds
of arbitrary holonomy.

\begin{table}\begin{center}
\begin{tabular}{ccc}
IIA Theory && M-Theory \\
\cline{1-3}
D4-brane & $\longrightarrow$ & M5-brane \\
String & $\longrightarrow$ & M2-brane \\
\end{tabular}\end{center}
\caption{M-theory lift of a D4-brane and a fundamental string,
which are allowed to end on D6-branes in Type IIA string theory.}
\end{table}

Now let us consider specific cases in more detail:

{\bf Domain Walls:}
In Type IIA string theory domain walls correspond to D4-branes
with boundary on the D6-brane world-volume, $\mathbb{R}^3 \times L$.
More precisely, the D4-brane world-volume is
$\mathbb{R}^2 \times D^{(3)} \subset \mathbb{R}^3 \times X/U(1)$,
such that:
\be
\partial D^{(3)} = \Sigma^{(2)} \subset L
\ee

In order for the D4-brane to be topologically stable,
$[\Sigma^{(2)}]$ should represent a non-trivial homology
class in $H_2(L;\mathbb{Z})$.
By Poincar\'e duality, the latter group is isomorphic to
$H^2_{\mathrm{cpct}} (L;\mathbb{Z})$, the cohomology
with compact support.
Therefore, we conclude that in a Type IIA background
with a D6-brane, domain walls are classified by the group:
\be
H_2(L;\mathbb{Z}) \cong H^2_{\mathrm{cpct}} (L;\mathbb{Z})
\label{h2l}
\ee
Since D4-branes lift to M5-branes, in M-theory every such
domain wall becomes a five-brane with world-volume:
\be
\mathbb{R}^2 \times \Sigma^{(4)} \subset \mathbb{R}^3 \times X
\ee
Now, $\Sigma^{(4)}$ must be a closed topologically non-trivial
4-cycle in $X$, for otherwise the domain wall would not be stable.
Hence, from the M-theory point of view, domain wall charges
take values in the group
\be
H_4(X;\mathbb{Z})
\label{h4x}
\ee
Since the spectrum of domain walls should be equivalent
in both pictures, we conclude that (\ref{h2l}) and (\ref{h4x})
should be isomorphic:
\be
H_2(L;\mathbb{Z}) \cong H_4(X;\mathbb{Z})
\ee

Furthermore, as we noted earlier, the first Chern class
of the $U(1)$ gauge bundle on the D6-brane in the Type IIA
model jumps by the dual cohomology class
$\widehat{[\Sigma^{(2)}]} \in H^2_{\mathrm{cpct}} (L;\mathbb{Z})$
when we cross such a domain wall \cite{GVW}.
On the other hand, different vacuum states on a D6-brane configuration
are classified by the first Chern class,
which takes values in the cohomology group
\be
H^2 (L;\mathbb{Z})
\label{c2l}
\ee
There is a natural map:
\be
f \colon H^2_{\mathrm{cpct}} (L; \mathbb{Z}) \to H^2 (L; \mathbb{Z})
\ee
which ``forgets'' that a cohomology class has compact support. In
general, this is
not an isomorphism when $L$ is non-compact. In fact, we may write down part of the long exact cohomology sequence for the pair $(L,F)$, where $F=\partial L$ is the boundary "at infinity" in $L$

\be
\ldots H^2(L,F;\mathbb{Z})\stackrel{f}{\longrightarrow}H^2(L;\mathbb{Z})\stackrel{i^*}{\longrightarrow}H^2(F;\mathbb{Z})\stackrel{\delta^*}{\longrightarrow}H^3(L,F;\mathbb{Z})\longrightarrow \ldots \label{exactLF}\ee

Here $i:F \hookrightarrow L$ denotes inclusion, and $H^2(L,F;\mathbb{Z})\cong H^2_{\mathrm{cpct}}(L;\mathbb{Z})$. By Poincar$\mathrm{\acute{e}}$ duality, $H^3(L,F;\mathbb{Z}) \cong H_1(L;\mathbb{Z})$. Hence, when $L$ is simply-connected, we see that the exact sequence (\ref{exactLF}) implies that different vacua, modulo those connected by domain walls, are classified by the group
\be
H^2 (F; \mathbb{Z}) = H^2 (L; \mathbb{Z}) / f( H^2_{\mathrm{cpct}} (L; \mathbb{Z})) \label{h2f}\ee

In fact, we shall find that $H_1(L;\mathbb{Z})=0$ for the examples in the present paper, so that the above formula applies.

Different vacua in M-theory are classified
by the flux of $G$, which in turn is classified\footnote{It is
actually the \emph{shifted} $G$-flux, $[G/2\pi]-\frac{1}{4}p_1(X)$, that may be
considered as an integral cohomology class, rather than the $G$-flux
itself. This fact will be extremely important in this paper, and will
be elaborated on further below.} (see (\ref{flux})) by the group
\be
H^4 (X; \mathbb{Z})
\label{c4x}
\ee
By a similar logic to before, the number of vacua should be the same
in the equivalent Type IIA and M-theory models, so that we obtain
another useful isomorphism:
\be
H^2 (L;\mathbb{Z}) \cong H^4 (X;\mathbb{Z})
\ee

Now, as in \cite{GVW}, one also finds that different vacua in M-theory, modulo those connected by domain walls, are classified by the group
\be
H^4 (Y; \mathbb{Z}) = H^4 (X; \mathbb{Z}) / f( H^4_{\mathrm{cpct}} (X; \mathbb{Z}))\label{h4y}
\ee
where $f$ is again the forgetful map and $Y=\partial X$. More precisely, this formula holds when $H^5(X,Y;\mathbb{Z})=0$. But in the present paper, $X$ is always a four-plane bundle over some four-manifold, $B$. Hence, by the Thom isomorphism, we have

\be
H^5(X,Y;\mathbb{Z})\cong H^1(B;\mathbb{Z})=0\end{equation}

since $B$ is simply connected (either $B={\bf S}^4$ or $B={\bf \mathbb{C}P}^2$). For consistency of both pictures, we must of course have

\be
H^2 (F; \mathbb{Z}) \cong H^4 (Y; \mathbb{Z})\ee

This is indeed consistent with the formulae (\ref{h2f}), (\ref{h4y}), together with the relations between the homology and cohomology groups of $L$ and $X$ derived so far in this section.

Finally, we note that all 8-manifolds $X$ with $\mathrm{Spin}(7)$
holonomy that we consider in this paper are simply-connected.
(Also, all \emph{compact} $\mathrm{Spin}(7)$-manifolds are simply-connected).
Therefore, for these manifolds there are no other domain walls,
in particular, there are no domain walls constructed from M2-branes.
In the Type IIA theory such a domain wall, if it existed,
would look like a D2-brane with boundary on a D6-brane.
In the case of multiple D6-branes this configuration would be
possible, and the boundary of a D2-brane would couple to
the second Chern class of the gauge bundle on the D6-branes.

{\bf Stable Particles:}
Again, we start in Type IIA theory with a D6-brane, where
stable particles in the three-dimensional effective field
theory correspond to either a string or a D4-brane with
boundary on a D6-brane.

The case of a D4-brane is very similar to what we considered
above. Namely, in order to represent a codimension two object
in 2+1 dimensions, a D4-brane must have world-volume
$\mathbb{R}^1 \times D^{(4)} \subset \mathbb{R}^3 \times X/U(1)$
such that:
\be
\partial D^{(4)} = \Sigma^{(3)} \subset L
\ee
And, following the above arguments, we conclude that
topologically stable particles are classified by
the homology group
\be
H_3 (L;\mathbb{Z})
\label{h3l}
\ee
Since a D4-brane on a 4-manifold $D^{(4)} \subset X/U(1)$
lifts to a five-brane on a closed 5-cycle $\Sigma^{(5)} \subset X$,
in M-theory such particles are classified by the group:
\be
H_5 (X;\mathbb{Z})
\label{h5x}
\ee
Identifying these objects particle-by-particle in Type IIA and
in M-theory, we conclude that (\ref{h3l}) and (\ref{h5x}) are
isomorphic:
\be
H_3 (L;\mathbb{Z}) \cong H_5 (X;\mathbb{Z})
\ee

There can be another kind of point-like stable object,
which in Type IIA corresponds to an open fundamental string
ending on a D6-brane. Its world-volume looks like the product
of a ``time direction'' and an interval in space.
Particles of this type can be stable only if
the ends of the string belong to different connected
components of the D6-brane locus, $L$.
Therefore, the charges of these particles form a lattice of
dimension $h_0 (L) - 1$.

In M-theory, every fundamental string ending on a D6-brane
lifts to a closed membrane, wrapped on a 2-cycle
$\Sigma^{(2)} \subset X$.
The dimension of these states is clearly $h_2 (X)$, and
must be the same as the dimension of the corresponding
stable particles in Type IIA theory:
\be
h_0 (L) -1 = h_2 (X)
\ee

{\bf Space-Filling Branes:}
Having established relations between $H_i (L; \mathbb{Z})$,
for $i=0,2,3$, and the corresponding homology groups of
the 8-manifold $X$, now we have to find a similar formula
for $H_1 (L; \mathbb{Z})$. We can obtain such a formula,
for example, by looking at D4-branes filling three-dimensional
space-time\footnote{There are other types of space-filling branes,
{\it e.g.} D2-branes, but they do not lead to new information
about the topology of $L$.}.
They have world-volume
$\mathbb{R}^3 \times D^{(2)} \subset \mathbb{R}^3 \times X/U(1)$,
where
\be
\partial D^{(2)} = \Sigma^{(1)} \subset L
\ee
Hence, the charges of such space-filling D4-branes take values in:
\be
H_1 (L;\mathbb{Z})
\ee
In M-theory, they lift to an M5-brane with world-volume
$\mathbb{R}^3 \times \Sigma^{(3)} \subset \mathbb{R}^3 \times X$
with $\Sigma^{(3)}$ a closed 3-cycle in $X$.
It follows that in M-theory the charges of the corresponding
space-filling branes take values in the group
\be
H_3 (X;\mathbb{Z})
\ee
Identifying the charges, as before, we find the last
isomorphism:
\be
H_1 (L;\mathbb{Z}) \cong H_3 (X;\mathbb{Z})
\ee

As we explain below, these groups are trivial in our models.
Hence the only space-filling branes which can occur are
D2-branes/membranes.

{\bf Instantons:}
Instanton effects play a very important role in the effective
${\cal N}=1$ three-dimensional gauge theory and will be discussed
further below. Here we just mention that they can come
either from string or D4-brane instantons which are completely
embedded in $X$ and have boundary on the D6-brane locus, $L$.
Since $L$ is non-compact, we have to consider only world-sheet
instantons.

The world-sheet string instantons with boundary on $L$ are
classified by\footnote{The reason we have $H_1 (L;\mathbb{Z})$,
rather than $\pi_1 (L)$ is that the `missing' elements of $\pi_1 (L)$,
which map to the zero element in $H_1 (L;\mathbb{Z})$, correspond
to bound states of multiple membrane instantons \cite{Triples}.
In this section we restrict ourselves only to basic instantons.}:
\be
H_1 (L;\mathbb{Z})
\ee
In M-theory these states correspond to membrane instantons,
classified by $H_3 (X;\mathbb{Z})$.
Therefore, lifting string instantons to M-theory we find
a relation between the corresponding homology groups that
we have seen in the previous example:
\be
H_1 (L;\mathbb{Z}) \cong H_3 (X;\mathbb{Z})
\ee

In the case of compactification on a Calabi-Yau manifold
or a manifold of $G2$ holonomy this would be the end of story.
However, if the dimension of $X$ is greater than or equal to eight,
we can discover a world-sheet anomaly related to the fact
that $L$ may fail to be a spin manifold \cite{wittenfreed}.
In order to see the anomaly, we start with a string
world-sheet, $\Sigma^{(2)}$, with boundary on the D6-brane,
and consider a one-parameter deformation of $\Sigma^{(2)}$
along a closed loop ${\bf S}^1$, such that we have an embedding
\be
\phi \colon \Sigma^{(2)} \times {\bf S}^1 \to \mathbb{R}^7,
\quad \quad
\phi(\partial \Sigma^{(2)} \times {\bf S}^1) \subset L
\ee
Then, going around the loop ${\bf S}^1$, the string world-sheet path
integral picks up a phase factor \cite{wittenfreed}:
\be
\mathrm{pfaff}(D) \to
(-1)^{\alpha} \cdot \mathrm{pfaff}(D)
\ee

where

\be
\alpha = \Big( \int_{\partial \Sigma^{(2)} \times {\bf S}^1} w_2 (L) \Big)
\ee

$w_2(L)$ is the second Stiefel-Whitney class of $L$,
and pfaff$(D)$ is the Pfaffian of the world-sheet Dirac operator $D$.
All oriented manifolds $L$ of dimension less than four are spin,
so that $w_2(L)=0$ automatically. However, in dimension four,
which is relevant to the present paper,
one can have a non-trivial class $w_2(L) \in H^2 (L;\mathbb{Z}_2)$.
As we shall see, this is precisely the case for the two
$\mathrm{Spin}(7)$ models discussed in the present paper.

If $w_2(L)$ happens to be non-zero, one can still have a consistent
D6-brane configuration, but in order to achieve this one needs to turn
on a non-trivial $U(1)$ "gauge field" on the D6-brane world-volume:
\be
\mathrm{pfaff}(D) \cdot
\exp \Big( i \oint_{\partial \Sigma^{(2)}} \mathcal{A} \Big)
\ee
Then, on going around the loop ${\bf S}^1$ the phases of the two factors can compensate each other if the field strength $\mathcal{F}$ obeys
the modified quantisation condition:
\be
\Big[ {\mathcal{F} \over 2 \pi} \Big] - {\nu \over 2} \in H^2(L;\mathbb{Z})
\label{fqcond}
\ee

Here $r(\nu)=w_2(L)$, under the reduction modulo two homomorphism
$r \colon H^2(L;\mathbb{Z}) \to H^2(L;\mathbb{Z}_2)$.

Now let us consider what happens when we lift this configuration
to M-theory. As we mentioned earlier, a string becomes a membrane
wrapped over a closed 3-cycle $\Sigma^{(3)} \subset X$.
The world-sheet fermion anomaly lifts to the membrane anomaly in M-theory.
Specifically, the way we detected the anomaly in Type IIA was
by looking at a one-parameter family of string world-sheets
parametrised by a circle ${\bf S}^1$.
In M-theory, $\Sigma^{(2)}$ lifts to $\Sigma^{(3)}$,
so that the process described above corresponds to studying
a one-parameter family of closed 3-cycles, such that
\be
\Sigma^{(3)} \times {\bf S}^1 \subset X
\ee
And, again, there are two dangerous factors in path integral in
the membrane world-volume theory, corresponding to
the Pfaffian of the Dirac operator and to the period
of the $C$-field. After going around the closed loop ${\bf S}^1$
these two factors pick up a phase \cite{witten1}:
\be
\exp \Big( i \int_{\Sigma^{(3)} \times {\bf S}^1} \pi \lambda + G \Big)
\ee
where the integral class $\lambda=p_1(X)/2\in{H}^4(X;\mathbb{Z})$
is canonically defined for a spin manifold $X$, since $p_1(X)$ is always divisible by two in this case\footnote{More fundamentally, $\lambda$ is the first obstruction class to the spin bundle of $X$ \cite{wittentop}.}.
The definition is very similar to the above definition of
the class $\nu$, but with $w_2 (L)$ replaced by $w_4 (X)$.
Specifically, $\lambda$ is congruent modulo two to $w_4 (X)$.
Note also that both $\nu$ and $\lambda$ come from fermionic
anomalies in the string/membrane world-volume theory,
so that it is natural to identify the two.
More precisely, we can express this as a map:
\be
\rho \colon \lambda \mapsto \nu
\ee
under the isomorphism:
\be
\rho \colon H^4(X; \mathbb{Z}) \to H^2(L; \mathbb{Z})
\ee

As evidence for the proposed identification, let's see
what happens if $\lambda/2$ is not an integral class.
In this case, the $G$-field in M-theory has to obey a shifted
quantization condition \cite{witten1}:
\be
\left[\frac{G}{2\pi}\right]-\frac{\lambda}{2}\in{H}^4(X;\mathbb{Z})\label{flux}
\ee
similar to the $\mathcal{F}$-field quantisation condition (\ref{fqcond}).
For example, when $\lambda$ (respectively $\nu$) is odd,
we need to turn on a half-integral $G$ (respectively $\mathcal{F}$) flux.
In particular, as expected both models
are (non-)anomalous at the same time.
This agrees with our general identification of degree two cohomology
elements in $L$ and degree four cohomology elements in $X$.

Altogether, we may summarise all of the above relations as follows:
\begin{eqnarray}
h_0 (L) & = & h_2 (X) + 1 \nonumber \\
H_i (L; \mathbb{Z}) & \cong & H_{i+2} (X; \mathbb{Z}), \quad i>0 \nonumber \\
H^2 (L;\mathbb{Z}) & \cong & H^4 (X;\mathbb{Z}) \nonumber \\
\nu & \stackrel{\rho}{=} & \lambda\label{genhom}
\end{eqnarray}

Applying similar arguments to other manifolds of special holonomy,
such as $G2$-manifolds or Calabi-Yau manifolds,
one can obtain the same universal result:
\be
H_{i} (L;\mathbb{Z}) = H_{i+2} (X;\mathbb{Z}),
\quad 0 < i < {\rm dim}_{\mathbb{R}} (L)
\label{hlgeneral}
\ee
This general formula is valid for all $i$,
except for the special case $i=0$ when
\be
h_0 (L) = h_2 (X) + 1
\label{hl0general}
\ee

Note, for example, that all of the D6-brane geometries dual to $G2$ conical
singularities, studied recently in \cite{AW}, satisfy this relation.

{\bf Example 1:}
In order to demonstrate how the above ideas work in practice,
let's take a $\mathrm{Spin}(7)$ metric (\ref{s4}) on $X = \Sigma_- {\bf S}^4$,
the total space of the negative chirality spinor bundle over ${\bf S}^4$.
At large distance this space looks like a cone over $Y$,
where $Y$ is a 7-sphere. In this case it follows from
Smith Theory that the fixed point set $F$ under a semi-free
circle action is an integral homology 3-sphere in $Y$.
More precisely, $F$ is a {\it knotted} homology
3-sphere\footnote{Such 3-spheres are related to
the ordinary knots in ${\bf S}^3$. It would be
interesting to understand the meaning of these knots
in the $\mathcal{N}=1$ three-dimensional effective theory
with Chern-Simons term that we discuss in section 6.} \cite{Levine},
so that when we build a cone over $F$, we obtain a singular
space $L$. Therefore, we conclude that M-theory on a $\mathrm{Spin}(7)$
manifold $X = \Sigma_- {\bf S}^4$ developing a conical singularity
is equivalent to Type IIA string theory on
$\mathbb{R}^{10} \cong \mathbb{R}^3 \times X/U(1)$
with a (singular) configuration of D6-branes on
\be
\mathbb{R}^3 \times L \subset \mathbb{R}^{10}
\ee

The space $L$, representing the locus of the D6-brane,
may be deformed to a smooth four-manifold which is a non-trivial $\mathbb{R}^2$ bundle over ${\bf S}^2$:
\be
L \cong \mathbb{R}^2 \times {\bf S}^2
\label{s4l}
\ee

Specifically, this bundle is the spin bundle of ${\bf
\mathbb{C}P}^1\cong {\bf S}^2$, which has first Chern class (or
equivalently Euler class) given by the generator of $H^2({\bf
S}^2;\mathbb{Z})\cong \mathbb{Z}$. That is, we have effectively blown
up the origin of $\mathbb{R}^4$ with a copy of ${\bf \mathbb{C}P}^1$. In fact, it is not hard to
construct explicitly the $U(1)$ action on $X$ such that the fixed
point set is $L$ above, with an orbit space that may be given the
differentiable structure of $\mathbb{R}^7$. Since we will not require
the details of this in the present paper, we shall leave the
precise description for a future publication \cite{sergeijames2}.

We may now use this example to verify the formulae (\ref{genhom}) in
this particular case. The only non-trivial homology groups of $X$ are
in dimension 0 and 4:
\be
H_0 (X; \mathbb{Z}) = H_4 (X; \mathbb{Z}) = \mathbb{Z}
\ee
Therefore, according to the above discussion, the non-trivial
homology groups of $L$ must be
\be
H_0 (L; \mathbb{Z}) = H_2 (L; \mathbb{Z}) = \mathbb{Z}
\ee
Clearly this is the case for $L$ of the form (\ref{s4l}). This example
therefore confirms the relations between the homology groups of $X$
and $L$ found in the general analysis above. Notice also that, as
expected in this example, $L$ is \emph{not} spin. As explained above,
this is precisely because $\lambda(X)$ is not divisible by
two.


{\bf Example 2:}
We can also consider $X$ to be an $\mathbb{R}^4$
bundle over ${\bf \mathbb{C}P}^2$, corresponding to the M-theory
lift of a D6-brane wrapped on the coassociative cycle
of the $G2$-space $\Lambda^- {\bf \mathbb{C}P}^2$.
The complete metrics of $\mathrm{Spin}(7)$ holonomy on
different bundles of this kind will be discussed in
the next section. In all cases we have
\be
H_i (X; \mathbb{Z}) =
\cases{\mathbb{Z} & $i=0,2,4$ \cr
0 & otherwise}
\ee

Now consider a reduction from M-theory on $X$ to Type IIA
string theory on $\mathbb{R}^{10}$ with D6-branes
wrapped on $L \subset X/U(1) \cong \mathbb{R}^{10}$. From
the identification of topologically stable objects
it follows that $L$ has non-trivial homology groups
\be
H_0 (L; \mathbb{Z}) = \mathbb{Z} \oplus \mathbb{Z}, \quad
H_2 (L; \mathbb{Z}) = \mathbb{Z}
\label{cplhom}
\ee
This agrees with the general result of \cite{MZ,Levine,Schultz}
that $F$ has to be a codimension 4 subspace of $Y = SU(3)/U(1)$.
In fact, a typical $U(1) \subset SU(3)$ has a fixed point
set $\{\mathrm{pt}\} \times {\bf \mathbb{C}P}^1$ inside ${\bf \mathbb{C}P}^2$.
Since $Y$ may be viewed as a homotopy 3-sphere
bundle over ${\bf \mathbb{C}P}^2$ it follows that the fixed point set must
be the union of a typical fibre and a circle bundle over ${\bf S}^2$.
In this example we find that
\be
L = \mathbb{R}^4 \cup \mathbb{R}^2 \times {\bf S}^2
\ee
which indeed has the homology groups (\ref{cplhom}). Again, the second
factor is more precisely given by the total space of the spin bundle of ${\bf S}^2$ \cite{sergeijames2}.


\sect{New Complete Non-Compact Spin(7) Manifolds}

In this section we present a new one-parameter family of complete metrics on the universal quotient bundle $\mathcal{Q}$ of
${\bf \mathbb{C}P}^2$. The method we use is a generalisation of the
procedure recently used to construct the hyper-K\"ahler Calabi
metrics on $T^*{\bf \mathbb{C}P}^n$, the cotangent bundle of
${\bf \mathbb{C}P}^n$ \cite{gary2}. In fact, the authors of that paper
also found a system of first order equations describing
$\mathrm{Spin}(7)$ metrics, and presented the solution (\ref{spinorb})
which lives on a $\mathbb{Z}_2$ orbifold of
the total space of $T^*{\bf \mathbb{C}P}^2$. As in section 2, one again assumes a
cohomogeneity one ansatz, with level surfaces this time taken to be the coset
space $SU(3)/U(1)$. There are an infinite number of distinct ways of embedding
the $U(1)$ in $SU(3)$, but the solutions in \cite{gary2} are described by the \emph{same} embedding. The new $\mathrm{Spin}(7)$ metrics presented
in this section correspond to a \emph{different} embedding of the circle
$U(1)$ in the group manifold $SU(3)$. Specifically, the manifold on
which the new family of $\mathrm{Spin}(7)$ metrics is
defined is the total space of the universal quotient bundle $\mathcal{Q}$.

The new $\mathrm{Spin}(7)$ metrics we have found here have recently been
generalised in \cite{garychris}, the results of which will be
summarised below. Their
local solutions that extend our solution (\ref{spin7}),
together with the global analysis of the next subsection, are
complementary. The upshot is that we obtain a set of
$\mathrm{Spin}(7)$ metrics defined on the total space of various cyclic quotients of any
$\mathrm{spin}^c$ structure over ${\bf \mathbb{C}P}^2$. We would like to thank the authors of
\cite{garychris} for sharing their results with us prior to publication.

\subsection{Aloff-Wallach Spaces}

We begin then with a discussion of the coset space $SU(3)/U(1)$. These
so-called Aloff-Wallach spaces were studied by various authors \cite{aloff}, mainly in the context of finding new Einstein
manifolds. We extend this work considerably, elucidating in
particular the global topology, and also make a connection between Aloff-Wallach spaces and the
space of $\mathrm{spin}^c$ structures over ${\bf \mathbb{C}P}^2$. These
results therefore describe the global topology of the local solutions
found recently in \cite{garychris}.

Let $G=SU(3)$. Every non-trivial circle in $G$ is of the form

\begin{equation}
T_{k,l}:e^{2\pi{i}t}\rightarrow\left\{\left[\begin{array}{ccc}e^{2\pi{i}kt} & 0 & 0 \\ 0 &
e^{2\pi{i}lt} & 0 \\ 0 & 0 &
e^{-2\pi{i}(k+l)t}\end{array}\right]\mid{t}\in\mathbb{R}/\mathbb{Z}\right\}\end{equation}

up to conjugation in $SU(3)$, where the integers $k$ and $l$ are not
both zero. For the action to be effective, which we assume, we require
that $k$ and $l$ be relatively prime, $\mathrm{gcd}(k,l)=1$. Then, by
definition, the Aloff-Wallach space $N_{k,l}$ is the quotient space

\begin{equation}
N_{k,l}\equiv SU(3)/U(1)=G/T_{k,l}\end{equation}

It will be necessary to consider various subgroups of $SU(3)$. In
particular, we define

\begin{equation}
H=\left\{\left[\begin{array}{cc}h & 0 \\ 0  &
\mathrm{det}h^{-1}\end{array}\right]\mid{h}\in{U}(2)\right\}\cong U(2)\end{equation}

so that\footnote{Note that it is $SU(3)/\mathbb{Z}_3$ and not $SU(3)$ that acts
effectively here.}

\begin{equation}
{\bf \mathbb{C}P}^2=SU(3)/U(2)=G/H\end{equation}

We also let $T^2 = U(1) \times U(1) \subset H \subset SU(3)$ be a maximal torus in $SU(3)$. Note that any two maximal tori are conjugate.

Various $N_{k,l}$ spaces are related by conjugation in
$SU(3)$. Specifically, it is the Weyl group of $SU(3)$ that permutes various Aloff-Wallach representations of the same manifold. Recall that for any compact connected Lie group $G$ the Weyl group may be defined as the centraliser of the maximal torus. For $G=SU(3)$, the maximal torus is $T^2$, and its centraliser is $\Sigma_3$, the group of permutations of three elements. The embedding of this group into $SU(3)$ is defined by the group of permutation matrices in $SU(3)$. For example, the element

\be
w_{(2)} \equiv -\left[\begin{array}{ccc} 0 & 1 & 0 \\ 1 &
0 & 0 \\ 0 & 0 &
1\end{array}\right]\in\Sigma_3\subset SU(3)\label{w2}\ee

has order two in $\Sigma_3$ and one may easily verify that

\be
w_{(2)}^{-1} \cdot T_{k,l} \cdot w_{(2)} = T_{l,k}\ee

Thus $w_{(2)} \in \Sigma_3$ maps $T_{k,l} \mapsto T_{l,k}$. Similarly, one can construct analogous order two elements that map $T_{k,l}$ to $T_{k,-k-l}$ and $T_{-l-k,l}$, respectively. There is also an element of order three, which one may take to be

\be
w_{(3)} \equiv \left[\begin{array}{ccc}0 & 1 & 0 \\ 0 &
0 & 1 \\ 1 & 0 &
0\end{array}\right]\in\Sigma_3\subset SU(3)\label{w3}\ee

and generates the group of even permutations $\mathbb{Z}_3 \subset \Sigma_3$. Applying $w_{(3)}$ repeatedly one finds

\be
\mathbb{Z}_3 \colon \quad \quad
\matrix{ && T_{k,l} && \cr
& \nearrow && \searrow & \cr
T_{-k-l,k} & & \longleftarrow & & T_{l,-k-l} }\label{cyclic}
\ee

The action of the Weyl group therefore permutes equivalent Aloff-Wallach spaces. Notice that the equivalence class of $N_{1,1}$ is in some sense degenerate, since $w_{(2)}$ maps $N_{1,1}$ into the same representation.

There is also a $\Sigma_2 = \mathbb{Z}_2$ group of outer
automorphisms that commutes with $\Sigma_3$ and acts by complex conjugation on $SU(3)$
(inducing complex conjugation on ${\bf \mathbb{C}P}^2$). On the Aloff-Wallach space $N_{k,l}$ this
group action induces an isomorphism:

\be
\Sigma_2 \colon \quad N_{k,l} \cong N_{-k,-l}
\ee

for every $k$ and $l$. The actions of $\Sigma_2$ and $\Sigma_3$ are
independent, except for the equivalence class of $N_{1,-1}$. In this
case, the action of complex conjugation introduces no new
representations. For example, the generator of $\Sigma_2$ acts in the same way as $w_{(2)}$ on $N_{1,-1}$.

Since the equivalence classes of $N_{1,1}$ and $N_{1,-1}$ are
exceptional, we refer to these as the exceptional Aloff-Wallach
spaces. The fact that the exceptional Aloff-Wallach spaces behave
differently under the symmetry groups will also show up in the
geometry, as we shall see later.

Consider the following tautological bundle diagram

\be
\matrix{ \hspace{1.2in} T^2 & & \cr \hspace{0.2in} \swarrow & \searrow & \cr H
\hspace{0.1in} \hookrightarrow & G \stackrel{p}{\longrightarrow} & G/H \cr
\downarrow &  \downarrow && \cr H/T^2 \hookrightarrow & G/T^2
\longrightarrow & G/H}\ee

The nested sequence $T_{k,l}\subset T^2\subset H \subset G$ allows us to factor through the diagram by $T_{k,l}$

\be
\matrix{ \hspace{1.2in} T^2/T_{k,l} & & \cr \hspace{0.2in} \swarrow & \searrow & \cr H/T_{k,l}
\hspace{0.1in} \hookrightarrow & G/T_{k,l} \longrightarrow & G/H \cr
\downarrow &  \downarrow && \cr H/T^2 \hookrightarrow & G/T^2
\longrightarrow & G/H}\ee

More concretely, this diagram reads

\be
\matrix{ \hspace{1.2in} U(1) & & \cr \hspace{0.2in} \swarrow & \searrow & \cr J_{k,l}
\hspace{0.1in} \hookrightarrow & N_{k,l} \stackrel{\pi}{\longrightarrow} & {\bf \mathbb{C}P}^2 \cr
\downarrow &  \downarrow && \cr {\bf S}^2 {\hookrightarrow} & \mathcal{S}\Lambda^-
\stackrel{P}{\longrightarrow} & {\bf \mathbb{C}P}^2}\label{diagram}\ee

where $\mathcal{S}\Lambda^-$ is the twistor space of
${\bf \mathbb{C}P}^2$, and we have defined the quotient space
$J_{k,l}=H/T_{k,l}$. Let us examine the diagram (\ref{diagram}) in more detail.

Reading
down the first column, we see that $J_{k,l}$ may be viewed as the total space of a
$U(1)$ bundle over ${\bf S}^2$. Such bundles are in 1-1 correspondence
with the first Chern number $\mathbb{Z}\cong{H}^2({\bf S}^2;\mathbb{Z})$,
the total space then being
the Lens space $L(1,N)={\bf S}^3/\mathbb{Z}_N$ for $N\in\mathbb{Z^+}$. Changing the
sign of $N$ simply reverses the orientation. When the first Chern class
is zero, one has the trivial bundle ${\bf S}^1\times {\bf S}^2$, which, for
convenience, we define to be the Lens space $L(1,0)$. Hence the
quotient manifold $J_{k,l}$ is a Lens space.

The second column says that $N_{k,l}$ may be viewed as a $U(1)$ bundle over the twistor space
of ${\bf \mathbb{C}P}^2$. This fact will be extremely important in determining the relation between
Aloff-Wallach spaces and the space of
$\mathrm{spin}^c$ structures on ${\bf \mathbb{C}P}^2$.

The second row of course describes the twistor space as the total
space of an ${\bf S}^2$ bundle over ${\bf \mathbb{C}P}^2$, with projection map
$P:\mathcal{S}\Lambda^-\rightarrow {\bf \mathbb{C}P}^2$.

Finally, the first row says that $N_{k,l}$ may also be considered as a bundle over ${\bf \mathbb{C}P}^2$
with fibre $J_{k,l}=H/T_{k,l}$ and structure group $H\cong{U}(2)$. More precisely, $\pi$ is the associated bundle, with fibre $H/T_{k,l}$, to
the $H$-principal bundle $p:G\rightarrow{G}/H$. Thus the Aloff-Wallach
space $N_{k,l}$ may be viewed as a Lens space bundle over
${\bf \mathbb{C}P}^2$.

Indeed, by conjugation in $SU(3)$, each $N_{k,l}$ may be viewed as
various different Lens space bundles over ${\bf \mathbb{C}P}^2$. We would
like to understand precisely which Lens spaces occur. To see this,
note that the embedding of $T_{p,q}\cong U(1)$ in $H \cong U(2)$ is defined by

\begin{equation}
e^{2\pi{i}t}\rightarrow\left\{\left[\begin{array}{cc}e^{2\pi{i}pt} & 0 \\ 0  &
e^{2\pi{i}qt}\end{array}\right]\subset{U}(2)\mid{t}\in\mathbb{R}/\mathbb{Z}\right\}\end{equation}

Now apply $S$ to the quotient $U(2)/T_{k,l}$. This gives
$SU(2)/S(T_{p,q})$. Now of course $SU(1)$ is trivial, but the $S$ here
refers to the $U(2)$ determinant. Clearly

\be
\det \left[\begin{array}{cc}e^{2\pi{i}pt} & 0 \\ 0  &
e^{2\pi{i}qt}\end{array}\right] = e^{2\pi i (p+q)t}\ee

and hence the subgroup of $T_{p,q}=U(1)\subset U(2)$ consisting of
matrices with unit determinant is given by $\{r/|p+q| \mid 0\leq r
<|p+q|, r \in \mathbb{Z}\}\cong \mathbb{Z}_{|p+q|}\subset U(1)$. Thus

\be
J_{p,q} = U(2)/T_{p,q} = SU(2)/S(T_{p,q}) = {\bf S}^3/\mathbb{Z}_{|p+q|} =
L(1,|p+q|)\ee

In order to specify an Aloff-Wallach space, one needs to give two (relatively prime) integers $k$ and $l$. However, in some sense it is more natural to give a triple $[k,l,-(k+l)]$. The manifold $N_{k,l}$ may then be viewed as an $L(1,N)$ bundle over ${\bf \mathbb{C}P}^2$, where

\be
N = |k|, |l|, \ \mathrm{or} \ |k+l|\label{triple}
\ee

the choices being permuted by the Weyl group $\Sigma_3$.

The reader should now find that some of the remarks at the end of section 2 have become more transparent. A generic $N_{k,l}$ space may be viewed as three
different Lens space bundles over ${\bf \mathbb{C}P}^2$ (\ref{triple}), permuted by the
cyclic group (\ref{cyclic}), whereas the exceptional cases have only
two bundle structures. Specifically, $N_{1,1}$ may be viewed as both an $L(1,1)={\bf S}^3$ and an
$L(1,2)={\bf S}^3/\mathbb{Z}^2={\bf \mathbb{R}P}^3$ bundle over
${\bf \mathbb{C}P}^2$. The $\mathrm{Spin}(7)$ metric (\ref{spinorb}) on
$T^*{\bf \mathbb{C}P}^2/\mathbb{Z}_2$ corresponds to the latter case,
whereas the former case describes the hyper-K\"ahler Calabi metric
(\ref{hyper}).

Similarly, the space $N_{1,-1}$ may be viewed as an
$L(1,1)={\bf S}^3$ and an $L(1,0)={\bf S}^1\times {\bf S}^2$ bundle over
${\bf \mathbb{C}P}^2$. The former case is relevant to the new $\mathrm{Spin}(7)$ metrics on the
universal quotient bundle $\mathcal{Q}$.

Finally, for future reference, we note that the fourth cohomology
group of $N_{k,l}$ is known \cite{aloff} to be

\be
H^4(N_{k,l};\mathbb{Z})\cong\mathbb{Z}_r \label{h4aloff}\end{equation}

where $r=|k^2+l^2+kl|$.

\subsection{The Relation to $\mathbf{\mathrm{Spin}^c}$ Structures}

In the construction of new special holonomy metrics, the Aloff-Wallach
spaces play the r$\mathrm{\hat{o}}$le of the level surfaces of a cohomogeneity one
metric. As we have just demonstrated, each Aloff-Wallach space may be viewed as various circle bundles over the twistor
space of ${\bf \mathbb{C}P}^2$ and also as various Lens space bundles over
${\bf \mathbb{C}P}^2$, related by the action of $\Sigma_3$. The Lens space bundles correspond to the boundary
of an $\mathbb{R}^4/\mathbb{Z}_N$ bundle over ${\bf \mathbb{C}P}^2$, which
we get by ``filling in'' the Lens space $L(1,N)$ with
$\mathbb{R}^4/\mathbb{Z}_N$ on each fibre. We
would like to understand precisely \emph{which} $\mathbb{R}^4$ bundles
arise in this way. In order to do so, we will first of all need to
recall some facts about $\mathrm{spin}^c$ structures.

Locally, one may always lift an $SO(n)$ bundle (or equivalently an oriented
$n$-plane bundle) to its double cover,
$\mathrm{Spin}(n)$. However, in general there is a global obstruction
to doing this, measured by a certain mod two cohomology class called the second
Stiefel-Whitney class. When this class vanishes, the bundle is said to
admit a spin structure. In the case of ${\bf \mathbb{C}P}^2$, one finds that
the second Stiefel-Whitney class $w_2=w_2(T{\bf \mathbb{C}P}^2)$ of the
tangent bundle is the generator of
$H^2({\bf \mathbb{C}P}^2;\mathbb{Z}_2)\cong\mathbb{Z}_2$, and hence there is
a global obstruction to lifting the tangent bundle
$SO(4)\rightarrow\mathrm{Spin}(4)\cong{S}U(2)\times{S}U(2)$. Consequently, one
may only define the $\mathrm{spin}$ bundle
$\Sigma=\Sigma_+\oplus\Sigma_-$ locally.

However, for a four-manifold $B$, there is always a complex line
bundle $L\rightarrow{B}$ over $B$ with first Chern class
$c_1=c_1(L)\in{H}^2(B;\mathbb{Z})$ such that $c_1$ reduces to $w_2(E)$,
modulo two, for any oriented vector bundle $E$. For example, in the
case of the tangent bundle, one may formally construct the
rank-two complex vector bundles

\be
\mathbb{V}_{\pm}=\mathbb{V}_{\pm}(L)=\Sigma_{\pm}\otimes{L}^{1/2}\ee

These are known as $\mathrm{spin}^c$ bundles. The point is that the
sign problems that one encounters in trying to consistently define the
transition functions of the spin bundles $\Sigma_{\pm}$ are precisely
cancelled by the ambiguity in taking the square roots of the
transition functions of $L$.

Of course, the choice of $L$ is not
unique here; we are free to tensor $\mathbb{V}_{\pm}$ with any complex line
bundle $M$ to obtain another $\mathrm{spin}^c$ bundle. This generates a free transitive action of the group
$H^2(B;\mathbb{Z})$ on the space of $\mathrm{spin}^c$ structures. In
particular, tensoring $\mathbb{V}_{\pm}(L)$ with $M$ shifts the first
Chern class $c_1\equiv c_1(L)=c_1(\mathbb{V}_{\pm})$ by

\be
c_1\rightarrow{c}_1+2a\ee

where $a=c_1(M)$. If $H^2(B;\mathbb{Z})$ is torsion-free (as is the
case for $B={\bf \mathbb{C}P}^2$), then the space of $\mathrm{spin}^c$
structures is in 1-1 correspondence with the set of cohomology classes
that reduce modulo two to the second Stiefel-Whitney class. Specialising this discussion to the case $B={\bf \mathbb{C}P}^2$, we see
that the space of $\mathrm{spin}^c$ structures on ${\bf \mathbb{C}P}^2$
correspond to the ``odd'' classes in
$H^2({\bf \mathbb{C}P}^2;\mathbb{Z})\cong\mathbb{Z}$.

What are some of these $\mathrm{spin}^c$ structures? First, note that
for any K\"ahler manifold $B$, there is a canonical choice for $L$, and
therefore a canonical choice of $\mathrm{spin}^c$ structure. Namely,
one may take $L=K^{-1}$, where $K=\Lambda^{2,0}B$ is the canonical line
bundle, and $\Lambda^{p,0}B$ denotes the bundle of holomorphic $p$-forms
over $B$. The first Chern class $c_1(K)$ is then also the first Chern class
of $B$. One finds that

\begin{eqnarray}
\mathbb{V}_+(K^{-1})=\Lambda^{0,0}\oplus\Lambda^{0,2}=1\oplus K^{-1}
\nonumber \\
\mathbb{V}_-(K^{-1})=\Lambda^{0,1}=T^+\end{eqnarray}

where in the last line $T^+=T^+B$ denotes the holomorphic tangent
bundle of $B$. For future reference, in the case of $B={\bf \mathbb{C}P}^2$,
we note that the total Chern class of this bundle is
$c(T^+)=(1+x)^3=1+3x+3x^2$ where $x$ generates the cohomology ring of
${\bf \mathbb{C}P}^2$. In particular, the Euler number of $T{\bf \mathbb{C}P}^2$
is $3x^2[{\bf \mathbb{C}P}^2]=3$.

Instead, we may now tensor the canonical $\mathrm{spin}^c$ structure
with the line bundle\footnote{In fact, this complex line bundle is precisely the universal subbundle $S$, defined below.} $M$ whose first Chern class is $-x$. We find in particular that

\be
\mathbb{V}_-=\mathcal{Q}\ee

is the universal quotient bundle of ${\bf \mathbb{C}P}^2$. This is defined
as follows \cite{bott}. Consider the tautological exact sequence over
${\bf \mathbb{C}P}^2$

\be
0\rightarrow{S}\rightarrow{\bf \mathbb{C}P}^2\times\mathbb{C}^3\rightarrow\mathcal{Q}\rightarrow0\ee

Here, $S$ is the universal subbundle, defined as

\begin{equation}
S=\{(l,z)\in{\bf \mathbb{C}P}^2\times\mathbb{C}^3\mid{z}\in{l}\}\end{equation}

That is, the fibre of $S$ above the point $l\in{\bf \mathbb{C}P}^2$ consists
of all points in $l$, where $l$ is now viewed as a complex line in
$\mathbb{C}^3$. We can think of $S$ as being obtained from
$\mathbb{C}^3$ by blowing up the origin, replacing it by a copy of
${\bf \mathbb{C}P}^2$. The dual bundle $S^*$ is known as the hyperplane
bundle. The Chern class of $\mathcal{Q}$ is easily computed to be
$c=1+x+x^2$. In particular, the Euler number of the underlying real
vector bundle is 1.

The reason for this detour on $\mathrm{spin}^c$ bundles is that every Aloff-Wallach space $N_{k,l}$ may be viewed as the boundary of some
negative $\mathrm{spin}^c$ bundle $\mathbb{V}_-(L)$, or rather a $\mathbb{Z}_N$
quotient of such a bundle. However, in order to see this
correspondence, we will have to make another slight digression.

Although the spin bundles $\Sigma_{\pm}$ do not in general exist
globally on a four-manifold $B$, their projectivisations do exist as
genuine bundles. In fact

\be
\mathbb{P}\Sigma_{\pm}=\mathbb{P}\mathbb{V}_{\pm}=\mathcal{S}\Lambda^{\pm}\ee

independently of the choice of line bundle $L$. Recall that for any
complex vector bundle $E$, one may define its projectivisation
$\mathbb{P}E$ by replacing each complex fibre
with the complex projective space one obtains by quotienting the complement of
zero by the natural action of $\mathbb{C}^*$ (essentially, we quotient by
the Hopf map on each fibre). The transition functions
are simply those naturally induced by this construction.

In particular, if $E$ has complex rank two, then its projectivisation has
fibre ${\bf \mathbb{C}P}^1={\bf S}^2$. It follows that the boundary of
$\mathbb{V}_-$, which is an ${\bf S}^3$ bundle over $B$, is the
total space of a $U(1)$ bundle over the twistor space
$\mathcal{S}\Lambda^-$ of $B$. The associated complex line bundle is known as
the hyperplane bundle\footnote{This is \emph{{\bf
not}} to be confused with the universal hyperplane bundle above, which
is a complex line bundle over ${\bf \mathbb{C}P}^2$.} $S^*$ of
$\mathbb{V}_-$. So, the boundary of the $\mathrm{spin}^c$ bundle
$\mathbb{V}_-$ is the total space of a $U(1)$ bundle over the twistor
space of ${\bf \mathbb{C}P}^2$. This makes the connection between
$\mathrm{spin}^c$ structures and Aloff-Wallach spaces - the latter may
also be viewed as circle bundles over the twistor space, as we see
from the rather useful diagram (\ref{diagram}). Let us
investigate this relation more carefully.

Complex line bundles over the twistor space are in 1-1 correspondence
with elements of the second cohomology group
$H^2(\mathcal{S}\Lambda^-;\mathbb{Z})$. For example, in the case of
$B={\bf \mathbb{C}P}^2$, this group is isomorphic to $\mathbb{Z}\oplus\mathbb{Z}$.
The two generators may be roughly thought of as the ${\bf \mathbb{C}P}^1$ that is linearly
embedded in ${\bf \mathbb{C}P}^2$ (and generates
$H_2({\bf \mathbb{C}P}^2;\mathbb{Z})$), and the ${\bf \mathbb{C}P}^1$ fibre. In general, if we denote $c_1(S^*)=y$ then
$y\in{H}^2(\mathbb{P}\mathbb{V}_-;\mathbb{Z})$, and the restriction
of $y$ to each ${\bf \mathbb{C}P}^1={\bf S}^2$ fibre of the twistor space generates the
cohomology of the fibre. It follows from the Leray-Hirsch Theorem
\cite{bott} that the cohomology ring of the twistor space
$H^*(\mathcal{S}\Lambda^-;\mathbb{Z})$ is a free module over
$H^*(B;\mathbb{Z})$. Specifically,

\be
H^*(\mathcal{S}\Lambda^-;\mathbb{Z})=H^*(\mathbb{P}\mathbb{V}_-;\mathbb{Z})=H^*(B;\mathbb{Z})[y]/(y^2+c_1y+c_2)\label{cring}\ee

where $c_i=c_i(\mathbb{V}_-)$ are the Chern classes of the rank two
complex vector bundle $\mathbb{V}_-$. This view of the Chern
classes of a complex vector bundle was originally due to
Grothendieck\footnote{See \cite{bott} for an excellent summary.}.

Now, the restriction of the
unit sphere bundle in $S^*$ to any ${\bf S}^2$ fibre of the twistor space of
course describes the Hopf map, $\mathcal{H}:{\bf S}^3\rightarrow{\bf S}^2$. Indeed, if
$P:\mathcal{S}\Lambda^-\rightarrow B$ denotes the
projection map for the twistor space, viewed as an ${\bf S}^2$ bundle over $B$, then

\be
P_*y=1\label{hopf}\ee

This equation in fact determines the cohomology class $y$ uniquely, up to
the addition of the pull-back under $P$ of an element of
$H^2(B;\mathbb{Z})$. But this is just the
choice of $\mathrm{spin}^c$ structure. To see this, note that tensoring $\mathbb{V}_-$ with a line bundle
$M$ with first Chern class $c_1(M)=a$ shifts

\begin{eqnarray}
c_1\rightarrow{c}_1+2a \nonumber \\
c_2\rightarrow{c}_2+c_1a+a^2\end{eqnarray}

which is the free transitive action of
$H^2(B;\mathbb{Z})$ on the space of $\mathrm{spin}^c$
structures. This is entirely equivalent to shifting the generator
$y\rightarrow{y}+a$, in the representation (\ref{cring}), and is
therefore equivalent to tensoring the hyperplane bundle
$S^*\rightarrow{S}^*\otimes{M}$, as one would expect. The latter action is
free and transitive on the space of circle bundles over the twistor space of
$B$ satisfying the condition (\ref{hopf}). Hence every such circle bundle (or equivalently
complex line bundle) arises in this way.

We are now ready to make the connection with the Aloff-Wallach
spaces. First, we consider a special case. Which $N_{k,l}$ spaces may
be viewed as ${\bf S}^3$ bundles over ${\bf \mathbb{C}P}^2$? We require $|k+l|=1$,
which implies that $k=\pm1-l$. Thus
$N_{k,l}=N_{l,-k-l}=N_{l,\mp1}=N_{\mp1,l}$. Without loss of
generality, we may take the plus sign, and hence we conclude that the
only Aloff-Wallach spaces that may be viewed as three-sphere bundles
over ${\bf \mathbb{C}P}^2$ are, up to conjugacy, of the form $N_{1,p}$ for
$p\in\mathbb{Z}$. Hence these spaces may be viewed as circle bundles
over the twistor space satisfying (\ref{hopf}), and therefore must be
the boundary of some $\mathrm{spin}^c$ structure.

In fact, it is not hard to see that \emph{every} $\mathrm{spin}^c$ structure arises in this way. Combining various formulae from above, it is easy to compute the Euler class of $\mathbb{V}_-(L)_{\mathbb{R}}$. Specifically, it is given by

\be
e(\mathbb{V}_-(L)_{\mathbb{R}})=c_2(\mathbb{V}_-(L))=x^2+\frac{1}{4}(c_1(L)-x)(c_1(L)+x)\ee

Hence the Euler number of the $\mathrm{spin}^c$ bundle of ${\bf \mathbb{C}P}^2$ with line bundle $L$ is

\be
\chi(\mathbb{V}_-(L)_{\mathbb{R}}) = 1+\frac{n^2-1}{4} \label{euler}\ee

where $c_1(L)=nx$ with $n$ an odd integer. Now examine the long exact cohomology sequence for the pair $(X,\partial X)$, where $X=\mathbb{V}_-(L)$ is the total space of some $\mathrm{spin}^c$ bundle. This reads

\be
\ldots H^4(X,\partial X;\mathbb{Z})\stackrel{f}{\longrightarrow}H^4(X;\mathbb{Z})\stackrel{i^*}{\longrightarrow}H^4(\partial X;\mathbb{Z})\stackrel{\delta^*}{\longrightarrow}H^5(X,\partial X;\mathbb{Z})\longrightarrow \ldots \label{exact}\ee

where $i:\partial X\hookrightarrow X$ denotes inclusion, and $H^4(X,\partial X;\mathbb{Z})\cong H^4_{\mathrm{cpct}}(X;\mathbb{Z})$ is the same as the compactly supported cohomology. The map $f$ "forgets" that a cohomology class has compact support. By the isomorphism of Thom \cite{milnor}, we have

\be
H^5(X,\partial X;\mathbb{Z}) \cong H^1({\bf \mathbb{C}P}^2;\mathbb{Z}) \cong 0\ee

since $X$ is the total space of a rank four bundle over ${\bf \mathbb{C}P}^2$. Hence the exact cohomology sequence (\ref{exact}) implies that

\be
H^4(\partial X;\mathbb{Z}) \cong H^4(X;\mathbb{Z})/f(H^4_{\mathrm{cpct}}(X;\mathbb{Z}))\ee

Now, the self-intersection number of ${\bf \mathbb{C}P}^2$ inside $X$ is given by the Euler number $\chi = [{\bf \mathbb{C}P}^2].[{\bf \mathbb{C}P}^2]$ \cite{bott}. The lattice $H^4_{\mathrm{cpct}}(X;\mathbb{Z})\cong H_4(X;\mathbb{Z}) \cong \mathbb{Z}$ is generated by $[{\bf \mathbb{C}P}^2]$, but the dual lattice $H^4(X;\mathbb{Z})$ is generated by $\frac{1}{\chi}[{\bf \mathbb{C}P}^2]$, since this has a scalar product of 1 with $[{\bf \mathbb{C}P}^2]$. Putting these facts together, it follows that

\be
H^4(\partial \mathbb{V}_-(L);\mathbb{Z})\cong \left(\frac{1}{\chi}\mathbb{Z}\right)/\mathbb{Z}\cong \mathbb{Z}_{\chi}\ee

Since the boundary of each $\mathrm{spin}^c$ bundle is supposed to be an Aloff-Wallach space of the form $N_{1,p}$, this is to be compared with the formula (\ref{h4aloff}), which reads

\be
H^4(N_{1,p};\mathbb{Z}) \cong \mathbb{Z}_r\ee

where $r=|1+p+p^2|$. One easily sees that the formula relating $n$ (which determines the $\mathrm{spin}^c$ bundle) to $p$ (which determines the Aloff-Wallach space $N_{1,p}$) is

\be
n = 2p+1\ee

Indeed, substituting this expression into (\ref{euler}) gives

\be
\chi = 1+\frac{n^2-1}{4} = 1 + \frac{(2p+1)^2-1}{4} = 1+p+p^2\ee

in agreement with (\ref{h4aloff}). This shows the 1-1 correspondence explicitly. We note in passing that $N_{1,0}$ ($n=1$) corresponds to the boundary of the universal quotient bundle $\mathcal{Q}$ and $N_{1,1}$ ($n=3$) corresponds to the cotangent bundle $T^*{\bf \mathbb{C}P}^2$.

It is now simple to pass to the general case. Each $N_{k,l}$ space may
be viewed as some $L(1,N)$ bundle over ${\bf \mathbb{C}P}^2$ for some
(various) $N$, given by (\ref{triple}). We may lift this bundle to its covering ${\bf S}^3$ bundle,
which, as we have just seen, is the boundary of some $\mathrm{spin}^c$
bundle. Hence, in general, $N_{k,l}$ is the boundary of some $\mathbb{Z}_N$
quotient of a $\mathrm{spin}^c$ bundle. Note that projectivising this cyclic quotient, we find that the Chern class of the hyperplane bundle is now given by $N$ times the generator $y$.

We conclude this subsection with some remarks about Einstein
metrics on Aloff-Wallach spaces \cite{aloff}. There exist two
inequivalent Einstein metrics on each of the unexceptional $N_{k,l}$
spaces. These all have isometry group $SU(3)\times{U}(1)$ and weak
$G2$ holonomy. The latter fact means that the cones over these spaces
are $\mathrm{Spin}(7)$ conifolds. The solutions found in
\cite{garychris} are resolutions of these conifolds. Of course,
as we have explained in this section, typically these resolutions have orbifold singularities, and so
strictly they are singular. The point is though that the orbifold singularity
is much milder than the curvature singularity one encounters at the
base of the conifolds. Indeed, the $A_{N-1}$ orbifold singularity will
later be interpreted in Type IIA string theory as $N$ coincident D6-branes, so
the singular nature of these solutions is actually rather desirable.

In contrast, $N_{1,-1} = N_{0,1}$ has one known Einstein
metric, which also has weak $G2$ holonomy and the same isometry group
as the unexceptional cases. Again, there is a resolution of this cone
\cite{garychris}. This is \emph{not} the same as the
$\mathrm{Spin}(7)$ metric presented in the next section, which is
asymptotically locally conical rather than asymptotically conical.

$N_{1,1}$ has
two Einstein metrics, both with isometry groups $SU(3)\times SU(2)$, being respectively weak $G2$ and
tri-Sasakian. This means that the cones over these spaces are
respectively $\mathrm{Spin}(7)$ and hyper-K\"ahler. Indeed, the
resolutions of these cones are nothing but the metrics
(\ref{spinorb}) and (\ref{hyper}), respectively.

\subsection{Construction of New Spin(7) Metrics}

We turn now to the construction of the metrics. This generalises the work of \cite{gary2}. We start by defining the
generators of $SU(3)$, together with the associated left-invariant
one-forms, which we denote $L_A^{\ \ B}$, satisfying the exterior algebra

\be
dL_A^{\ \ B}=iL_A^{\ \ C}\wedge L_C^{\ \ B}\label{ext}\ee

One must then split the generators into those that lie in the coset
$SU(3)/U(1)$ and those that lie in the denominator $U(1)$. In
particular, one must specify the $U(1)$ generator, $Q$. In
the main text of \cite{gary2}, the choice of $U(1)$ generator corresponds to the
Aloff-Wallach space $N_{1,1}$. The more general case was briefly mentioned
in their Appendix C, although we shall use different notations. The
generalisation to $N_{k,l}$ is obtained by setting

\be
Q\equiv \sqrt{\frac{2}{k^2+l^2}}\left(kL_1^{\ \ 1}+lL_2^{\ \ 2}\right)\ee

where the normalisation is chosen to coincide with Appendix C of
\cite{gary2}. There is one other $U(1)$ generator, $\lambda$, which lives in the coset
space $SU(3)/U(1)$. This is orthogonal to $Q$, and together $Q$ and
$\lambda$ generate the maximal torus of $SU(3)$. Specifically, we take

\be
\lambda\equiv \sqrt{\frac{2}{k^2+l^2}}\left(lL_1^{\ \ 1}-kL_2^{\ \ 2}\right)\ee

The remaining generators of $N_{k,l}$ are then

\be
\sigma\equiv L_1^{\ \ 3}, \quad \Sigma\equiv L_2^{\ \ 3}, \quad
\nu\equiv L_1^{\ \ 2}\ee

These are all complex, so one may split them into real and imaginary
parts

\be
\sigma\equiv \sigma_1+i\sigma_2, \quad \Sigma\equiv \Sigma_1+i\Sigma_2,
\quad \nu\equiv \nu_1+i\nu_2\ee

The exterior algebra (\ref{ext}) then reduces to

\begin{eqnarray}
d\sigma_1 & = & -\alpha(2l-k)\lambda\wedge\sigma_2 - \alpha(2k+l)Q\wedge\sigma_2 - \nu_1\wedge\Sigma_2 - \nu_2\wedge\Sigma_1  \nonumber \\
d\sigma_2 & = & \alpha(2l-k)\lambda\wedge\sigma_1 + \alpha(2k+l)Q\wedge\sigma_1 + \nu_1\wedge\Sigma_1 -
\nu_2\wedge\Sigma_2 \nonumber \\
d\Sigma_1 & = & -\alpha(l-2k)\lambda\wedge\Sigma_2 - \alpha(k+2l)Q\wedge\Sigma_2 - \nu_1\wedge\sigma_2 +
\nu_2\wedge\sigma_1 \nonumber \\
d\Sigma_2 & = & \alpha(l-2k)\lambda\wedge\Sigma_1 + \alpha(k+2l)Q\wedge\Sigma_1 + \nu_1\wedge\sigma_1 +
\nu_2\wedge\sigma_2 \nonumber \\
d\nu_1 & = & -\alpha(k+l)\lambda\wedge\nu_2 -\alpha(k-l)Q\wedge\nu_2 - \sigma_2\wedge\Sigma_1 +
\sigma_1\wedge\Sigma_2 \nonumber \\
d\nu_2 & = & \alpha(k+l)\lambda\wedge\nu_1 + \alpha(k-l)Q\wedge\nu_1
+ \sigma_1\wedge\Sigma_1 +
\sigma_2\wedge\Sigma_2 \nonumber \\
d\lambda & = & 4\alpha\left[(k+l)\nu_1\wedge\nu_2 +
l\sigma_1\wedge\sigma_2 - k\Sigma_1\wedge\Sigma_2\right] \nonumber \\
dQ & = & 4\alpha\left[(k-l)\nu_1\wedge\nu_2 + k\sigma_1\wedge\sigma_2 +l\Sigma_1\wedge\Sigma_2\right]\label{algebra}
\end{eqnarray}
where the constant $\alpha=\alpha(k,l)=1/\sqrt{2(k^2+l^2)}$.


It is easy to see how the Weyl group $\Sigma_3$ acts on these forms. For example, the element $w_{(2)}\in \Sigma_3$ defined by (\ref{w2}) permutes the labels "1" and "2" and hence induces

\begin{eqnarray}
\nu & \mapsto & \bar{\nu} \nonumber \\
\sigma & \mapsto & \Sigma \nonumber \\
\Sigma & \mapsto & \sigma\end{eqnarray}

On the other hand, the order three element $w_{(3)}$ defined by (\ref{w3}) cyclically permutes (up to complex conjugation) the three one-forms:

\begin{eqnarray}
\nu & \mapsto & \bar{\sigma} \nonumber \\
\sigma & \mapsto & \bar{\Sigma} \nonumber \\
\Sigma & \mapsto & \nu\end{eqnarray}

One may check that performing this transformation twice more brings one back to the initial ordering. On the other hand, the $\mathbb{Z}_2$ group of outer automorphisms of $SU(3)$ acts naturally by complex conjugation: $\Sigma \to \bar \Sigma$, {\it etc.}

The metric ansatz we take is the cohomogeneity one ansatz
\be
ds^2=dt^2+f^2\lambda^2+a^2(\sigma_1^2+\sigma_2^2)
+b^2(\Sigma_1^2+\Sigma_2^2)+c^2(\nu_1^2+\nu_2^2)\label{ansatz}\ee
where $a,b,c$ and $f$ are functions of the ``radial'' variable
$t$. The following system of first-order equations describe a solution
of $\mathrm{Spin}(7)$ holonomy

\begin{eqnarray}
\frac{\dot{a}}{a} & = &
\frac{b^2+c^2-a^2}{abc}-\frac{2\alpha lf}{a^2} \nonumber \\
\frac{\dot{b}}{b} & = &
\frac{a^2+c^2-b^2}{abc}-\frac{2\alpha kf}{b^2}
\nonumber \\
\frac{\dot{c}}{c} & = &
\frac{a^2+b^2-c^2}{abc}+\frac{2\alpha (k+l)f}{c^2}
\nonumber \\
\frac{\dot{f}}{f} & = & -\frac{2\alpha f(k+l)}{c^2}+\frac{2\alpha lf}{a^2}
+\frac{2\alpha kf}{b^2}\label{generalsys}\end{eqnarray}

where $\dot{a}=\frac{da}{dt}$,
etc. To see this, we introduce the orthonormal basis

\begin{eqnarray}
e^1 = \frac{dr}{f}, \quad e^2 = f\lambda, \quad e^3 = a\sigma_1, \quad e^4 = a\sigma_2, \nonumber \\
e^5 = c\nu_1, \quad e^6 = c\nu_2, \quad e^7 = b\Sigma_1,
\quad e^8 = b\Sigma_2
\end{eqnarray}

The metric (\ref{ansatz}) now becomes

\be
ds^2 = \sum_{i=1}^8 e^i \otimes {e}^i\ee

Denoting $e^{ijkl}=e^i\wedge{e}^j\wedge{e}^k\wedge{e}^l$, we find that

\begin{eqnarray}
\Phi = -e^{1234} + e^{1256} + e^{1278} - e^{1367} - e^{1358} - e^{1468} + e^{1457} \nonumber \\
+ e^{2368} - e^{2357} - e^{2467} - e^{2458} -
e^{3456} - e^{3478} + e^{5678}
\label{somega}
\end{eqnarray}

is self-dual, and imposing $d\Phi=0$ (which is equivalent to $\Phi$ being harmonic) precisely reproduces the
first-order system (\ref{generalsys}). The self-dual harmonic four-form $\Phi$ is precisely the Cayley form that determines the reduction of the structure group from $\mathrm{Spin}(8)$ to $\mathrm{Spin}(7)$.

Notice that setting $f\equiv0$ and $a\equiv{b}$ leads to the
following consistent truncation

\begin{eqnarray}
\dot{a} & = & \frac{c}{a} \nonumber \\
\dot{c} & = & 2-\frac{c^2}{a^2}\end{eqnarray}

independently of $k$ and $l$. These equations describe the known metric of $G2$ holonomy on the
bundle of anti-self-dual two-forms over ${\bf \mathbb{C}P}^2$, (\ref{G2}). This
is hardly surprising. Setting $f=0$ roughly corresponds to removing
the twisting due to the D6-branes (this will be explained in the next section), and $a=b$ yields the standard
Fubini-Study metric on the ${\bf \mathbb{C}P}^2$ base space.

\subsection{New Spin(7) Metrics on $\mathcal{Q}$}

Now consider setting $k=0$, $l=1$. Defining the new radial
coordinate $dr=Fdt$, where $F=\sqrt{2}f$, the general system (\ref{generalsys}) becomes

\begin{eqnarray}
a^{\prime} & = & \frac{b^2+c^2-a^2}{bcF}-\frac{1}{a} \nonumber \\
b^{\prime} & = & \frac{a^2+c^2-b^2}{acF} \nonumber \\
c^{\prime} & = & \frac{a^2+b^2-c^2}{abF}+\frac{1}{c} \nonumber \\
F^{\prime} & = & \frac{F}{a^2}-\frac{F}{c^2}\label{des}\end{eqnarray}

where a prime denotes derivative with respect to $r$. These equations are remarkably similar to the first-order equations
describing $G2$ metrics on the spin bundle of ${\bf S}^3$ found in
\cite{sergei}. In fact, the system possesses a similar $\mathbb{Z}_2$
symmetry

\be
r\rightarrow-r, \quad {a}\leftrightarrow{c}, \quad {b}\rightarrow-b \label{z2}\ee

We have been able to find the following explicit solution
\begin{eqnarray}
a^2 & = & \left(r+\frac{2}{3}\right)\left(r-\frac{4}{3}\right) \nonumber \\
b^2 & = & r^2 \nonumber \\
c^2 & = & \left(r-\frac{2}{3}\right)\left(r+\frac{4}{3}\right) \nonumber \\
F^2 & = & \frac{\left(r-\frac{4}{3}\right)\left(r+\frac{4}{3}\right)}{\left(r-\frac{2}{3}\right)\left(r+\frac{2}{3}\right)}\end{eqnarray}

This in fact gives us a one-parameter family of solutions. To see
this, note that rescaling the metric $g\rightarrow{\frac{9a^2}{4}}g$ and then
scaling $r\rightarrow{2r}/3a$ also gives us a metric of
$\mathrm{Spin}(7)$ holonomy, given explicitly by

\begin{eqnarray}
ds^2 & = & \frac{(r-a)(r+a)}{(r-2a)(r+2a)}dr^2+\frac{9a^2}{8}\frac{(r-2a)(r+2a)}{(r-a)(r+a)}\lambda^2+(r+a)(r-2a)(\sigma_1^2+\sigma_2^2)+\nonumber \\
&& +r^2(\Sigma_1^2+\Sigma_2^2)+(r-a)(r+2a)(\nu_1^2+\nu_2^2)\label{spin7}\end{eqnarray}

This is a complete metric on the universal quotient bundle
$\mathcal{Q}$ of ${\bf \mathbb{C}P}^2$,
for either $r>2a$ or $r<-2a$. As in \cite{sergei}, the
two solutions are interchanged by the $\mathbb{Z}_2$ symmetry. We
choose to take $r>2a$.

To see that this is a smooth complete metric on $\mathcal{Q}$, set $\rho^2=3a(r-2a)$. Then, near $\rho=0$, the metric approaches

\be
ds^2=\left[d\rho^2+\rho^2\left(\frac{1}{2} \lambda^2+\sigma_1^2+\sigma_2^2\right)\right]+4a^2(\nu_1^2+\nu_2^2+\Sigma_1^2+\Sigma_2^2)\ee

The metric in square brackets is in fact the standard Euclidean metric on
$\mathbb{R}^4$. This may be seen by rescaling the variables

\begin{eqnarray}
\eta_1 \equiv 2 \sigma_1 \nonumber \\
\eta_2 \equiv 2 \sigma_2 \nonumber \\
\eta_3 \equiv \sqrt{2}\lambda\end{eqnarray}


The three one-forms $\eta_i$ may now be written as $\eta_i=s_i+\ldots$
where the $s_i$ are a set of left-invariant one-forms on the $SU(2)\cong
{\bf S}^3$ fibres, and ``$\ldots$'' denotes additional terms that
describe the twisting of the fibres over the base ${\bf
\mathbb{C}P}^2$. These latter terms show up in the exterior algebra
(\ref{algebra}). The metric in square brackets is now

\be
d\rho^2+\frac{\rho^2}{4}(\eta_1^2+\eta_2^2+\eta_3^2)\label{r4}\ee

which is clearly the usual metric on $\mathbb{R}^4$ in spherical polars. The induced metric on $\rho=0$ is the standard
Fubini-Study metric on ${\bf \mathbb{C}P}^2$, up to a scale factor. Thus the
principal $N_{0,1}=SU(3)/U(1)$ orbits collapse smoothly down to a
${\bf \mathbb{C}P}^2$ bolt. One must be slightly cautious when interpreting the global structure of (\ref{r4}) - the periodicity of $\lambda$ may be such that we obtain an orbifold $\mathbb{R}^4/\mathbb{Z}_N$, with $N>1$, rather than $\mathbb{R}^4$. However, since the level surfaces are $N_{0,1}$, we know from our discussion of Aloff-Wallach spaces that $N_{0,1}=N_{1,0}$ may indeed be viewed as the boundary of an $\mathbb{R}^4$ bundle over ${\bf \mathbb{C}P}^2$ - specifically, a $\mathrm{spin}^c$ structure. As we demostrated in a previous section, this bundle is precisely the universal quotient bundle $\mathcal{Q}$.

At large distances ($r\rightarrow\infty$), the
metric asymptotes to the geometry

\be
ds^2=\frac{9a^2}{8}\lambda^2+dr^2+r^2(\sigma_1^2+\sigma_2^2+\nu_1^2+\nu_2^2+\Sigma_1^2+\Sigma_2^2)\ee

Thus the $U(1)$ fibres tend to a constant length while the other
directions expand linearly with radius. This behaviour is somewhat
similar to that of the Taub-NUT metric, which of course describes a
flat D6-brane in flat space. In the case at hand, the metric
asymptotes to ${\bf S}^1\times{C}(\mathcal{S}\Lambda^-)$ where $C(\mathcal{S}\Lambda^-)$
is the cone over the ${\bf \mathbb{C}P}^2$ twistor space
$\mathcal{S}\Lambda^-=SU(3)/T^2$. However, as for the $G2$ metric
on $\Lambda^-$, the twistor space metric asymptotes to
the squashed Einstein metric, rather than the usual one,
\cite{gary}. Our metric is therefore asymptotically locally conical,
or ALC. Notice that the functional form of the metric (\ref{spin7}) is extremely similar to that of the ALC $\mathrm{Spin}(7)$ metric on $\Sigma_-{\bf S}^4$ (\ref{alcs4}).

\sect{Flux Quantisation in M-Theory on Spin(7) Manifolds}

In this section we study in detail various M-theoretic aspects of our models. We find in particular that the M-theory lift of a Type IIA configuration of D6-branes wrapping a coassociative cycle is always described by the total space of a $\mathrm{spin}^c$ bundle. We study also the quantisation of the M-theory four-form
$G$ on the various $\mathrm{Spin}(7)$ manifolds of interest. We find that in all cases one must turn on a half-integral $G$-flux in order that the M-theory solutions be consistent.

\subsection{D6-branes Wrapping Coassociative Cycles}

We begin this subsection by describing the lift to M-theory of a configuration of D6-branes wrapping a coassociative cycle of a $G2$-manifold $M^7$. The mathematical description of this lift ties in very closely with our previous discussion of circle bundles over twistor spaces, and also with the work of McLean on the deformability of supersymmetric cycles \cite{mclean}.

Suppose then that the Type IIA manifold takes the form of a metric
product $\mathbb{R}^3\times{M^7}$ of Minkowski three-space with $M^7$, and
consider wrapping $N$ D6-branes over the submanifold
$W=\mathbb{R}^3 \times B$. As usual, $M^7$ is one of the non-compact $G2$ manifolds of section 2, and $B$
is a coassociative submanifold of $M^7$. By supersymmetry, the lift
of this configuration to M-theory should be described locally by a
manifold (or, more precisely, an orbifold) of $\mathrm{Spin}(7)$ holonomy \cite{gomis}.

From \cite{mclean} we know that $NB$ must be the bundle of anti-self-dual
two-forms $\Lambda^-B$ over $B$. Indeed, the explicitly known $G2$ metrics
presented in section (2.2) are defined on the total space of such a
bundle. Since we are primarily interested in the local physics near the
D6-branes, we may therefore restrict our attention to this case. Notice that $\Lambda^-B$ is $\mathrm{spin}^c$ since any oriented vector
bundle over a four-manifold $B$ is $\mathrm{spin}^c$ \cite{bryant}. This fact is important since $W$ must be $\mathrm{spin}^c$ in order that wrapping branes
on $W$ makes sense. Otherwise, one encounters
global anomalies on the string worldsheet \cite{wittenfreed}, and
there is no K-theoretic interpretation of D-brane charge
\cite{wittenK}.

The D6-branes are viewed, at the IIA
level, as probe branes. That is, one ignores the gravitational
back-reaction of the branes on the geometry. The presence of the branes implies that
the RR two-form $G_2$ has a delta-function singularity on $W$

\be
[dG_2] = N \delta(W)\label{mag}\ee

where $\delta(W)$ is Poincar\'e dual to the worldvolume $W$, and has support on $W$. Equation (\ref{mag}) is the statement that the D6-branes act as a magnetic source for $G_2$. However, on the complement of $W$,
$G_2$ is closed. We may then interpret $[G_2/2\pi]$ as the first Chern class of the
"M-theory circle bundle" over the complement of $W$. Thus $[\frac{G_2}{2\pi}]=c_1(\mathcal{L})$, where
$\mathcal{L}$ is a complex line bundle over $\mathbb{R}^3 \times M^7 \setminus W\equiv M^{10}_0$.

The total M-theory space $M^{11}$ is a degenerate circle bundle over the Type IIA manifold $M^{10}=\mathbb{R}^3 \times M^7$, with the circle fibres collapsing to zero on a copy of the D6-brane worldvolume $W$. More precisely, the complement of this "lift" of $W$ is the total space of the bundle of unit vectors $\mathcal{S}\mathcal{L}$ in $\mathcal{L}$, which in a tubular neighbourhood of $W$ is also an $L(1,N)$ bundle over $W$. This Lens space bundle is then "filled in" with the associated $\mathbb{R}^4/\mathbb{Z}_N$ bundle, whose zero section is the copy of $W$. On occasion, it will be important to distinguish logically between the brane worldvolume $W\subset M^{10}$ in Type IIA and its lift $W\subset M^{11}$ in M-theory. For example, in a later subsection we will refer to the latter as $\widehat{W}$. Hopefully the context should make it clear as to which submanifold we are referring.

Now, going back to Type IIA, the RR two-form $G_2$ satisfies

\be
P_*\left(\frac{G_2}{2\pi}\right) = N\label{numbranes}\ee

where $P:\mathcal{S}NW\rightarrow{W}$ denotes the $\epsilon$-sphere bundle of the normal bundle of $W$ in $\mathbb{R}^3 \times M^7$. This may be thought of as consisting
of all points which
are at a distance\footnote{For $W$ non-compact, one would generally
need to take $\epsilon$ to be a positive function on $W$ in order that
the image of $\mathcal{S}NW$ under the exponential map is an
embedding.} $\epsilon>0$ from $W$. Equation (\ref{numbranes}) determines $G_2$ up to a shift $G_2\rightarrow{G}_2+2\pi{P}^*(a)$ where
$a\in H^2(W;\mathbb{Z})\cong H^2(B;\mathbb{Z})$. Shifting by different values of $a$ results in a
different lift to M-theory, since $[G_2/2\pi]$ is the first Chern class
of the M-theory circle bundle $\mathcal{S}\mathcal{L}$. Thus in order to describe the lift of the D6-brane configuration to M-theory, one must specify the flux of $G_2$ over $B$.

Now, the twistor space $\mathcal{S}\Lambda^-B$ may be described in
terms of the projectivisation of the negative $\mathrm{spin}^c$
bundles

\be
\mathcal{S}\Lambda^-\cong\mathbb{P}\mathbb{V}_-(L)\cong\mathbb{P}(\Sigma_-\otimes{L}^{1/2})\ee

The ambiguity in the choice of $G_2$ flux over $B$ is \emph{precisely} the
ambiguity in the choice of $\mathrm{spin}^c$ structure, just as in section
4. Hence we conclude that the lift to M-theory of a configuration of D6-branes wrapped over a coassociative submanifold $B$ is always described locally by a $\mathrm{spin}^c$ bundle over $B$, and, moveover, this correspondence is actually 1-1. This gives the fact that we have a choice of $\mathrm{spin}^c$
bundle for our $\mathrm{Spin}(7)$ manifold, which in turn describes the M-theory manifold, a more physical meaning - it is just the choice of $G_2$ flux over $B$. For the case $B={\bf \mathbb{C}P}^2$, we also know that this choice of $G_2$
flux corresponds to a choice of Aloff-Wallach space that bounds the
total space of the appropriate $\mathrm{spin}^c$ bundle.

The case of $B={\bf S}^4$ is more
straightforward. Since $H^2({\bf S}^4;\mathbb{Z})=0$, there is no flux of
$G_2$ over ${\bf S}^4$, and therefore the lift to M-theory is entirely
determined by the number of D6-branes. From our above remarks, the lift of a
single D6-brane wrapped on the coassociative ${\bf S}^4$ of $\Lambda^-{\bf S}^4$ is a $\mathrm{Spin}(7)$ metric on the bundle of negative chirality spinors $\Sigma_-$
over ${\bf S}^4$. The choice of $\mathrm{spin}^c$ structure is
unique in this case since the line bundle $L$ must be trivial. Specfically, the $\mathrm{spin}^c$ bundle is just the spin bundle. The appropriate
$\mathrm{Spin}(7)$ metric is of course (\ref{alcs4}). These facts fit
in nicely with the work of McLean \cite{mclean}, where it was shown that the normal bundle of
a Cayley ${\bf S}^4$ is precisely the bundle $\Sigma_-$. Hence the
results of \cite{mclean} are consistent with the lift to M-theory
described in this section. This lift was also constructed explicitly
in the context of gauged supergravity in \cite{hernandez}, where the
AC Spin(7) metric (\ref{s4}) was obtained by lifting to M-theory a
$D=8$ supergravity solution describing D6-branes wrapped on the
coassociative ${\bf S}^4$ of $\Lambda^-{\bf S}^4$. Of course, for $N>1$ branes, one
simply takes the $\mathbb{Z}_N$ quotient of the spin bundle $\Sigma_-$.

Notice that this reasoning actually \emph{implies} the existence of the $\mathrm{Spin}(7)$ spaces found recently in \cite{garychris}, and gives them a physical interpretation within string theory.

\subsection{Flux Quantisation}

At the classical level, that is all there is to say. The M-theory
manifold satisfies the classical (eleven-dimensional supergravity) field equations with $G=0$, and preserves a
certain fraction (namely $\frac{1}{16}$) of the (maximal) supersymmetry of the
vacuum. However, as we saw in the last section, in the quantum theory
things are more subtle. In particular, if $w_4(M)$ is not zero, the
periods of $G/2\pi$ must be half-integral in order that the membrane path
integral make sense. It
follows that if $\lambda$ is not divisible by two, one \emph{must}
turn on a half quantum of $G$-flux in order that the M-theory solution
be consistent. We shall find that this is the case for our examples.

Consider first the case $X=\Sigma_-{\bf S}^4$. Since $p_1(T{\bf S}^4)=0$ and
$p_1(\Sigma_-{\bf S}^4)=-2u$, we have\footnote{Theorem 20.8 of \cite{milnor} states that, up to two-torsion (which is irrelevant here), $p(TE)= \pi^*(p(E)p(TB))$ where $\pi:E\rightarrow B$ is a smooth vector bundle over $B$.} $p_1(TX)|_{{\bf S}^4}=-2u$, where $u$ is a generator
of $H^4({\bf S}^4;\mathbb{Z})\cong\mathbb{Z}$. Hence
the restriction of $\lambda(TX)=p_1(TX)/2$ to ${\bf S}^4$ generates the
cohomology of ${\bf S}^4$. Notice that this is also the Euler class of the
bundle $\Sigma_-{\bf S}^4$. In particular though, $\lambda$ is \emph{not} divisible by two.

We turn to $X=\mathbb{V}_-(L)$ with $L$ an ``odd'' complex line bundle
over ${\bf \mathbb{C}P}^2$ - that is, $c_1(L)$ is an odd integer in
$H^2({\bf \mathbb{C}P}^2;\mathbb{Z})\cong\mathbb{Z}$. In this case,
$p_1(T{\bf \mathbb{C}P}^2)=3x^2$ and $p_1(\mathbb{V}_-(L)_{\mathbb{R}})=-x^2
+ \frac{1}{2}(c_1(L)-x)(c_1(L)+x)$, where $x$ generates
$H^2({\bf \mathbb{C}P}^2;\mathbb{Z})$. Hence

\be
p_1(TX)\mid_{{\bf \mathbb{C}P}^2} = 3x^2 -
x^2+\frac{1}{2}(c_1(L)-x)(c_1(L)+x)\ee

and so

\be
\lambda(TX)\mid_{{\bf \mathbb{C}P}^2} = x^2 +
\frac{1}{4}(c_1(L)-x)(c_1(L)+x)\ee

Of course, $(c_1(L)-x)(c_1(L)+x)$ is always divisible by eight, so that
$\lambda$ is an integral class. But we also see from this that
$\lambda$ is \emph{not} divisible by two. Indeed,  on ${\bf \mathbb{C}P}^2$ $\lambda(TX)$ restricts to
the Euler class of $\mathbb{V}_-(L)_{\mathbb{R}}$,
which mod two generates $H^4({\bf \mathbb{C}P}^2;\mathbb{Z}_2)$.

Hence the fourth Stiefel-Whitney class of the tangent bundle of
$X$, restricted to $B$, generates
$H^4(B;\mathbb{Z}_2)$, in all cases. In particular, it cannot be zero,
and hence $\lambda(X)$ is not divisible by two. Thus
M-theory on either the total space of $\Sigma_-{\bf S}^4$, or any
$\mathrm{spin}^c$ bundle $\mathbb{V}_-$ over ${\bf \mathbb{C}P}^2$ is consistent only if one turns on a
half-integral $G$-flux. But our D6-brane configuration contained no
such $G$-flux! We can only conclude that the D6-brane configuration
we started with must have been inconsistent in some way. Indeed, this
is the correct conclusion. One is forced to turn on half-integral
$G_4$ in Type IIA.

The meaning of this non-zero $G_4$ is at first slightly
mysterious. For M-theory on the product $M^{11}=S^1\times{M^{10}}$,
non-zero $w_4(M^{11})$ is equivalent to non-zero $w_4(M^{10})$. Indeed, the
shift in quantisation of $G$ in \cite{witten1} was derived by
analysing global anomalies on the M2-brane worldvolume. The analysis
for $G_4$ goes through in precisely the same way, with D2-branes in Type IIA
string theory, rather than M2-branes in M-theory. It is easy to check\footnote{The reader may verify this using the Theorem quoted in the last footnote, together with the results derived in Appendix A.}
that $w_4(TM^7)=0$ with $M^7$ either of the total spaces of the bundle of
anti-self-dual two-forms over ${\bf S}^4$ or ${\bf \mathbb{C}P}^2$. So if we were to
consider M-theory on either of these spaces, we would find no shift in
the quantisation law for $G$. Hence in the recent studies of M-theory on
G2 manifolds \cite{AW} no such shift in the $G$-flux would have been found. In fact, $w_4(M^7)=0$ is \emph{always} zero for
a compact spin seven-manifold $M^7$. To see this, note that one can write
the fourth Stiefel-Whitney class as

\be
w_4(M^7)=\sum_{i+j=4}\mathrm{Sq}^i(V_j)\ee

where $\mathrm{Sq}^i$ denotes a certain mod 2 cohomology operation
known as the $i^{\mathrm{th}}$ Steenrod square,
and $V_j\in{H}^j(M^7;\mathbb{Z}_2)$ denotes the $j^{\mathrm{th}}$ Wu
class of $M^7$ \cite{milnor}. Now, $V_4=0$ on dimensional grounds, and the
Steenrod squares annihilate the $V_1$ and $V_2$ terms for a similar
reason. Hence we are left with

\be
w_4(M^7)=\mathrm{Sq}^1(V_3)+\mathrm{Sq}^2(V_2)\ee

Now, $V_2=w_1^2(M^7)+w_2(M^7)$, and so this is zero as $M^7$ is
spin. Similarly, $V_3$ may be written as a polynomial in
Stiefel-Whitney classes of $M^7$, and so $V_3$ also vanishes for the same reason. Hence we
conclude that $w_4(M^7)=0$. Note that this argument breaks down for spin eight-manifolds, as $V_4$
is no longer zero in general.

So, where does this half integral flux come from? The key is that the
presence of the D6-branes implies that the M-theory manifold is not
just a product of the Type IIA manifold with a circle.
In fact, this contribution to $G_4$ is K-theoretic in nature, as we shall now explain.

\subsection{G-Flux from K-Theory}

In this subsection we will find that the shift in quantisation law for $G$
found in the last subsection is related to the K-theory classification of
RR-fields in Type IIA string theory. A detailed knowledge of K-theory
will not be required in order to follow the argument, provided one is
willing to accept certain facts taken from the literature without proof.

In the absence of branes, RR fields in Type IIA string theory on\footnote{The notation
$M^{10}_0$ may seem a little cumbersome, but the space to which we shall eventually apply these facts will be denoted in this way. It is merely the complement of the D6-brane worldvolume $W$.}
$M^{10}_0$ are, roughly speaking,
classified by an element of the K-group $K(M^{10}_0)$. Specifically, given an
element $v\in K(M^{10}_0)$, one has

\be
\frac{G_*}{2\pi}=\sqrt{\hat{A}}\mathrm{Ch}(v+\theta/2)\ee

where $G_*=G_0+G_2+G_4+\ldots$ is the total RR-form field
strength, $\hat{A}$ is the usual Dirac genus, which will not enter our
discussion here, and

\be
\mathrm{Ch}:K(M^{10}_0)\rightarrow{H}^{\mathrm{even}}(M^{10}_0;\mathbb{Q})\ee

is the Chern character, which maps an element of the K-group to an
even cohomology class. The $\theta$ characteristic lives in the
space $\Gamma_1/2\Gamma_1$ where the lattice
$\Gamma=K(M^{10}_0)/K(M^{10}_0)_{\mathrm{tors}}$ splits as a direct sum
$\Gamma_1\oplus\Gamma_2$. The defining property of the sublattice
$\Gamma_1$ is that it is maximally Langrangian; that is, it is a
maximal sublattice such that $\omega(x,y)=0$, $\forall{x},y\in\Gamma_1$ where $\omega$ is a
certain natural symplectic form on $\Gamma$. The interested reader is referred to
\cite{E8} and references therein for more details.

Fortunately, we do not need to know much about $\theta$ in the present
paper. We extract the following formulae from \cite{E8}

\be
G_0(\theta)=G_2(\theta)=0\ee

and

\be
\frac{G_4(\theta)}{2\pi}=-\lambda\quad(\mathrm{mod}\ 2)\ee

The mod two is inserted to remove from $G_4(\theta)$ various
terms that won't interest us in the present discussion. Note that it
is the contribution of the $\theta$ characteristic that gives the
shift in $G_4$ analagous to the shift of $G$ in M-theory. Since
$\lambda=0$ modulo two for spin seven-manifolds, we may ignore the $\theta$
characteristic from here on.

With this {\it proviso}, and assuming also that the
class $v$ satisfies $G_0=0$, we have

\be
[G_2/2\pi]=c_1(v)\ee

\be
[G_4/2\pi]=\frac{1}{2}c_1(v)^2-c_2(v)\label{G4}\ee

where the last formula receives corrections from $\theta$ that will not
interest us (as they are integral).
We interpret $[G_2/2\pi]=c_1(v)=c_1(\mathcal{L})$ in terms of the first Chern class of a complex line bundle $\mathcal{L}$, and $-c_2(v)$ is identified
with the characteristic class of an $E_8$ bundle over $M^{10}_0$. This class
specifies the $E_8$ bundle uniquely, up to bundle isomorphism
\cite{wittentop}. Again, we shall not make use of this fact.

Now, $[G_2/2\pi]$ is an integral class, as it is a Chern
class. Similarly, $c_2(v)$ is integral. However, the
fact that RR fields live in K-theory and not cohomology means that,
when one projects from K-theory to cohomology via the Chern character, the fields become
mixed. In particular, as we see from equation (\ref{G4}), $G_4$ depends on $G_2$. If $c_1(v)^2$ is not
divisible by two, then (\ref{G4}) implies that $[G_4/2\pi]$ is
half-integral, and, in particular, one cannot set it to zero.

These comments apply to IIA string theory in the \emph{absence}
of branes. Consider our general setup, with $N$ D6-branes wrapped over some
oriented $\mathrm{spin}^c$ submanifold $W$ of $M^{10}$, which is oriented and spin. Note also
that $M^{10}$ spin implies that $w_2(TW)=w_2(NW)$. A single D6-brane
(corresponding to $N=1$) lifts to the manifold $M^{11}$ in M-theory, with $W$ lifting to $\widehat{W}$, and $NW_0$
lifting to $N\widehat{W}_0$ (see the discussion in section 5.1. A subscript zero always denotes the complement of (the) zero (section)). The case $N>1$ is obtained by taking the obvious
$\mathbb{Z}_N$ quotient. The presence of the D6-branes implies that $G_2$ has
a delta-function singularity on $W$ (\ref{mag}), but is otherwise closed on the complement of $W$ in $M_{10}$.

Consider now excising the brane worldvolume $W$ from $M^{10}$ to yield
the space $M^{10}_0$ (for technical purposes, one may actually
want to remove a small tubular neighbourhood of $W$ from $M^{10}$ to form
$M^{10}_0$, and later take a limit). $G_2$ is closed on $M^{10}_0$ and there are no longer
any branes. However, the fact that there \emph{was} a brane where $W$ used to
be is encoded by the equation

\be
P_*\left(\frac{G_2}{2\pi}\right) = N\ee

This is in fact the same statement as (\ref{numbranes}).

We are left with RR-flux on $M^{10}_0$ without any brane
sources. Our main technical assumption is that this configuration
corresponds to some K-theory class $v\in K(M^{10}_0)$. This is in
spirit with the discussion of the K-theory classification of RR-fields in
\cite{wittenmoore}. Indeed, the presence of a brane on $M^{10}$ induces a
RR-field $K(\partial{M}^{10})$ on the ``boundary at infinity'',
$\partial{M}^{10}$. In the present situation,
$\partial{M}^{10}$ is a deformation retraction of $M^{10}_0$. K-groups are homotopy invariant, and so one may equivalently consider the brane as
inducing a K-Theory class in $K(M^{10}_0)$.

With $G_0=0$ (this corresponds to no D8-brane flux,
which seems natural as we started with D6-branes. Also, there is no
M-theory interpretation of $G_0$, so we cannot lift this to compare
with our analagous discussion in M-theory) it follows that

\be
\left[\frac{G_2}{2\pi}\right]=c_1(v)=c_1(\mathcal{L})\ee

and there is a possible half-integral contribution to $[G_4/2\pi]$ of the
form\footnote{Recall that $\mathcal{L}$ is a complex
line bundle over $M^{10}_0$, whose bundle of unit vectors coincides
with the ``M-theory circle direction''.}
$\frac{1}{2}c_1(\mathcal{L})^2$. It is precisely this contribution to
$G_4$ that we now focus on, ignoring any other integral part.

We would like to lift the present situation to M-theory. $G_4$ lifts directly to $G$, the
M-theory four-form. We have already seen in the last subsection that in some cases $[G/2\pi]$ must be half-integral in the quantum theory. Following \cite{E8}, we must shift $G/2\pi$ by
a representative of $\frac{1}{2}c_1(\mathcal{L})^2$. This class, although non-trivial in
$M^{10}_0$ is trivialised tautologically when pulled back to
$M^{11}_0$, the total space of $S\mathcal{L}$. To see this, one
introduces the one-form $\omega=(d\tau+A)/2\pi$ where $\tau$
parametrises the M-theory circle direction, and $A$ is a connection on
$\mathcal{L}$. $\omega$ integrates to $1$ on each ${\bf S}^1$ fibre, and
satisfies $d\omega=F/2\pi$ where $F$ is the
curvature of $\mathcal{L}$. Indeed, the $\omega$ of
\cite{E8} is in fact a global angular form on $S\mathcal{L}$, and its
exterior derivative is therefore topologically trivial. The point is that a gauge
transformation $\tau\rightarrow\tau+\phi$ is cancelled by the
corresponding gauge transformation of the connection,
$\mathcal{A}\rightarrow\mathcal{A}-d\phi$, and hence $\omega$ is globally defined, even
though its constituents are not.

One now shifts $C$ by $\delta{C}=\pi\omega\wedge d\omega$, and correspondingly
$G/2\pi=dC/2\pi$ is shifted by $\delta{G}/2\pi=\frac{1}{2}F\wedge F/(2\pi)^2$. Thus
$G/2\pi$ has been shifted by the pull-back to $M^{11}_0$ of
$\frac{1}{2}c_1(\mathcal{L})^2$, which is topologically trivial.

Now, our shift in $G$ was only defined on $M^{11}_0$, not on $M^{11}$. Recall that the former is obtained from the latter by excising $\widehat{W}$. However, the fact that
the shift in cohomology class is trivial on $M^{11}_0$ does not
imply that the shift is trivial when extended to $M^{11}$. This is just as well,
otherwise we could not explain the half-integral of flux we found in
M-theory using K-theory!

To see this, consider the shift in the $C$
field, $\delta{C}=\pi\omega\wedge d\omega$. This is a globally defined
three-form on
  $N\widehat{W}_0$, where recall that $\widehat{W}$ is the lift of the D6-brane worldvolume $W$ to $M^{11}$. Consider
integrating this form over a Lens
space $L(1,N)$
sitting in some normal space to $\widehat{W}$ in $M^{11}$. The one-form $\omega$
integrates to $1$ on the circle direction, and $d\omega=F/2\pi$ integrates to
$N$ on the two-sphere base
(the bundle is the $N$th power of the Hopf map $\mathcal{H}:{\bf S}^3\rightarrow{\bf S}^2$). Hence $\delta{C}$ integrates to
$\pi{N}$ over $L(1,N)$.

For simplicity, let us now specialise to the case $N=1$, where the geometry is smooth. If the unit three-sphere bundle of $N\widehat{W}$ is
denoted $\Pi:\mathcal{\bf S}^3\rightarrow \widehat{W}$, then we have

\be
\Pi_*(\delta{C}/\pi)=1\ee

That is, $\delta{C}/\pi$ integrates to $1$ over a unit three-sphere fibre. Now, on any $n$-sphere bundle, there exists an $n$-form $\psi$, the global angular form \cite{bott}, whose restriction to each fibre generates the cohomology of the fibre. This form also satisfies

\be
d\psi= -\Pi^*e\ee

That is, the exterior derivative is minus the pull-back of the Euler class of the sphere-bundle. Hence we may write

\be
\delta C/\pi = \psi + \ldots\ee

Since we know that $\delta{C}/\pi$ is a global form that integrates to $1$ over each fibre, it follows that the terms $"\ldots"$ integrate to $0$ over each fibre, and will not interest us. Thus, ignoring these for the moment, the shift in $C$ induces

\be
\delta{G}/2\pi=d\delta{C}/2\pi=\frac{1}{2}d\psi=-\Pi^*e(N\widehat{W})/2\label{kshift}\ee

The pull-back of the Euler class to $N\widehat{W}$ will in general be
non-trivial. We have thus
shown that
the half-integral contribution to $[G/2\pi]$ contributes a term given by the Euler
class of the normal bundle, which ties in with the comments at the end
of the previous subsection. In our case-studies, we have shown
explicitly that this Euler class is equal to the restriction of
$\lambda$ to $\widehat{W}$. Thus (\ref{kshift}) reads

\be
\delta{G}/2\pi=-\lambda/2\ee

Hence this K-theory shift is \emph{precisely} the
usual shift due to $\lambda$. We now go on to prove this in the
general case, modulo two (which is all that matters). The reader may
wish to skip the remainer of this subsection, which is rather technical,
as the argument merely confirms previous results, rather than adding
anything new.

Thus, reducing modulo two, and restricting to $\widehat{W}$, we obtain

\be
[\delta{G}/\pi]=w_4(N\widehat{W})\quad (\mathrm{mod}\ 2)\ee

since for
a rank four oriented vector bundle the Euler class reduces mod two to the fourth Stiefel-Whitney class. The Whitney product formula gives us

\be
w_4(TM\mid_{\widehat{W}})=w_4(T\widehat{W})+w_1(T\widehat{W})w_3(N\widehat{W})+w_2(T\widehat{W})w_2(N\widehat{W})+w_3(T\widehat{W})w_1(N\widehat{W})+w_4(N\widehat{W})\ee

Now, $M^{11}$ is assumed spin, and so $w_2(T\widehat{W})=w_2(N\widehat{W})$. $\widehat{W}\cong W$ is assumed
orientable (in order to couple the D-brane to RR-forms). Hence
$w_1(T\widehat{W})=0$. Moreover, $\widehat{W}$ must also be $\mathrm{spin}^c$, and so $w_3(T\widehat{W})=0$. We are left with

\be\lambda(TM\mid_{\widehat{W}})\stackrel{\mathrm{mod}2}{=}w_4(TM\mid_{\widehat{W}})=w_4(T\widehat{W})+w_2(T\widehat{W})^2+w_4(N\widehat{W})\ee

Now, $\widehat{W}$ is an oriented $\mathrm{spin}^c$ six-manifold. Hence,
arguing as we did in the last section, by Wu's Theorem

\be
w_4(\widehat{W})=\mathrm{Sq}^1(V_3)+\mathrm{Sq}^2(V_2)\ee

$V_3=0$ as $\widehat{W}$ is oriented and $\mathrm{spin}^c$, and $V_2=w_2(\widehat{W})$
as $w_1(\widehat{W})=0$. Since $\mathrm{Sq}^2(w_2)=w_2^2$, we conclude that

\be
w_4(\widehat{W})+w_2^2(\widehat{W})=0\ee

Applying this to our formula above, we see that

\be
\lambda(TM\mid_{\widehat{W}})=w_4(N\widehat{W})\quad (\mathrm{mod}\ 2)\ee

We conclude that, in the general case, the K-theory shift we have
found in this section is equivalent to the half-integral shift in $G$
found from M2-brane worldvolume anomalies.


\sect{Effective $\mathcal{N}=1$ Three-Dimensional Field Theory}

In this section we discuss the effective three-dimensional field theory
obtained from M-theory compactification on a (singular) manifold
$X$ of $\mathrm{Spin}(7)$ holonomy.
After discussing general aspects of such compactifications,
we then go on to study in detail two examples obtained from
compactification of M-theory on\footnote{In this section we use the
notation $\mathbb{B}_8\equiv \Sigma_-{\bf S}^4$ to denote
the total space of the bundle of negative chirality spinors
over ${\bf S}^4$, as in \cite{gary7}.} $\mathbb{B}_8$ and $\mathcal{Q}$.
We describe some dynamical aspects
of both these models and also explain the
relation between the anomalous shift of
the background $G$-flux and the Chern-Simons coefficient in
the $\mathcal{N}=1$ three-dimensional effective field theory.

\subsection{Compactification on a General Spin(7) Manifold}

In general, the effective field theory is expected to be ${\cal N}=1$
supersymmetric gauge theory in three dimensions.
If for a moment we assume that $X$ is non-singular
(although it may still be non-compact)
and gently curved, then one can deduce the spectrum of the massless
modes from the Kaluza-Klein reduction of eleven-dimensional supergravity.
As a result, in the compact case one finds a total number of $(b_2(X) + b_3(X) + b_4^-(X) + 1)$
bosonic modes and the same number of fermionic modes, which complete
the ${\cal N}=1$ supersymmetry multiplets \cite{PT}.
In a theory without a superpotential or Chern-Simons couplings,
all of these bosonic modes may be thought of as scalar fields
(due to vector-scalar duality in three dimensions).

However, as will be explained below, many models typically
have both Chern-Simons and superpotential terms, which prevents
us from dualizing scalar and vector fields into each other.
Therefore, we have to distinguish between vector modes
and scalar field modes.
Bearing this in mind, from the Kaluza-Klein reduction
we find $b_2$ abelian vector fields, $\mathcal{A}^i$,
which come from the modes of the three-form field $C$,
and $(b_3 + b_4^- + 1)$ scalars, $\phi^a$.
Some of these scalar fields, namely $b_3$ of them,
come from the $C$-field, whereas the others correspond
to deformations of the $\mathrm{Spin}(7)$ structure.
If $X$ is non-compact, instead of the Betti numbers $b_k$
one should use the dimension of the space of $L^2$-normalisable
$k$-forms on $X$.

Taking into account Chern-Simons and superpotential terms,
we may write the complete supersymmetric action at a generic
point in the moduli space of $X$:
\begin{eqnarray}
S_{3d} &= \int d^3 x \Big[
\sum_{k} {1 \over 4 g_k^2} \big( \mathcal{F}_{\mu \nu}^k \mathcal{F}^{k~ \mu \nu}
+ \bar \psi^k i \Gamma \cdot D \psi^k \big)
+ {1 \over 2} \sum_{a,b} g_{ab} \big(
\partial_{\mu} \phi^a \partial^{\mu} \phi^b
+ \bar \chi^a i \Gamma \cdot D \chi^b \big) - \nonumber \\
& - {1 \over 2} \sum_{a,b} g^{ab} \big(
{\partial {\cal W} (\phi) \over \partial \phi^a}
{\partial {\cal W} (\phi) \over \partial \phi^b}
- {1 \over 2}{\partial^2 {\cal W} (\phi) \over \partial \phi^a \partial \phi^b}
\bar \chi^a \chi^b \big) \Big]
- \sum_{i,j} {i k_{ij} \over 4 \pi} \int \big( {\cal A}^i \wedge d {\cal A}^j
+  \bar \psi^i \psi^j \big)
\label{effaction}
\end{eqnarray}
Here, $\psi^i$ are the gaugino fields, $\chi^a$ represent the fermionic
superpartners of the scalar fields $\phi^a$, $g^i$ are the gauge couplings,
and $g^{ab}$ denotes the scalar field metric.
Since we are mainly interested in non-compact $\mathrm{Spin}(7)$
manifolds, in this Lagrangian we omit the terms corresponding
to interactions with supergravity.

If the space $X$ develops a singularity, one should also expect
some non-abelian gauge fields coming from the singularity.
In the models that admit a description in terms of
$D6$-branes -- such as $X=\mathbb{B}_8$ and $X=\mathcal{Q}$
discussed in this paper -- one can derive non-abelian degrees
of freedom from the corresponding D-brane models.
The effective action for the non-abelian fields
can be written as ${\rm Tr} ( \ldots )$.
In particular, the Chern-Simons terms take the form
\be
I_{\mathrm{CS}}=\frac{k}{4\pi}\int\mathrm{Tr}
\left(\mathcal{A}\wedge{d\mathcal{A}}+\frac{2}{3}
\mathcal{A}\wedge\mathcal{A}\wedge\mathcal{A}
+ \bar \psi \psi \right)
\ee
where $\mathcal{A}$ is a gauge connection in
the adjoint representation of the gauge group.

Now, following \cite{GVW}, let us discuss vacua
in the resulting theory, and domain walls which connect them.
As we explained in section 3, microscopically a domain wall
in the three-dimensional ${\cal N}=1$ field theory can be described
as an M5-brane wrapped over a topologically non-trivial 4-cycle:
\be
\Sigma^{(4)} \in H_4(X;\mathbb{Z})
\ee
These domain walls, classified by elements of
the homology group $H_4 (X; \mathbb{Z})$, interpolate between
vacua corresponding to different values of the $G$-flux.
The latter, in turn, are classified\footnote{Again, we stress that it
is the shifted $G$-flux (including the quantum correction due to
$\lambda$) that is integral.} by $H^4 (X; \mathbb{Z})$.
On a compact manifold these two groups are isomorphic,
by Poincar\'e duality, so that in such a model
all vacua can be connected by domain walls.

In the present paper we are interested in the case of a non-compact
space $X$, where Poincar{$\acute{\mathrm{e}}$} duality asserts that
$H_4(X;\mathbb{Z})$ is isomorphic to cohomology with compact support:
\be
H_4(X;\mathbb{Z})\cong{H}^4_{\mathrm{cpct}}(X;\mathbb{Z})
\ee
Then, from the long exact sequence for the pair $(X,\partial{X})$
it follows that different vacua, modulo those which can be connected
by domain walls, are classified by the cohomology of the boundary
7-manifold $Y=\partial{X}$ \cite{GVW}:
\be
H^4 (Y; \mathbb{Z}) = H^4 (X; \mathbb{Z}) / f( H^4_{\mathrm{cpct}} (X; \mathbb{Z}))
\label{hqfour}
\ee
where $f( H^4_{cpct} (X; \mathbb{Z}))$ is the image of
the cohomology with compact support under the natural map:
\be
f \colon H^4_{\mathrm{cpct}} (X; \mathbb{Z}) \to H^4 (X; \mathbb{Z})
\ee
Therefore, we conclude that different models are classified
in part by $H^4 (Y; \mathbb{Z})$.

The other data needed to completely specify the compactification
is the value of the flux at infinity \cite{GVW}:
\be
\Phi_{\infty} =
N_{M2} - {\chi (X) \over 24}
+ {1 \over 2} \int {G \over 2 \pi} \wedge {G \over 2 \pi}
\label{anom}
\ee
Here $N_{M2}$ is the number of membranes filling
three-dimensional space-time, and $\chi(X)$ is
the integral of the Euler density over $X$. Note also that the anomaly
cancellation condition requires $\Phi_{\infty} =0$
for a compact space $X$ \cite{SVW,witten1}.

If $X$ is non-compact, the $\chi (X)$ that enters the global
anomaly condition (\ref{anom}) is defined
as an integral of the Euler density over $X$.
This may not agree with the topological Euler number of $X$.
There is an effective way to compute $\chi(X)$,
provided that an equivalent D6-brane model is available.
Indeed, let us assume that M-theory on the $\mathrm{Spin}(7)$ manifold $X$
can be viewed as the lift of some D6-brane configuration
on a $G2$ space $M^7$ to eleven dimensions.
Both of our models may be realised
in this way with $M^7$ being either
$M^7 = \Lambda^- B$ or (topologically) $\mathbb{R}^7$.
In the first case, a D6-brane is wrapped on the non-trivial
coassociative 4-cycle\footnote{We remind the reader that
the main two examples discussed in this paper correspond
to $B={\bf S}^4$ and $B={\bf \mathbb{C}P}^2$.}
$B$ in the $G2$ space $M^7 = \Lambda^- B$, as in section 5.
In the second case, discussed in section 3, the D6-brane has
world-volume $\mathbb{R}^3 \times L \subset \mathbb{R}^{10}$.
For simplicity, let us assume\footnote{Of course, this cannot occur in our models, but the conclusions are in any case independent of this assumption.} that $G=0$, so the anomaly condition (\ref{anom}) reads:
\be
{\chi (X) \over 24} = N_{M2} - \Phi_{\infty}
\label{anomchi}
\ee
The reason we decided to write the anomaly condition in
this form is that the right-hand side of (\ref{anomchi})
represents the effective M2-brane charge, whereas the left-hand
side is the anomaly term (obtained by integrating $R^4$
terms in the eleven-dimensional action).
After reduction to Type IIA theory the effective membrane
charges become the effective charge of space-filling D2-branes.
What is the Type IIA interpretation of the left-hand
side of the anomaly formula (\ref{anomchi})?

Since from the Type IIA perspective the three-dimensional effective
theory is obtained by compactification on a seven-dimensional
$G2$ manifold $M^7$, there is no contribution to the D2-brane
charge from the bulk.
However, in Type IIA theory we also have a space-filling
D6-brane wrapped on the coassociative 4-cycle $B$ inside $M^7$.
Due to the non-trivial embedding of the D6-brane world-volume
in space-time, the Ramond-Ramond fields couple to the
gauge field strength $\mathcal{F}$ as \cite{green,MM,CY}:
\be
I_{\mathrm{WZ}}=\int_{\mathbb{R}^3 \times B}
{C_*} \wedge \mathrm{Ch}(\mathcal{F})
\wedge \sqrt{{\hat A ({T}_B) \over \hat A ({N}_B)}}
\label{anomwz}
\ee
Here ${T}_B$ (respectively ${N}_B$) denotes the tangent
(respectively normal) bundle of $B$ inside $M^7$ (not inside $X$ !),
and the Dirac genus $\hat A$ can be expressed in terms of
the Pontryagin classes as follows \cite{lawson}:
\be
\hat A = 1 - {p_1 \over 24} + {7p_1^2 - 4 p_2 \over 5760} + \ldots
\ee
If we now compare the $C_3$ coupling on the right-hand side
of (\ref{anomwz}) with the left-hand side of the formula
(\ref{anomchi}), we obtain a relation:
\be
{\chi (X) \over 24} =
\int_{B} \mathrm{Ch}(\mathcal{F})
\wedge \sqrt{{\hat A ({T}_B) \over \hat A ({N}_B)}}
\label{chiaroof}
\ee
In particular, if the gauge bundle on the D6-brane is trivial, from
the total D2-brane charge induced by the anomaly we find:
\be
\chi (X) =
{N \over 2} \int_{B} \Big( p_1({N}_B) - p_1({T}_B) \Big)
\label{chipp}
\ee
We should stress here that the right-hand side of this formula
is computed on a $G2$ manifold $M^7$, whereas the left-hand side
is computed on the corresponding 8-manifold $X$ of $\mathrm{Spin}(7)$
holonomy. Thus, we are able to compute $\chi(X)$ (which in the compact
case is the Euler number of $X$) by computing locally the two-brane
charge which is induced on the D6-branes. In fact, a similar picture
arises in F-theory \cite{SVW}. In the latter reference it was argued
that the Euler number of a (non-singular) elliptically fibred
Calabi-Yau four-fold could be computed by calculating the local
three-brane charge which is induced on the seven-branes, the latter
having a worldvolume given by the discriminant of the fibration (times
flat Minkowski four-space). This three-brane charge is again induced by WZ terms
involving Pontryagin classes, and therefore the situation described in
the present paper may be regarded as an M-theory analogue of the
F-theory picture outlined in \cite{SVW}.

Now, let us briefly mention the r\^ole of supersymmetry.
In M-theory on a manifold $X$ of $\mathrm{Spin}(7)$ holonomy,
vanishing of the supersymmetry variations of the gravitino
fields implies that the covariantly constant spinor $\eta$
obeys \cite{becker7,gary7}:
\be
G_{mpqr} \gamma^{pqr} \eta = 0\ee

This can be expressed in terms of the Cayley 4-form $\Phi$:
\be
G_{mpqr} \Phi_n^{pqr} =0
\label{susycond}
\ee
This condition implies that in a supersymmetric vacuum the
$G$-flux must be self-dual:
\be
G = * G
\ee
This explains, for example, why no solutions with
anti-self-dual four-form flux have been found in \cite{gary7}.
Moreover, from the self-duality of $G$ and the anomaly
condition (\ref{anom}) it follows that on a compact $\mathrm{Spin}(7)$
manifold there can be only finitely many supersymmetric vacua.
In fact, we can rewrite (\ref{anom}) as:
\be
\Phi_{\infty} + {\chi (X) \over 24} =
N_{M2} + {1 \over 2} \int {G \over 2 \pi} \wedge * {G \over 2 \pi}
\label{anomplus}
\ee
Since a given choice of the covariantly constant
spinor $\eta$ is compatible with membranes of only
one orientation\footnote{Membranes with the opposite
orientation can also be interpreted as anti-membranes.},
supersymmetry also requires:
\be
N_{M2} \ge 0
\ee
It follows that in a supersymmetric configuration
the right-hand side of (\ref{anomplus}) is always
non-negative, and because the $G$-flux is quantised
it can take only finitely many values, for $X$ compact.
In particular, if $\chi(X)<0$ there are no supersymmetric
vacua at all.

Note also that the right-hand side of (\ref{anomplus})
can be interpreted as the energy density in a given vacuum.
Since for a finite tension domain wall the energy densities
in the two vacua connected by the wall should be the same,
it follows that the sum of the two terms on the right-hand
side of (\ref{anomplus}) is invariant in a given model
(although individual terms are not), {\it cf.} \cite{GVW}.
This is another way to argue that $\Phi_{\infty}$ is
an invariant of the dynamics.

Finally, let us point out that, at least in the supergravity
approximation, the position of the membranes is arbitrary in $X$,
so that the three-dimensional effective theory has $8 \cdot N_{M2}$
moduli (super)fields, which parametrise $N_{M2}$ copies of $X$.
After reduction to Type IIA theory these membranes turn into
D2-branes, localised at arbitrary points of the internal space.
However, if we have a configuration of multiple D6-branes,
$N>1$, it is natural to consider a `Higgs branch', where
the D2-branes dissolve in the D6-branes. In fact, every such
D2-brane looks like an instanton on $B$ in the world-volume
theory of the D6-branes. Hence, $N_{M2}$ dissolved D2-branes
correspond to a configuration of $N_{M2}$ instantons on $B$.
Classically, scalar fields in the effective $\mathcal{N}=1$
three-dimensional theory parametrise the moduli space
$\mathcal{M}_{N_{M2}} (B)$, of $N_{M2}$ instantons on $B$.
In particular, the dimension of this branch is given by:
\be
\mathrm{dim} (\mathcal{M}) = 4 N_{M2} N - (N^2 - 1) (1 + b_2^+ (B))
\ee
In our models we have $b_2^+ ({\bf S}^4)=0$ and
$b_2^+ (\mathbb{C}{\bf P}^2)=1$.

This picture agrees nicely with the fact\footnote{We thank N.~Nekrasov for discussions on this point.}
that the internal part of the D6-brane world-volume theory
is $\mathcal{N}=4$ topologically twisted Yang-Mills theory
on the four-manifold $B$ \cite{BSV}. Specifically, it is
the $\mathcal{N}=4$ topological theory such that the fundamental
representation of the $SU(4)$ R-symmetry group decomposes into
$({\bf 1}, {\bf 2}) \oplus ({\bf 1}, {\bf 2})$
of the twisted $SU(2) \times SU(2)$ Lorentz group.
This theory enjoys $SL(2,\mathbb{Z})$ duality symmetry \cite{VW}.
In particular, one generator of this group corresponds
to the shift of the background $C$-field:
\be
T \colon \int_{M^3} C \to \int_{M^3} C + 2 \pi
\ee

This is in fact a symmetry of M-theory, since $G=dC$ is invariant under both local gauge
transformations, $C\rightarrow{C}+d\nu$, with $\nu$ a two-form, and also
``large'' gauge transformations, $C\rightarrow{C}+\mu$, where $\mu$ is a
closed three-form with $2\pi$ periods. This latter transformation acts as follows on the coupling
constant in the $\mathcal{N}=4$ topological theory:
\be
T \colon \tau \to \tau + 1
\ee
The partition function of this theory can be written
as a series in $q = \exp(2 \pi i \tau)$:
\be
Z = \mathrm{const} \sum_n \chi (\mathcal{M}_n (B)) q^n
\ee
It would be interesting to study further the relation between
this topological theory and the $\mathcal{N}=1$ effective theory
in three dimensions.

In the rest of this section we will mainly be interested
in three-dimensional vacua which have a mass gap.
So, we assume that $N_{M2}=0$, unless otherwise stated.

To summarise, the models with mass gap may be microscopically
classified by the value of $\Phi_{\infty}$, and by $[G/ 2\pi]$
in a given coset (\ref{hqfour}), so that (\ref{anom}) is obeyed.


\subsection{Chern-Simons Terms Induced by $G$-Flux}

As we have mentioned earlier, M-theory on a $\mathrm{Spin}(7)$
manifold $X$ can often be understood as the lift of
a certain D6-brane configuration, and under such
a duality some non-abelian degrees of freedom are
mapped to gauge fields on multiple D-branes.
Thus, in both of our examples we can start with $N$
D6-branes wrapped on the coassociative 4-cycle $B$
in the manifold $M^7 = \Lambda^- B$ of $G2$ holonomy.
Clearly, in both cases, corresponding to $B={\bf S}^4$
and $B={\bf \mathbb{C}P}^2$, we obtain a $U(N)$ factor
in the three-dimensional gauge group\footnote{In the case
of $B={\bf \mathbb{C}P}^2$ one finds an extra $U(1)$ factor from
the bulk fields. It will be discussed in section 6.4.}.

In three dimensions, in addition to the usual
Maxwell-Yang-Mills term, one may also include
in the Lagrangian a level $k$ supersymmetric
Chern-Simons term:
\be
I_{\mathrm{CS}}=\frac{k}{4\pi}\int\mathrm{Tr}
\left(\mathcal{A}\wedge{d\mathcal{A}}+\frac{2}{3}
\mathcal{A}\wedge\mathcal{A}\wedge\mathcal{A}
+ \bar \psi \psi \right)
\ee
Our goal in this subsection will be to understand
how such terms can be generated by classical and
quantum effects in M-theory.

The Chern-Simons coupling $k$ is quantised topologically \cite{DJT}.
However, integrating out the gluino fields $\psi$
generates a shift in the effective value of $k$.
This shift is exact at one loop.
Specifically, the level is shifted to \cite{KLL,kogan}:
\be
k_{\mathrm{eff}}=k-\frac{N}{2}\mathrm{sign}(k)
\label{keff}\ee
for gauge group $SU(N)$.
The level $k_{U(1)}$ of the $U(1)$ gauge factor
is \emph{not} renormalised, {\it i.e.} $k_{\mathrm{eff}} = k$.

It is $k_{\mathrm{eff}}$ that is now quantised
in a purely bosonic effective field theory. We may as well assume
that $k>0$, and so the quantisation condition becomes:
\be
k-\frac{N}{2}\in\mathbb{Z}
\ee
Hence, for $N$ odd, the shift in $k$ is half-integral, and, in
particular, $k$ cannot be zero. Since the theory conserves parity
only for $k=0$, this is often called
the parity anomaly \cite{AGW,redlich,witten1}.

In order to see how such Chern-Simons terms can be generated
by the background $G$-flux in our models,
notice that on the worldvolume of the D6-branes,
one has the following Wess-Zumino term \cite{green}:
\be
I_{\mathrm{WZ}}=\int_{M^{3}\times B}
\mathrm{Ch}(\mathcal{F})\wedge{C_*}
=\frac{1}{2}\int_{M^{3}\times B}
\mathrm{Tr}\left(\frac{\mathcal{F}}
{2\pi}\wedge\frac{\mathcal{F}}{2\pi}\right)\wedge{C_3}
\label{wzterm}
\ee
where $C_3$ is the pull-back to $M^{3} \times B$
of the RR three-form potential, and $\mathcal{F}$ is the curvature
of the gauge field on the branes.
Here $B$ is either the ${\bf S}^4$ or $\mathbb{C} {\bf P}^2$
bolt of $\mathbb{B}_8$ or $\mathcal{Q}$, respectively.
Integrating by parts, we have:
\be
I_{\mathrm{WZ}}=\frac{1}{4\pi}\int_{M^{3}\times B}
\mathrm{Tr}\left(\mathcal{A}
\wedge{d\mathcal{A}}+\frac{2}{3}\mathcal{A}\wedge\mathcal{A}\wedge
\mathcal{A}\right)\wedge\frac{G_4}{2\pi}
\ee
Integrating over $B$ and taking into account
the fermionic superpartners
of $\mathcal{A}$, we obtain the following Chern-Simons term
in the effective supersymmetric gauge theory:
\be
I_{\mathrm{CS}}= \frac{k}{4\pi} \int_{M^{3}}\mathrm{Tr}
\left(\mathcal{A}\wedge{d\mathcal{A}}+\frac{2}{3}\mathcal{A}
\wedge\mathcal{A}\wedge\mathcal{A} + \bar \psi \psi \right)
\label{csthreedim}
\ee
where:
\be
k = \int_{B} \frac{G}{2\pi}
\ee
Note that $k$ does not have to take integer values.
It is $k_{\mathrm{eff}}$ which should always be integer, and
the corresponding value of $k$ takes either integer or
half-integer values in order to produce a consistent model.

The world-volume theory on a D6-brane contains fermions,
which might also cause a shift in the Chern-Simons level.
In fact, integrating out a massive fermion of mass $M$
in a representation $R$ of the gauge group, one finds
a finite one-loop renormalisation \cite{KLL}:
\be
k \to k + {d_2(R) \over 2} \mathrm{sign}(M)
\label{csfermionshift}
\ee
where $d_2(R)$ is the quadratic Casimir of $R$.
This is precisely what happens in the Type IIB supergravity
dual of $\mathcal{N}=1$ three-dimensional gauge theory,
proposed recently by Maldacena and Nastase \cite{MN}.
Specifically, the model of \cite{MN} consists of $N$
five-branes in Type IIB theory wrapped on ${\bf S}^3$.
As in the present setup, there is a certain amount of $H$-flux through the three-sphere,
which via the Wess-Zumino coupling (\ref{wzterm}) determines
the Chern-Simons level in the effective theory on the D5-branes.
The value of the $H$-flux is not shifted, but there is still
a finite renormalisation of $k$ that comes from summing
over all massive chiral multiplets.
In fact, as shown in \cite{MN}, all the massive fermionic
modes on ${\bf S}^3$ have partners with opposite mass, except
for one mode, which leads to the finite shift in $k$.
In our case, there is no contribution like (\ref{csfermionshift}) from massive fermions
on the four-dimensional manifold $B$ since all the modes
with positive and negative mass are paired
up\footnote{We thank J. Maldacena for explanations
and very helpful discussions on these points.}.


Now, let us focus on the case $N=1$, where the geometry
of $X$ is smooth, and the effective theory in three dimensions is abelian. From the last section, we know that the flux of the $G$-field in M-theory is shifted from standard Dirac quantisation\footnote{Note that we have three different $k$s here.
The value of $k_0$ stands for the starting value of the $G$-flux
in the eleven-dimensional supergravity theory
(without quantum correction $\lambda$),
whereas $k$ and $k_{\mathrm{eff}}$ denote the Chern-Simons coefficients
in supersymmetric three-dimensional theory and its bosonic
low-energy description, respectively.}:

\be
k_0 = \int_{B} \Big( \frac{G}{2\pi}- \frac{\lambda}{2} \Big)
\in\mathbb{Z}
\label{fluxthrub}
\ee

This shift results in a shift of the Chern-Simons coefficient in the three dimensional gauge theory:
\be
k = k_0 + \delta k
\ee
where
\be
\delta k = \int_{B} \frac{\lambda}{2}
\label{deltakb}
\ee

In both of the cases $X=\mathbb{B}_8$ and $X=\mathcal{Q}$, $k_0$ is therefore shifted to $k=k_0 + 1/2$ by quantum corrections
in M-theory. However, since the three-dimensional gauge theory is abelian,
we have:
\be
k_{\mathrm{eff}} = k = k_0 + \frac{1}{2}
\label{kkk}
\ee
At first sight this might seem to be a contradiction because
$k_{\mathrm{eff}}$ is half-integer for integer $k_0$.
However, a special property of abelian Chern-Simons theory -- that
string theory seems to know about -- is that it can be consistently
defined for both integer \emph{and} half-integer values of $k$.
Moreover, it can be consistently defined on any closed
oriented three-dimensional manifold $M^3$ \cite{wittenfb}.

The construction may be described as follows \cite{wittenfb}.
Let us start with
an arbitrary closed oriented three-dimensional manifold
$M^3$ (so far, we mainly focused on the cases
where $M^3=\mathbb{R}^3$ or $M^3=T^3$).
Since any oriented three-dimensional manifold is spin, we can consider
$\mathcal{N}=1$ supersymmetric gauge theory on $M^3$.
As before, let $\mathcal{F}$ denote the curvature on
the $U(1)$ gauge bundle $\mathcal{L}$ over $M^3$.
Such bundles are classified by maps to ${\bf \mathbb{C}P}^{\infty}$.
As pointed out in \cite{wittenfb}, the spin bordism group is
\be
\Omega^{\mathrm{Spin}}_3 ({\bf \mathbb{C}P}^{\infty}) =0
\ee
This implies that both the spin structure and the gauge
bundle $\mathcal{L}$ may be extended over some oriented
four-manifold $M^4$ with boundary $M^3$. One can then define
the level $k$ Chern-Simons term as \cite{wittenfb}:
\be
I_{\mathrm{CS}}= \frac{k}{4\pi} \int_{M^{4}}
\mathcal{F} \wedge \mathcal{F}
\label{halflevelcs}
\ee

{\it A priori}, this definition depends on the choice of four-manifold, $M_4$. However, suppose that one may find another such four-manifold $M_4^{\prime}$. One may glue $M_4$ and $M_4^{\prime}$ along their common boundary $M_3$
\be
Y_4\equiv M_4 \bigcup_{M_3} -M_4^{\prime}
\ee
to form the \emph{closed} four-manifold $Y_4$. It follows that the difference between the Chern-Simons terms defined by $M_4$ and $M_4^{\prime}$ is given by
\be
 \frac{k}{4\pi} \int_{Y_4}
\mathcal{F} \wedge \mathcal{F}
\ee

Since $Y_4$ is spin, the intersection form on $H^2(Y_4;\mathbb{Z})$ is even, and hence this last formula is always an integral multiple of $4 \pi k$.
Hence, $I_{\mathrm{CS}}$ defined by (\ref{halflevelcs}) is
well-defined, up to the addition of some multiple of $2 \pi$, even for $k \in \mathbb{Z} + \frac{1}{2}$. It follows that the path-intergral factor $e^{iI_{\mathrm{CS}}}$ is well-defined.
This ensures consistency of Type IIA D-brane configurations
dual to M-theory on the $\mathrm{Spin}(7)$ manifolds $\mathbb{B}_8$
and $\mathcal{Q}$ disscussed in the present paper.
In both models one finds Chern-Simons terms at half-integer level. Notice that the definition of the Chern-Simions term, $I_{\mathrm{CS}}$ with $k \in \mathbb{Z} + 1/2$, depends on the choice of a spin structure on $M^3$.

Let us now look at the M-theory picture. The presence of fluxes
in M-theory or string theory typically leads to superpotential
and Chern-Simons terms in the lower-dimensional theory \cite{GVW,sergei2}.
The explicit expression for the Chern-Simons coupling can be derived
from the analogous Chern-Simons term in the eleven-dimensional
effective action
\be
\frac{I^{(11)}_{CS}}{2\pi} = \frac{1}{6\cdot (2\pi)^3} \int C
\wedge G \wedge G \label{cselevendim}
\ee

In flat space, the $U(1)$ gauge field on the D6-branes derives from a Kaluza-Klein reduction of the M-theory three form potential $C$ along the $L^2$-normalisable harmonic two-form of Taub-NUT space, $\omega^{(2)}$. Thus one makes an ansatz of the form $C=\mathcal{A}\wedge \omega^{(2)}$, which gives rise to a dynamical gauge field $\mathcal{A}$ upon reduction. Since the eight-manifolds of $\mathrm{Spin}(7)$ holonomy discussed here also represent the M-theory lift of D6-branes in Type IIA theory, one would expect them to possess $L^2$-normalisable exact harmonic two-forms, in order to reproduce the correct spectrum of gauge fields in three dimensions. However, as in \cite{sergei}, we find that exact harmonic two-forms $\omega^{(2)}$ do indeed exist on these spaces, but they are not  $L^2$-normalisable. This would imply that the corresponding gauge fields are non-dynamical.

Bearing this puzzle in mind, in Type IIA theory
each D6-brane carries a $U(1)$ gauge field ${\cal A}$,
which corresponds to a Kaluza-Klein mode of the $C$-field
in M-theory, $C = {\cal A} \wedge \omega^{(2)}$.
In the case of several harmonic two-forms we obtain
\be
C = \sum_i \mathcal{A}_i \wedge \omega_i^{(2)}
\ee
Substituting this into (\ref{cselevendim}) and integrating over
the eight-manifold $X$, we get
\be
I_{CS} = \frac{k_{ij}}{4\pi}\int_{\mathbb{R}^{3}}
\mathcal{A}_i \wedge {d\mathcal{A}_j}
\ee
where the Chern-Simons coefficients, $k_{ij}$, are given by:
\be
k_{ij} = \int_{X} \frac{G}{2\pi}
\wedge \omega_i^{(2)} \wedge \omega_j^{(2)}
\label{csviaflux}
\ee

For example, in the case $X=\mathbb{B}_8=\Sigma_-{\bf S}^4$
we expect a single harmonic two-form $\omega^{(2)}$,
whose square determines the anomalous shift in $k$:
\be
\delta k = {1 \over 2} \int_{X} \lambda \wedge \omega^{(2)}
\wedge \omega^{(2)}\label{mtheoryk}
\ee

One would like to reconcile this formula with the Type IIA result (\ref{deltakb}). In order to see this relation, consider for example the ALC $\mathrm{Spin}(7)$ metric (\ref{alcs4}). This describes the uplift of a D6-brane wrapped on the ${\bf S}^4$ of $\Lambda^-{\bf S}^4$, with an asymptotically finite value of the string coupling constant. This solution has a $U(1)$ Killing vector field, $\partial/\partial\phi$, that generates rotations of the asymptotically finite sized ${\bf S}^1$. As is well-known, on a Ricci-flat manifold the one-form dual to such a $U(1)$ Killing vector is harmonic. Specifically, we may write

\be
\eta^{(1)} = f^2(r)\sigma\ee

and

\be
\omega^{(2)} = d\eta^{(1)}\ee

where the function

\be
f^2(r) = \frac{(r-3a)(r+a)}{(r-a)^2}\ee

and $\sigma = d\phi+{A}$ where $A$ is a connection on the $U(1)$ bundle. It is easy to verify that $\eta^{(1)}/2\pi$ is harmonic, and integrates to one over the ${\bf S}^1$ at infinity. Indeed, $\sigma/2\pi$ may be regarded as a global angular form on the $U(1)$ bundle over the twistor space ${\bf \mathbb{C}P}^3$ of ${\bf S}^4$. The total space of this bundle is of course diffeomorphic with the boundary of $\mathbb{B}_8$. The form $\sigma$ is only defined on the complement of the zero section of $\Sigma_-{\bf S}^4$, but the form $\eta^{(1)}$ is in fact a global form on $\mathbb{B}_8$ since $f^2(r)$ vanishes at the zero-section $r=3a$. Moreover, the derivative of $\sigma$ is the curvature $F$ of the $U(1)$ bundle. When projected down to Type IIA, this by definition means that $d\sigma$ coincides with the RR two-form $G_2$. The formula (\ref{mtheoryk}) therefore gives

\begin{eqnarray}
\delta k & = & {1 \over 2} \int_X \lambda \wedge d\eta^{(1)}
\wedge \omega^{(2)} \nonumber \\
& = & {1 \over 2} \int_{\partial{X}} \lambda \wedge \eta^{(1)}
\wedge \omega^{(2)} \nonumber \\
& = & {1 \over 2} \int_{\partial{X}/U(1)} \lambda \wedge \frac{G_2}{2\pi} \nonumber \\
& = & {1 \over 2} \int_{{\bf S}^4}  \lambda
\end{eqnarray}

where in the last step we have used the fact that $G_2/2\pi$ integrates to one over the ${\bf S}^2$ fibre of the twistor space ${\bf S}^2\hookrightarrow {\bf \mathbb{C}P}^3 \rightarrow {\bf S}^4$, since there is one D6-brane present. Hence we recover the Type IIA formula (\ref{deltakb}). Precisely the same reasoning goes through for the case $X=\mathcal{Q}$, with the new $\mathrm{Spin}(7)$ metric (\ref{spin7}).

In fact, with some care this argument may be used to determine the shift in $k$ for $N>1$. It is natural to expect that in the case of non-abelian
gauge theory $\delta k$ is related to the one-loop shift
of $k$ in the IR bosonic theory, (\ref{keff}). However, for $N>1$ the M-theory "manifold" is singular; specifically, it contains a $\mathbb{Z}_N$ orbifold singularity at the D6-brane locus. Thus one must be careful when using the results of \cite{witten1} to determine the shift in the quantisation of the $G$-flux. However, in the smooth case, we know that the form $G/2\pi$ gets shifted by

\be
\frac{1}{16\pi^2}\mathrm{Tr}R\wedge R\label{RR}\ee

where $R$ is the curvature of some smooth metric on $X$. The cohomology class of this differential form is associated with the integral characteristic class $\lambda(X) = p_1(X)/2$ of the tangent bundle of $X$.

Now suppose that we take a $\mathbb{Z}_N$ quotient of $X$, such that the fixed point set is a codimension four submanifold of $X$. This is identified with the D6-brane locus in Type IIA. If we denote the complement of the fixed point set as $X_0$, then $X_0/\mathbb{Z}_N$ is a smooth manifold, and we are free to use the quantisation law for $G$ found in \cite{witten1}. That is, $G/2\pi$ is shifted in the quantum theory by the gravitational correction term (\ref{RR}). Thus it would seem natural that the shift in $G/2\pi$ as a differential form on $X_0/\mathbb{Z}_N$ is the same as it is for $N=1$. We may therefore perform the integration just as for $N=1$:

\begin{eqnarray}
\delta k & = & {1 \over 2} \int_{\partial{X}/U(1)} \frac{1}{16\pi^2}\mathrm{Tr}(R\wedge R) \wedge \frac{G_2}{2\pi} \nonumber \\
& = & {N \over 2} \int_{{\bf S}^4} \frac{1}{16\pi^2}\mathrm{Tr}(R\wedge R) \nonumber \\
& = & {N \over 2}\end{eqnarray}

Here we have used the fact that the RR two-form field strength
$G_2/2\pi$ integrates to $N$ over the two-sphere fibre of the twistor
space. This is equivalent to the assertion that
there are $N$ D6-branes present.
Thus this argument would seem to imply that the shift in quantisation
of $G$ is indeed related to the one-loop shift in the value of
the Chern-Simons coupling. It would be interesting to investigate
this further \cite{sergeijames2}.

Apart from the Chern-Simons term, the background $G$-flux also generates
an effective superpotential, therefore giving a mass to the scalar fields.
In M-theory compactification on a manifold $X$ of $\mathrm{Spin}(7)$
holonomy, one can construct a $\mathrm{Spin}(7)$ singlet by taking
a wedge product of the $G$-flux with the Cayley calibration $\Phi$:
\be
\mathcal{W}=\frac{1}{2\pi}\int_X\Phi\wedge{G} \label{weff}
\ee
It is natural to identify this function of the moduli fields
with the effective superpotential induced by the $G$-flux \cite{sergei2}.
Indeed, one can show that the supersymmetry constraints,
which follow from the effective superpotential $\mathcal{W}$
are equivalent to the supersymmetry conditions (\ref{susycond})
in the eleven-dimensional supergravity \cite{AXG}.

The combined effect of the Chern-Simons and superpotential terms
is to produce a mass for the vector and scalar fields, and also
to lift some vacua in the three-dimensional effective theory.
Note that without the Chern-Simons and superpotential terms, {\it i.e.}
in a theory where bosonic fields are massless, one can dualise vector
fields (respectively multiplets) into scalar fields (respectively multiplets),
and vice versa.
Since we find that M-theory on either $\mathbb{B}_8$
or $\mathcal{Q}$ does not lead to propagating scalar fields
in three dimensions, we will not discuss superpotential terms
further in this paper.

\subsection{M-Theory on $\mathbb{B}_8$}

In this subsection we decribe some general features of the model based on $X =\mathbb{B}_8 = \Sigma_-{\bf S}^4$,
and its $\mathbb{Z}_N$-quotient, corresponding to $N$ D6-branes
in Type IIA string theory on $M^7 = \Lambda^-{\bf S}^4$. A more complete treatment of the dynamics will be presented in \cite{sergeijames2}.
The asymptotically locally conical
$\mathrm{Spin}(7)$ metric (\ref{alcs4}) relevant to this model was found recently in \cite{gary7},
and reviewed in section 2 of the present paper.

We would like to study M-theory in the case when $X$ develops
an isolated conical singularity. Since we are interested
in the behaviour near the singularity, for the main part
of this section we can think of $X$ as a smooth eight-manifold
asymptotic to a cone over a seven-sphere:
\be
Y = SO(5)/SO(3) \cong {\bf S}^7
\label{sfoury}
\ee
The relevant AC $\mathrm{Spin}(7)$ metric is given by \cite{gary}:
\be
ds^2 = \left(1-\left(\frac{a}{r}\right)^{10/3}\right)^{-1}dr^2
+\frac{9}{100}r^2\left(1-\left(\frac{a}{r}\right)^{10/3}\right)
(\sigma_i-A^i)^2+\frac{9}{20}r^2d\Omega_4^2\label{s42}
\ee
As the parameter $a\rightarrow0$, $X$ develops an isolated conical singularity at $r=0$. A small deformation of the metric $\delta g/g \sim r^{-10/3}$
(corresponding to a change in the parameter $a$)
is not $L^2$-normalizable since the following norm
\be
|\delta g|^2 = \int d^8 x \sqrt{g} g^{ac} g^{bd} \delta g_{ab} \delta g_{cd} \sim \int d^8 x \sqrt{g} (\delta g /g)^2 \sim \int dr \ r^{1/3}
\ee
diverges. This means that the corresponding scalar field has zero kinetic
energy and, therefore, its dynamics is frozen. In this sense,
it is a true modulus (or, rather, a coupling constant).
For each expectation value of this field,
the models in question have finitely many vacua.

Another way to see that the scalar mode corresponding to the change
in the parameter $a$ is non-dynamical is to remember that compactification
of M-theory on $X$ is equivalent to Type IIA string theory on
a $G2$ manifold $M^7 = \Lambda^- {\bf S}^4$ with a D6-brane
wrapped around the coassociative cycle $B \cong {\bf S}^4$.
In this picture, the effective ${\cal N}=1$ three-dimensional
theory is the theory on a space-filling D6-brane.
In particular, light scalar fields come from normal
deformations of the D6-brane world-volume inside the $G2$ manifold.
According to McLean \cite{mclean},
such deformations of a coassociative submanifold $B$
are unobstructed and correspond to harmonic anti-self-dual
two-forms on $B$. Moreover, he proves that the moduli space
of coassociative manifolds is locally a smooth manifold
of dimension:
\be
b^-_2 (B) = \mathrm{dim} H^2_- (B; \mathbb{R})
\ee
Since $b^-_2 ({\bf S}^4)=0$, we conclude that
the coassociative 4-sphere is rigid in this example and, therefore,
that there are no light scalar fields on the D6-brane world-volume.

Let us now examine the global symmetries of the simplest
$N=1$ model.
As explained in \cite{AW}, global symmetries that
come from the $C$-field are gauge transformations
$\delta C = d \Lambda$ that have $d \Lambda =0$ at infinity.
Hence these transformations are classified by the group
\be
H^2(Y;U(1))
\label{cysymm}
\ee
Moreover, unbroken symmetries are those which extend over $X$.
In other words, unbroken symmetries are classified by
\be
H^2(X;U(1))
\label{cxsymm}
\ee
In the present example both (\ref{cysymm})
and (\ref{cxsymm}) are trivial.
So, we proceed to examine the geometrical symmetries.

Since $Y$ can be realised as a homogeneous space (\ref{sfoury})
of the form $G/H$, geometrical symmetries of $Y$ consist
of left actions by the elements of $G$, and also from right
actions by elements $w \in G$ that centralise $H$:
\be
w^{-1} h w \in H, \quad \forall h \in H
\ee
Since $SO(5)=Sp(2)/ \mathbb{Z}_2$ and
$SO(3)=Sp(1)/ \mathbb{Z}_2$, the space $Y$
may also be viewed as the following quotient:
\be
Y = Sp(2) / Sp(1)
\ee
For the natural action of $Sp(1)$ on $Sp(2)$
the induced Einstein metric is that of the "round" seven-sphere.
However, this is not the metric we want.
The metric on $Y$ relevant to our problem is
a weak $G2$ metric (since we require the cone on $Y$ to be $\mathrm{Spin}(7)$), and is often described as a squashed seven-sphere.
This may be constructed explicitly as follows.
Consider $Sp(1)_A \times Sp(1)_B \subset Sp(2)$, and define $G = Sp(2) \times Sp(1)_C$, together with the subgroups $K = Sp(1)_A \times Sp(1)_B \times Sp(1)_C$ and $H = Sp(1)_D \times Sp(1)_B$, where $Sp(1)_D = Sp(1)_{A+C}$ is the diagonal subgroup of $ Sp(1)_A \times Sp(1)_C$. Then the four-cycle $B$ is given by

\be
B = G / K
\ee

and we can represent our seven-manifold $Y$ as:
\be
Y = G/ H\label{coset}
\ee

It follows that $Y$ fibres over $B$. The fibres themselves are copies of $K/H \cong Sp(1) \cong SU(2) \cong {\bf S}^3$. This is of course the quaternionic Hopf fibration. The induced metric on $Y$ is a weak $G2$ metric
and is in fact the squashed seven-sphere that we need. Up to homothety, the metric is given explicitly by

\be
ds^2_7 = \mu^2 (\sigma_i-A^i)^2 + d\Omega_4^2\label{squash}
\ee

where the "squashing parameter" $\mu^2 = 1/5$ for the weak G2 squashed seven-sphere (whereas the round sphere is given by $\mu^2=1$).
This is to be compared with the Spin(7) metric (\ref{s42}), which may be regarded as a resolution of the cone over this weak G2 manifold. We briefly remind the reader that $d\Omega_4^2$ is the round metric on the unit ${\bf S}^4$, $\sigma_i$ are a set of left-invariant one-forms on $SU(2)\cong {\bf S}^3$, and $A_i$ is a connection for the $SU(2)$ Yang-Mills instanton on ${\bf S}^4$. Another way of realising the metric (\ref{squash}) is as the distance sphere in ${\bf \mathbb{H}P}^2$. That is, one takes a point in the quaternionic projective space ${\bf \mathbb{H}P}^2$ with its Fubini-Study metric, and considers the hypersurface consisting of all points a geodesic distance $\chi$ from that point, where $\mu = \cos\chi$. Clearly, for small $\chi$, the resulting hypersurface is topologically ${\bf S}^7$; the induced metric is given by (\ref{squash}), up to homothety.

Now we are ready
to identify the symmetries of this metric. Since $Y$ may be viewed as the coset space (\ref{coset}), it is manifestly invariant under the left action of $G = Sp(2) \times Sp(1)_C$. There are also symmetries of $Y$ that come from the right action by elements $w \in G$ that centralise $H$. In this case, there is only one such non-trivial element. It may be constructed as follows. Consider the following element of $SU(2) \cong Sp(1)$

\be
\omega \equiv \left(\begin{array}{cc}0 & 1 \\ -1 &
0\end{array}\right)\ee

Then $w \equiv (\omega, 1, \omega^{-1}) \in Sp(1)_A \times Sp(1)_B \times Sp(1)_C = K \subset G$. Notice that since $\omega^{-1} = -\omega$, $w$ is not an element of $H$. The action $w^{-1} H w$ simply complex conjugates the first copy of $Sp(1)$ in $H = Sp(1)_D \times Sp(1)_B$. Notice that $w$ acts trivially on $B = {\bf S}^4$ since $w \in K$. We conclude that the symmetry group of $Y$ is

\be
\left[\left(Sp(2) \times SU(2)\right)/\mathbb{Z}_2\right] \times \mathbb{Z}_2\ee

with the last factor generated by $w$. It is clear that this symmetry group extends to the resolution $X$ given by the $\mathrm{Spin}(7)$ metric (\ref{s42}), and hence there is no spontaneous symmetry breaking of this type.


Let us now examine the Type IIA dual of this solution, where a D6-brane is wrapped on the coassociative
four-cycle $B={\bf S}^4$ inside the $G2$ holonomy manifold
\be
M^7 = \Lambda^- {\bf S}^4
\ee
From this perspective it is clear that the three-dimensional
effective theory on the D6-brane is simply ${\cal N}=1$
supersymmetric $U(1)$ gauge theory with a Chern-Simons term:
\be
L_{3d} = {1 \over 4 g^2} \int d^3 x (\mathcal{F}_{\mu \nu} \mathcal{F}^{\mu \nu}
+ \bar \psi i \Gamma \cdot D \psi)
+ {i(k_0 + {1 \over 2}) \over 4 \pi} \int ({\cal A} \wedge d {\cal A}
+  \bar \psi \psi )
\ee
As we discussed in the previous subsection, the value of
the Chern-Simons coefficient $k = k_0 + 1/2$ is half-integer
and defined by the $G$-flux through the 4-sphere,
{\it cf.} (\ref{fluxthrub}):
\be
\int_{{\bf S}^4} \Big[ {G \over 2 \pi} \Big] = k_0 + {1 \over 2}
\label{fluxthrus}
\ee
On the other hand, the gauge coupling constant $g$
is related to the volume of ${\bf S}^4$:
\be
{1 \over g^2} = \mathrm{Vol} ({\bf S}^4) \sim a^4
\label{ggaugesfour}
\ee

Of course, since the Chern-Simons coefficient $k$ is half-integer, in particular it cannot be zero. This implies that parity symmetry is broken in this theory. In a sense, this breaking of parity symmetry is spontaneous. Classically, one may turn off the Chern-Simons term, so that the theory is parity-invariant. However, in M-theory one is forced to choose a physical state in which $G$ is non-zero, and therefore parity is violated in the effective three-dimensional theory due to this choice of state, rather than explicitly. Of course, even classically one may choose $G$ non-zero, and therefore violate parity, but the point is that in the quantum theory this is unavoidable.


There is another useful D6-brane configuration dual to this model, which was
discussed in section 3. It is obtained from M-theory on
$X$ via reduction on a circle ${\bf S}^1 \cong U(1)$,
such that $X/U(1) \cong \mathbb{R}^7$, $Y/U(1) \cong {\bf S}^6$,
and the $U(1)$ action has a fixed point set in codimension four.
Following \cite{AW}, we denote $F \subset Y$ and $L \subset X$
to be the set of fixed points on $Y$ and $X$, respectively.
Then, according to our analysis in section 3,
$F$ must be a homology 3-sphere, knotted inside $Y \cong {\bf S}^7$.
The fixed point set $F$ is isomorphic to an ${\bf S}^1$ bundle over ${\bf S}^2$
(with Euler number 1), so that $L$, which is a cone on $F$, is an
$\mathbb{R}^2$ bundle over ${\bf S}^2$:
\be
L \cong {\bf S}^2 \times \mathbb{R}^2
\ee
The singular configuration, where the ${\bf S}^2$ zero-section shrinks to zero size,
corresponds to a point in the moduli space where $X$ develops
a conical singularity. At this point the four-manifold $L$ degenerates to $\mathbb{R}^4$.


For each value of $k_0$ the theory is expected to be
infrared-free and have only one massive vacuum.
To see how this follows from M-theory,
note that $H^4 (Y; \mathbb{Z})$ is trivial.
Therefore, one would expect the model to be completely
specified by the value of the flux at infinity (\ref{anom}):
\be
\Phi_{\infty} = \frac{1}{12} + {1 \over 2} (k_0 + {1 \over 2})^2
\ee
Here we used the relation (\ref{chipp}) to compute
$\chi (X) = (-4-0)/2=-2$, and also the fact that
the self-intersection number of the ${\bf S}^4$ is equal to 1.
Note that the value of $k_0$ in the formula for $\Phi_{\infty}$
is related to the value of $k_0$ in the Chern-Simons term
because the Chern-Simons coupling is induced by the $G$-flux,
according to our discussion in the previous subsection.
One can check that when $k_0$ is shifted by an integer,
$\Phi_{\infty}$ also changes by an integer.

We have just explained that M-theory predicts only one
supersymmetric vacuum in the $U(1)$
Chern-Simons theory on $\mathbb{R}^3$ at half-integer level.
In particular, it does not depend on the value of
the Chern-Simons coefficient.
This is to be compared with the number of ground states in
the $U(1)$ Chern-Simons theory, say, on $\mathbb{R} \times T^2$
at integer level $k$. Let us call this number $I'(k)$.
Then, $I'(k)$ can be computed, for example, via quantisation
of ground states\footnote{
After we reduce the $2+1$-dimensional theory on $T^2$,
we obtain supersymmetric quantum mechanics on $E$,
where $E= (T^2)^{\vee}$ is the moduli space of flat
$U(1)$ connections on $T^2$. The quantum Hilbert space of this
model is the space of spinors with values in $\mathcal{L}^k$,
or equivalently the space of $(0,q)$-forms valued in $\mathcal{L}^k$.
Moreover, the supersymmetry generators can be identified with the operators
$\bar \partial$ and $\bar \partial^{\dagger}$,
whereas the Hamiltonian can be identified with
$H = \{ \bar \partial, \bar \partial^{\dagger} \}$ \cite{wittengauge}.
It follows that supersymmetric ground states correspond to
the elements of $H^0 (E, \mathcal{L}^k) \oplus H^1 (E, \mathcal{L}^k)$.
Without loss of generality we may assume $k>0$, and,
using the Riemann-Roch formula and Serre duality, compute:
$I_{U(1)}' (k) = h^0 (\mathcal{L}^k) + h^1 (\mathcal{L}^k )
= \mathrm{deg} (\mathcal{L}^k) + 2 h^1 (\mathcal{L}^k)
= k + 2 h^0 (\mathcal{L}^{-k}) = k$.} as in \cite{wittengauge}:
\be
I'(k)= \vert k \vert
\label{uoneind}
\ee
Note that the supersymmetric index in the $U(1)$ theory is equal
to zero due to the presence of the fermionic zero mode.
Excluding the fermionic zero mode one can define
$I'(k)$, which is not zero, and gives non-trivial information
about the number of vacua \cite{AV,MN}.

To summarise, in the case $N=1$, $N_{M2}=0$ we find that M-theory
on $X=\mathbb{B}_8$ gives a $U(1)$ effective gauge theory
with half-integer Chern-Simons term $k = k_0 + {1 \over 2}$.
The theory is parametrised by one real parameter -- the gauge
coupling constant.
There is only one classical limit,
corresponding to large $\mathrm{Vol}({\bf S}^4)$. For all values of
$k_0 \in \mathbb{Z}$ this theory is expected to be infrared-free
and to have only one massive vacuum.

One can consider various generalizations of this model
corresponding to $N > 1$ and/or $N_{M2} > 0$.
Let us briefly comment on the models with $N>1$.
In Type IIA theory they correspond to configurations
of multiple D6-branes wrapped on the coassociative four-cycle
$B={\bf S}^4$ inside the $G2$ space $M^7=\Lambda^- {\bf S}^4$.
Now, the effective field theory on the D6-branes is
$U(N)$ gauge theory with Chern-Simons coefficient
congruent to $N/2$ mod $\mathbb{Z}$:
\be
L_{3d} =
{1 \over 4 g^2} \int d^3 x \mathrm{Tr} (\mathcal{F}_{\mu \nu} \mathcal{F}^{\mu \nu})
+ {i(k_0 + {N \over 2}) \over 4 \pi} \int_{\mathbb{R}^{3}}\mathrm{Tr}
\left(\mathcal{A}\wedge{d\mathcal{A}}+\frac{2}{3}\mathcal{A}
\wedge\mathcal{A}\wedge\mathcal{A} \right) + \mathrm{fermions}
\ee
Like in the $N=1$ case, the gauge coupling in this theory
is related to the volume of the four-sphere, whereas $k_0$
is determined by the $G$-flux through the ${\bf S}^4$,
{\it cf.} (\ref{fluxthrus}):
\be
\int_{{\bf S}^4} \Big[ {G \over 2 \pi} \Big] = k_0 + {N \over 2}
\ee

In M-theory this non-abelian theory is expected to come from
compactification on a singular $\mathrm{Spin}(7)$ manifold $X$,
which is an $\mathbb{R}^4 / \mathbb{Z}_N$ bundle over ${\bf S}^4$.
The boundary of $X$ is a seven-manifold $Y = {\bf S}^7 / \mathbb{Z}_N$.
Since $\mathbb{Z}_N$ acts freely on ${\bf S}^7$, $Y$ is smooth, and we have:
\be
H^4(Y; \mathbb{Z}) \cong \mathbb{Z}_N
\ee
Therefore, the model is specified now by $\Phi_{\infty}$ and
a (half)-integer number $k=k_0 + \frac{N}{2}$ defined modulo $N$.
At least for large values of $\Phi_{\infty}$, where
the IR dynamics is dual to AdS$_4 \times Y$, the theory
is expected to flow to a non-trivial $\mathcal{N}=1$
superconformal field theory.

For a given value of $k$, this effective theory
(compactified on an extra $T^2$) is expected
to have $I_{U(N)}'(k)$ supersymmetric vacua, where:
\be
I_{U(N)}'(k) = {I'(k) \cdot (k + N/2 -1)! \over N! (k-N/2)!}
\ee
In order to compute $I_{U(N)}'(k)$ one can think of the $U(N)$ gauge group as
a product $U(1) \times SU(N) / \mathbb{Z}_N$, {\it cf.} \cite{AV,MN}.
Then, $I_{U(N)}'(k)$ is a product of the number of ground states in
the $U(1)$ gauge theory and the number of vacua in the $SU(N)$ gauge theory,
divided by $N$.
We remind the reader that in ${\cal N}=1$ three-dimensional
gauge theory with gauge group $SU(N)$ the supersymmetric index
is \cite{wittengauge}:
\be
I_{SU(N)} = {(k + N/2 -1)! \over (N-1)! (k-N/2)!}
\ee
where $\vert k \vert > N/2$. By definition, $I_{SU(N)}$ gives the number
of bosonic ground states minus the number of fermionic ground states.
However, it was argued in \cite{wittengauge} that all ground states
in three-dimensional $SU(N)$ super-Yang-Mills theory are bosonic,
so that $I_{SU(N)}$ gives the actual number of ground states.
On the other hand, the supersymmetric index in $U(N)$ gauge theory
is zero due to the fermionic zero mode in the ``central'' $U(1)$,
as we remarked ealier.

In \cite{wittengauge}, by analysing the supersymmetric index it was
shown that supersymmetry is unbroken in this quantum theory for
$k \geq N/2$, whereas strong evidence was given to support
the conjecture that supersymmetry is dynamically broken for $k < N/2$.
It would be interesting to study this problem in eleven-dimensional
supergravity, constructing an explicit solution with non-zero $G$-flux,
and to understand the relation of such a solution to the octonionic superstring
soliton constructed in \cite{HS}.


\subsection{M-Theory on $\mathrm{Spin}^c$ Bundles over ${\bf \mathbb{C}P}^2$}

The dynamics of the models based on the total space of ($\mathbb{Z}_N$ quotients of) $\mathrm{spin}^c$ bundles over ${\bf \mathbb{C}P}^2$ is
more interesting and subtle, even in the simple case $N=1$.
In fact, instead of a single parameter $N$ we have two integer
numbers, $k$ and $l$, which parametrise different types of complete
$\mathrm{Spin}(7)$ metrics on $X$. The corresponding manifolds are
asymptotically locally conical with principal orbits
\be
Y = SU(3)/T_{k,l} \cong SU(3)/U(1)
\label{ysuthree}
\ee
Here, following our notations in section 4, $k$ and $l$
parametrise different $U(1)$ actions on $SU(3)$.
For example, for $k=-l=-1$ the explicit metric was
constructed in (\ref{spin7}), and describes a complete metric of $\mathrm{Spin}(7)$ holonomy on $\mathcal{Q}$:

\begin{eqnarray}
ds^2 & = & \frac{(r-a)(r+a)}{(r-2a)(r+2a)}dr^2+\frac{9a^2}{8}\frac{(r-2a)(r+2a)}{(r-a)(r+a)}\lambda^2+(r+a)(r-2a)(\sigma_1^2+\sigma_2^2)+\nonumber \\
&& +r^2(\Sigma_1^2+\Sigma_2^2)+(r-a)(r+2a)(\nu_1^2+\nu_2^2)\label{spin77}\end{eqnarray}

In general, a $\mathrm{Spin}(7)$ metric on a ($\mathbb{Z}_N$ quotient of a) $\mathrm{spin}^c$ bundle over ${\bf \mathbb{C}P}^2$ can be written in the form (\ref{ansatz}):
\be
ds^2=dt^2+f^2\lambda^2+a^2(\sigma_1^2+\sigma_2^2)
+b^2(\Sigma_1^2+\Sigma_2^2)+c^2(\nu_1^2+\nu_2^2)
\ee
where $a$, $b$, $c$, and $f$ are certain functions of the radial
variable $t$. Other examples of $\mathrm{Spin}(7)$ metrics on such $\mathbb{R}^4/\mathbb{Z}_N$ bundles over ${\bf \mathbb{C}P}^2$ can be found
in \cite{gary2} and also in the recent paper \cite{garychris}. Indeed, in the latter reference, explicit asymptotically locally conical metrics were found for all values of $k$ and $l$, generalising our solution (\ref{spin77}). They also found evidence for the existence of asymptotically conical versions of these solutions, one of which is given by the AC metric on $T^*{\bf \mathbb{C}P}^2/\mathbb{Z}_2$ (\ref{spinorb}).
When the size of the Cayley ${\bf \mathbb{C}P}^2$ goes to zero, these manifolds
develop a conical singularity, and our goal in this section will be
to understand the behaviour of M-theory near such a singularity.
For these purposes, we can often take $X$ to be (asymptotically)
a cone over $Y$.

Following \cite{AW}, let us study the global symmetries of these models.
The classical symmetries of the three-dimensional theory
correspond to geometrical symmetries of $Y$ and gauge
symmetries of the $C$-field, classified by $H^2(Y;U(1))$.
The symmetries which extend to the entire eight-manifold $X$
can be identified with the symmetries of the quantum theory.

First, let us describe the geometrical symmetries.
Since $Y$ can be represented as a quotient space (\ref{ysuthree}),
it is invariant under the left action of $SU(3)$.
Moreover, since the maximal torus of $SU(3)$ is two-dimensional,
there is a $U(1)_K \subset SU(3)$, such that
$U(1)_K$ centralises $T_{k,l}$.
Therefore, $U(1)_K$ is also a symmetry of this model.


Let us now describe the symmetries associated with
the gauge transformations of the $C$-field. The spectral sequence for the fibration $U(1)\hookrightarrow SU(3) \rightarrow Y$ gives:
\be
H^2 (Y; U(1)) = H^1 (U(1); U(1)) = U(1)_C
\ee
Therefore, $H^2 (Y;U(1)) = U(1)_C$ is a classical
symmetry of the three-dimensional $\mathcal{N}=1$
effective field theory. In fact this symmetry is
unbroken in the quantum theory since $X$ is contractible
to ${\bf \mathbb{C}P}^2$ and $H^2 ({\bf \mathbb{C}P}^2; U(1)) = U(1)_C$.

To summarise, the effective three-dimensional
$\mathcal{N}=1$ gauge theory has the following
classical symmetry group:
\be
SU(3) \times U(1)_K \times U(1)_C
\ee
In the case $k=l$ we have an additional discrete $\Sigma_2$ symmetry.

Let us now examine the equivalent D6-brane description of
the same model, discussed in section 3.
Namely, we found that there is a semi-free $U(1)$ action
on $X=\mathcal{Q}$, such that $X/U(1) \cong \mathbb{R}^7$
and the fixed point set is a union of two disconnected
four-manifolds:
\be
L = \mathbb{R}^4 \cup \mathbb{R}^2 \times {\bf S}^2
\subset \mathbb{R}^7
\ee
Therefore, M-theory on the $\mathrm{Spin}(7)$ space $X$ can equivalently be described as a D6-brane on $\mathbb{R}^3 \times L$
in Type IIA string theory. The point in the moduli space
where $X$ develops a singularity corresponds in this
language to the singular configuration of two intersecting
D6-branes on $L = \mathbb{R}^4 \cup \mathbb{R}^4$.
Note that at this point a string stretched between
two connected components of the D6-brane world-volume
is lifted to a membrane wrapped on a topologically
non-trivial two-cycle in M-theory on $X$.

As in the earlier sections, we also find very useful
the Type IIA description of the same system in terms of D6-branes
wrapped on the coassociative four-cycle $B={\bf \mathbb{C}P}^2$ inside
the $G2$-holonomy manifold
\be
M^7 = \Lambda^- {\bf \mathbb{C}P}^2
\ee
The number of D6-branes is determined by the topology of $X$.
Namely, we have:
\be
N = \vert k \vert, \ \vert l \vert, \ \mathrm{or} \ \vert k+l \vert
\ee
depending on how the Aloff-Wallach space $N_{k,l}$ is resolved. In this Type IIA description one can
easily see the spectrum of massless modes.
It suffices to count only bosonic fields since fermionic
modes will complete the ${\cal N}=1$ supermultiplets.
The only bosonic field on the D6-branes is a $U(N)$ gauge field,
which, after reduction on $B$, gives a $U(N)$ gauge field
in three dimensions. Indeed, since $B$ is simply-connected and
$b_2^-(B)=0$ there are no extra bosonic fields in three dimensions.

However, there might be some bulk modes. Specifically,
one might have a $U(1)$ gauge field ${\cal A}_0$ from
the decomposition of the 3-form field $C$ along
the generator $\omega^{(2)}_0$ of $H^2(M^7;\mathbb{Z})$.
Furthermore, there could also be a scalar field from
the deformation of the $\mathrm{Spin}(7)$ metric on $X$.
However, these bulk fields are not dynamical since
the corresponding modes are not $L^2$ normalisable \cite{AW}.
For example, the $L^2$ norm, $\vert \delta g \vert^2$,
of the deformation of the $\mathrm{Spin}(7)$ metric on $\mathcal{Q}$
corresponding to a change in the parameter $a$ is divergent, just as in the $\mathbb{B}_8$ case.

From this description we can also derive the effective
action of the three-dimensional theory.
For simplicity, let us focus on the case $N=1$.
Then, the effective $\mathcal{N}=1$ theory is a $U(1)$ gauge
theory with Chern-Simons coupling of the form (\ref{csviaflux}).
The model can be viewed as a limit $g_1 \to 0$ of the
$U(1)_0 \times U(1)_1$ gauge theory with the following Lagrangian:
\be
L_{3d} = \int d^3 x \Big(
{1 \over 4 g_0^2} \mathcal{F}_0^2
+ {1 \over 4 g_1^2} \mathcal{F}_1^2
\Big)
- \int \Big(
{i k_{00} \over 4 \pi} {\cal A}_0 \wedge d {\cal A}_0
+ {i k_{11} \over 4 \pi} {\cal A}_1 \wedge d {\cal A}_1
+ {i k_{01} \over 4 \pi} {\cal A}_0 \wedge d {\cal A}_1 \Big)
\label{effcpaction}
\ee

The "off-diagonal" Chern-Simons coefficient $k_{01}$
is proportional to the flux of the $U(1)$ gauge field
on the D6-brane through the basic two-cycle
${\bf S}^2 \in H_2({\bf \mathbb{C}P}^2;\mathbb{Z})$, {\it cf.} (\ref{wzterm}):
\be
k_{01} = \int_{{\bf S}^2} {\mathcal{F} \over 2 \pi}
\label{ffluxcptwo}
\ee
As we discussed in section 3, following \cite{wittenfreed},
this flux is half-integer and, in particular, cannot be zero.
Namely, it is convenient to introduce an integer number
$\tilde k_{01} \in \mathbb{Z}$, such that:
\be
k_{01} = \tilde k_{01}+ {1 \over 2}
\ee
This D-brane description suggests that this flux is
also related to the choice of $\mathrm{spin}^c$ structure
on ${\bf \mathbb{C}P}^2$, which in turn is related to the flux
of the RR two-form field strength $G_2$.

As in gravity duals of gauge theories with larger amount of
supersymmetry \cite{KStrassler, Polchinski},
the gauge coupling constant $g_0$ in the Lagrangian
(\ref{effcpaction}) is proportional to the D6-brane tension:
\be
{1 \over g^2} \sim \int_{{\bf \mathbb{C}P}^2} d^4x
\sqrt{- \mathrm{det}(g_{mn} + 2 \pi \alpha' \mathcal{F}_{mn})}
\label{ggaugecptwo}
\ee
Therefore, in the limit $\mathrm{Vol} ({\bf \mathbb{C}P}^2) \to 0$
the value of the coupling constant remains finite
(determined by the $\mathcal{F}$-flux through
${\bf S}^2 \subset {\bf \mathbb{C}P}^2$):
\be
{1 \over g^2} \sim (\tilde k_{01}+ {1 \over 2})^2
\ee
This suggests that, even if the manifold $X$ develops
a geometrical singularity, the dynamics of M-theory on $X$
is still non-singular.

The Chern-Simons couplings $k_{00}$ and $k_{11}$
are determined by the background $G$-flux.
Specifically, $k_{11}$ is given by (\ref{csviaflux}),
whereas $k_{00}$ depends also
on the $G_2$-flux and $\mathcal{F}$-flux through the
${\bf S}^2 \in H_2 ({\bf \mathbb{C}P}^2;\mathbb{Z})$:
\be
k_{00} =k_0 + {1 \over 2}
+ (\tilde k_{01} + {1 \over 2})^2
- (\tilde k_{01} + {1 \over 2}) \cdot
\int_{{\bf S}^2} {G_2 \over 2 \pi}
\ee
Here we follow the normalisation of (\ref{fluxthrub}):
\be
\int_{{\bf \mathbb{C}P}^2} \Big[ {G \over 2 \pi} \Big] = k_0 + {1 \over 2}
\label{fluxthrucp}
\ee

In order to classify the models and, in particular,
to compute the restriction of the $C$-field to $Y=\partial X$,
we need to know $H^4(Y;\mathbb{Z})$. For a general space ${X}_{k,l}$ over
${\bf \mathbb{C}P}^2$ this group is:
\be
H^4(Y;\mathbb{Z}) = \mathbb{Z}_r, \quad
\mathrm{where} \quad r = |k^2 + l^2 + kl|
\ee
In particular, if we restrict ourselves to $N=1$
(corresponding to the set of $\mathrm{spin}^c$ bundles over ${\bf \mathbb{C}P}^2$), we end up with
only one parameter, $p$, which labels the $\mathrm{spin}^c$
structure on ${\bf \mathbb{C}P}^2$.
The corresponding Aloff-Wallach spaces look like
$Y = N_{1,p}$, and $H^4(Y;\mathbb{Z})\cong \mathbb{Z}_{r(p)}$ where
\be
r(p) = 1 + p + p^2
\ee
In particular, in the case $p=0$ we get the $\mathrm{Spin}(7)$
manifold (\ref{spin7}), while in the case $p=1$ we
get the cotangent bundle of ${\bf \mathbb{C}P}^2$ with
$H^4(Y;\mathbb{Z}) \cong \mathbb{Z}_3$.
The vacua in the latter model which preserve $\mathcal{N}=3$
supersymmetry in three dimensions were studied in \cite{GVW}.
This model is related to the corresponding $\mathrm{Spin}(7)$
manifold in question by means of orientifold projection; see appendix B for details.
For general $r(p)$ we find that different models
are classified by an integer $k_0$ mod $r$,
and also by the value of $\Phi_{\infty}$:
\be
\Phi_{\infty} + \chi (X) = {1 \over 2 r} (k_0 + {r \over 2})^2
\label{phicptwo}
\ee
Here we have used the fact that the self-intersection number of
$B={\bf \mathbb{C}P}^2$ is equal to $r$, and also the following
convention for the integral of the $G$-flux, {\it cf.}
(\ref{fluxthrucp}):
\be
\int_{{\bf \mathbb{C}P}^2} \Big[ {G \over 2 \pi} \Big] = k_0 + {r \over 2}
\ee
A consistency check on the formula (\ref{phicptwo})
is that $\Phi_{\infty}$ shifts by an integer when $k_0$
is shifted by a multiple of $r$.

All models with $p=0$ have a unique massive vacuum.
Just like the model based on $X = \mathbb{B}_8$ that
we discussed in the previous subsection,
they are uniquely specified by the value of $\Phi_{\infty}$.

The same is true for most of the models with $p>0$,
except for those where $\Phi_{\infty}$ is related to $r(p)$
as follows:
\be
\Phi_{\infty} = {r \over 8} - \chi (X)
= {1+p+p^2 \over 8} - \chi (X)
\ee
These models have two vacua at $k_0=0$ and $k_0 = - r(p)$.

The value of $\chi(X)$ in the equation (\ref{phicptwo})
can again be computed by comparing this model to Type IIA
theory with a D6-brane wrapped on the coassociative 4-cycle
$B={\bf \mathbb{C}P}^2$.
However, in this case the calculation is more subtle,
since there is a non-trivial flux (\ref{ffluxcptwo}) on the brane,
and one has to use the more general formula (\ref{chiaroof}).
Specifically, from the relation (\ref{chiaroof}) we get:
\be
{\chi (X) \over 24} = {1 \over 48}
\Big( p_1({N}_{{\bf \mathbb{C}P}^2})
- p_1({T}_{{\bf \mathbb{C}P}^2}) \Big)
+ {1 \over 2} (\tilde k_{01} + {1 \over 2})^2
\ee
The first two terms in this expression are computed in
Appendix A. Their total contribution to $\chi(X)$ is $-1/8$.
Note that the second terms looks very similar to
the right-hand side of (\ref{phicptwo}) which suggests
a relation between the $F$-flux on the D6-brane, parametrised by $k_{01}$,
with the $G$-flux in M-theory on $X$, parametrised by $k_0$.
This identification agrees with other arguments we found
in the present section and in section 3.

For large values of $\Phi_{\infty}$ we expect that the model can
be described by M-theory on AdS$_4 \times Y$ where $Y = SU(3)/U(1)$
is the Aloff-Wallach space discussed in section 4.
Therefore, at least for large $\Phi_{\infty}$ we expect our models
to flow to a non-trivial $\mathcal{N}=1$ superconformal fixed point
in the infra-red.

\medskip

\centerline{\bf Acknowledgments}
\noindent

We wish to thank K.~Costello, G.~Gibbons, C.~Herzog, J.~Maldacena,
N.~Nekrasov, C. N\'u\~nez, C.~Pope, E.~Rabinovici, S.~Schafer-Nameki, A.~Strominger,
C.~Vafa, and E.~Witten for useful discussions.
This research was partially conducted during the period S.G.
served as a Clay Mathematics Institute Long-Term Prize Fellow.
The work of S.G. is also supported in part by grant RFBR No. 01-02-17488,
and the Russian President's grant No. 00-15-99296.


\appendix

\renewcommand{\theequation}{\Alph{section}.\arabic{equation}}

\section{Computation of Pontryagin classes}

\label{app:A}

\setcounter{equation}{0}

In this Appendix, we compute the first Pontryagin class of the bundle of anti-self-dual two-forms over $B$, where $B$ is either ${\bf S}^4$ or ${\bf \mathbb{C}P}^2$. This is a fairly standard calculation involving Chern classes. One may find an explanation of most of the tools used here in \cite{bott}.

The key point in the calculation is to note (see, for example, \cite{besse}) that the
complexification of $\Lambda^-\equiv\Lambda^-B$ satisfies

\begin{equation}
\Lambda^-_{\mathbb{C}} \equiv \Lambda^-\otimes \mathbb{C} \cong S^2 \Sigma_-\label{symm}\end{equation}

where $\Sigma_-$ is the spin bundle of $B$, and $S^2E$ denotes the second symmetric power product of the
(complex) vector bundle E. This is just like the exterior power $\Lambda$, but instead we take the symmetrised tensor product, rather than the antisymmetrised
product. Counting dimensions, we see that, according to (\ref{symm})
$\Lambda^-_{\mathbb{C}}$ is a rank three complex vector bundle over $B$,
which is indeed correct.

By definition, we have

\begin{equation}
p_1(\Lambda^-)\equiv -c_2(\Lambda^-_{\mathbb{C}}) = -c_2(S^2
\Sigma_-)\end{equation}

We therefore need to compute the Chern class $c(S^2E)$ for $E$ a complex rank
two vector bundle. This is a fairly standard calculation. It is easy to verify that

\begin{equation}
c(S^2E)=\prod_{1\leq i_1\leq i_2 \leq 2}(1+x_{i_1}+x_{i_2})\end{equation}

where $x_1,x_2$ are the first Chern classes of the line bundle into
which $E$ splits when pulled back to the splitting manifold i.e.

\be
c(E)=\prod_{i=1}^2(1+x_i)\ee

This gives

\begin{equation}
c(S^2E) = (1+2x_1)(1+x_1+x_2)(1+2x_2)\end{equation}

Since this is symmetric in $x_1,x_2$, we must be able to write it in
terms of Chern classes. One obtains

\begin{equation}
c(S^2E) = 1+3c_1 + 2c_1^2 + 4c_2 + 4c_1c_2\end{equation}

and so we pick out

\begin{equation}
c_2(S^2E) = 2c_1^2 + 4c_2\end{equation}

Thus, putting everything together, we get

\begin{equation}
p_1(\Lambda^-) = - (2c_1^2 + 4c_2)\end{equation}

where $c = c(\Sigma_-)$ are the Chern classes of $\Sigma_-$.

We may apply this directly to ${\bf S}^4$. $c_1$ is obviously zero, and

\begin{equation}
c_2(\Sigma_-) = e(\Sigma_{-,{\mathbb{R}}}) = u\end{equation}

where $u$ generates $H^4({\bf S}^4;\mathbb{Z})\cong \mathbb{Z}$, and
so

\begin{equation}
p_1(\Lambda^-{\bf S}^4) = -4u\end{equation}

For ${\bf \mathbb{C}P}^2$, we have to work a little bit harder, since
$\Sigma_-$ doesn't exist. However, we \emph{can} pick a $\mathrm{spin}^c$
bundle

\be
\mathbb{V}_-(L) \equiv \Sigma_-\otimes L^{1/2}\ee

with complex line bundle $L$ which has first Chern
class $c_1(L)=c_1(\mathbb{V}_-(L))=nx$ where $n$ is an odd integer and
$x$ generates the cohomology ring of ${\bf \mathbb{C}P}^2$. Then the symmetric product bundle $S^2\Sigma_-$ \emph{does} exist, and satisfies

\begin{equation}
S^2\Sigma_- \cong L^{-1}\otimes S^2\mathbb{V}_-(L)\end{equation}

We first set $F=\mathbb{V}_-(L)$ so that $c_1(F)=nx$ and
$c_2(F)=x^2+\frac{c_1(L)^2-x^2}{4}$. Thus we have

\begin{eqnarray}
c_1(S^2F) & = & 3c_1(F) = 3nx \nonumber \\
c_2(S^2F) & = & 2c_1(F)^2 + 4c_2(F) = 2n^2x^2 + 4x^2 + (n^2-1)x^2 =
3(n^2+1)x^2 \nonumber \\
c_3(S^2F) & = & 4c_1(F)c_2(F) = 4nx\left(1+\frac{n^2-1}{4}\right)x^2
\end{eqnarray}

Then

\begin{equation}
c(S^2\Sigma_-) = c(L^{-1}\otimes S^2\mathbb{V}_-(L))\end{equation}

\begin{equation}
=\sum_{i=0}^3 c_i(S^2F) (1-nx)^{3-i}\end{equation}

After some algebra, one eventually finds

\begin{equation}
c_2(S^2\Sigma_-) = 3n^2x^2-6n^2x^2+3n^2x^2+3x^2 = 3x^2\end{equation}

The fact that $n$ drops out here is of course a good check on the
calculation, since the ``spin bundle'' $\Sigma_-$ doesn't depend on
$n$. So, we get

\begin{equation}
p_1(\Lambda^-{\bf \mathbb{C}P}^2) = -c_2(S^2\Sigma_-) = -3x^2\end{equation}




\section{Joyce Construction of Spin(7) Manifolds}

\label{app:B}

\setcounter{equation}{0}

In this appendix we explain how a new manifold of $\mathrm{Spin}(7)$ holonomy can be constructed from the cotangent
bundle of ${\bf \mathbb{C}P}^2$, with the Ricci-flat Calabi
metric \cite{calabi}. Our goal here will be to find a suitable
anti-holomorphic involution $\tau$ on $T^* {\bf \mathbb{C}P}^2$,
such that $\tau$ does not have fixed points.
Then, according to \cite{JoyceCY}, the quotient space
$X = T^* {\bf \mathbb{C}P}^2 / \tau$ is a manifold of
$\mathrm{Spin}(7)$ holonomy.

Let us start with the construction of $T^* {\bf \mathbb{C}P}^2$ itself.
We can describe this manifold in terms of two sets of complex
variables $B_i$ and $C_i$ which, respectively, have charges $+1$ and $-1$
under the action of the $U(1)$ symmetry group \cite{GVW}.
Both $B_i$ and $C_i$ transform as the ${\bf 3}$ of the $SU(3)$ global
symmetry group that will also be a symmetry of the quotient
space $X$. Furthermore, $T^* {\bf \mathbb{C}P}^2$ is invariant
under $SU(2)$, under which $(B_i, \bar C_i)$ transform as a doublet
for every $i=1,2,3$.

In these variables the space in question is described by one real
and one complex equation:
\be
\sum_i |B_i|^2 - |C_i|^2 = d_r
\label{sudterm}
\ee
and
\be
\sum_i B_i C_i = d_c
\label{sufterm}
\ee
where $d_r \in \mathbb{R}$ and $d_c \in \mathbb{C}$.
After we divide by the action of $U(1)$, we obtain
an eight-dimensional manifold, $T^* {\bf \mathbb{C}P}^2$,
asymptotic to a cone over $N_{1,1} = SU(3)/U(1)$.

Now, consider the following anti-holomorphic involution:
\be
\tau \colon B_i \to \bar C_i
\ee
which manifestly preserves the $SU(3)$ symmetry group.

The real equation (\ref{sudterm}) is compatible with this
involution only if $d_r=0$, so that we can write it as:
\be
\sum_i |B_i|^2 = \sum |C_i|^2
\label{appbone}
\ee
On the left-hand side of the other equation (\ref{sufterm})
the involution $\tau$ acts by complex conjugation.
Hence, $d_c$ must be real; we denote it simply $d$.
Then, from the equation (\ref{sufterm}), we find:
\be
\sum_i B_i C_i = d
\label{appbtwo}
\ee

To summarise, $X = T^* {\bf \mathbb{C}P}^2 / \tau$
is locally described by two equations
(one real equation (\ref{appbone}) and one complex equation (\ref{appbtwo})),
divided by the action of $U(1)$ and $\tau$.
According to \cite{JoyceCY}, the quotient space $X$
is a (singular) $\mathrm{Spin}(7)$ manifold.
We wish to focus on the simple case where the involution
$\tau$ has no fixed points\footnote{In general, if $\tau$
has isolated fixed points of a suitable kind, one may resolve them following the construction of Joyce \cite{joyce}.}.


The fixed points of the involution are at $B_i = \bar C_i$.
Substituting this into (\ref{appbtwo}) we get:
\be
\sum_i |B_i|^2 = d
\ee
So, the set of fixed points is a copy of ${\bf \mathbb{C}P}^2$
for $d>0$, and an empty set for $d<0$.
We wish to focus on the second possibility,
and examine the topology of the resulting space.
It is convenient to introduce new complex variables
$M_i$ and $N_i$, such that:
\be
B_i = M_i + N_i, \quad
C_i = \bar M_i - \bar N_i
\ee
The new variables $M_i$ and $N_i$ both have charge $+1$
under the action of $U(1)$ and transform in the following
way under the action of $\tau$:
\begin{eqnarray}
M_i & \to M_i \nonumber \\
N_i & \to - N_i
\end{eqnarray}

We can rewrite equation (\ref{appbone}) as:
\be
\sum_i (M_i \bar{N}_i + \bar{N}_i M_i)=0
\label{appba}
\ee
On the other hand, from (\ref{appbtwo}) we get two equations,
corresponding to the real and imaginary parts, respectively:
\be
\sum_i (|M_i|^2 - |N_i|^2)=d,
\label{appbb}
\ee
\be
\sum_i (M_i \bar{N}_i - \bar{N}_i M_i)=0
\label{appbc}
\ee
{}From equations (\ref{appba}) and (\ref{appbc}) we find:
\be
\sum M_i \bar{N}_i = 0
\label{appbd}
\ee
which is very simlar to equation (\ref{appbtwo}).
Together with (\ref{appbb}) it describes the space
with the expected topology, cf. \cite{GVW}. Indeed, since $d$ is
assumed to be negative, one can introduce a new variable:
\be
Z_i = \frac{N_i}{\sqrt{ - d + \sum_i |M_i|^2}}
\ee
which takes values in ${\bf S}^5$.
After dividing by the $U(1)$ symmetry we get
a copy of ${\bf \mathbb{C}P}^2$. In principle,
one should also divide out by the action of $\tau$, which acts
as $Z_i \to - Z_i$. However, on the ${\bf S}^5$ this transformation
is equivalent to a $U(1)$ gauge transformation.
Finally, from equation (\ref{appbd}) it follows that $X$ is
an $\mathbb{R}^4$ bundle over ${\bf \mathbb{C}P}^2$.

One would expect that the new manifold $\mathbb{C}_8$ of
$\mathrm{Spin}(7)$ holonomy found recently in \cite{garychris}
may also be constructed from an $\mathcal{O} (-4)$ bundle
over ${\bf \mathbb{C}P}^3$, after dividing by a suitable
anti-holomorphic involution, as in \cite{JoyceCY}.

\end{document}